\documentclass[12pt,a4paper]{article}           
\pdfoutput=1               
\usepackage[skins,theorems]{tcolorbox}
\tcbset{highlight math style={enhanced,
  colframe=red,colback=white,arc=0pt,boxrule=1pt}}
\usepackage[bookmarksopen, bookmarksnumbered, bookmarksopenlevel=2]{hyperref}
 \usepackage{tikz}
 \usepackage{tikz-3dplot}
 \usetikzlibrary{calc}
 \usetikzlibrary{decorations} %
\usepackage[UKenglish]{babel}
\usepackage[toc,page]{appendix}
\usepackage{amsmath}
\usepackage{amssymb}
\usepackage{graphicx}
\usepackage{hhline}
\usepackage[bf]{caption}
\usepackage{cite}
\usepackage[vcentermath]{youngtab}
\usepackage{geometry}
\usepackage{slashed}
\usepackage{color}
\usepackage{stackrel}
\usepackage{tikz}
\usepackage{tikz-cd} 
\usepackage{empheq}

\DeclarePairedDelimiter\floor{\lfloor}{\rfloor}

\usepackage{multirow}

\newcommand{\bqa}{\begin{eqnarray}}
\newcommand{\eqa}{\end{eqnarray}}

\newcommand{\nn}{\nonumber}

\newcommand{\IR}{\mathbb{R}}

\hypersetup{
    pdftitle={},
    pdfauthor={},
    pdfsubject={}
}
\numberwithin{equation}{section}
\numberwithin{table}{section}

\makeatletter

\renewcommand{\sl}[1]{{\color{black} #1}}

\newtheorem{claim}{Assertion}

\newcommand{\be}{\begin{equation}}
\newcommand{\ee}{\end{equation}}
\newcommand{\beq}{\begin{equation}}
\newcommand{\eeq}{\end{equation}}
\newcommand{\ba}{\begin{aligned}}
\newcommand{\ea}{\end{aligned}}

\newcommand{\bea}{\begin{eqnarray}}
\newcommand{\eea}{\end{eqnarray}}

\newcommand{\cO}{\mathcal{O}}

\newcommand{\cN}{\mathcal{N}}

\newcommand{\cF}{\mathcal{F}}
\newcommand{\cI}{\mathcal{I}}

\def\tr{\mathop{\mathrm{tr}}\nolimits}

\def\unit{{1\kern-.65ex {\rm l}}}
\def\1{{1\kern-.65ex {\rm l}}}

\def\IZ{\mathbb{Z}}
\def\IP{\mathbb{P}}

\newcount\hour \newcount\minute
\hour=\time \divide \hour by 60
\minute=\time
\def\now{%
\ifnum \hour<13
  \ifnum \hour=0 \advance \hour by 12 \number\hour:\else \number\hour:\fi%
     \ifnum \minute<10 0\fi%
     \number\minute%
\ A.M.%
\else \advance \hour by -12 \number\hour:%
  \ifnum \minute<10 0\fi%
  \number\minute%
  \ P.M.%
\fi%
}

\def\fnote#1#2{\begingroup\def\thefootnote{#1}\footnote{#2}
     \addtocounter{footnote}{-1}\endgroup}
     
\def\otherK3{{{\cal K}}   }
\def\Kcone{{\bf K}}
\def\Mcone{{\bf M}}

\makeatother

\begin{document}

\begin{flushright}
{\tt\normalsize CERN-TH-2018-190}\\
\end{flushright}

\vskip 40 pt
\begin{center}
{\huge \bf Tensionless Strings and the \vspace{4mm}\\Weak Gravity Conjecture}

\vskip 7 mm

Seung-Joo Lee${}^{1}$, Wolfgang Lerche${}^{1}$,
and Timo Weigand${}^{1,2}$

\vskip 7 mm

\small ${}^{1}${\it CERN, Theory Department, \\ CH-1211 Geneva 23, Switzerland} \\[3 mm]
\small ${}^{2}${\it Institut f\"ur Theoretische Physik, Ruprecht-Karls-Universit\"at, \\
Philosophenweg 19, 69120 Heidelberg, Germany}

\fnote{}{seung.joo.lee, wolfgang.lerche,  
timo.weigand @cern.ch}

\end{center}

\vskip 7mm

\begin{abstract}
We test various conjectures about quantum gravity for six-dimensional string compactifications
in the framework of F-theory.
Starting with a gauge theory coupled to gravity, we analyze the limit in K\"ahler moduli space where the gauge coupling tends to zero while gravity is kept dynamical. We show that such a limit must be located at infinite distance in the moduli space.
As expected, the low-energy effective theory breaks down in this limit due to 
a tower of charged particles becoming  massless. These are the excitations of an asymptotically tensionless 
string, which is shown to coincide with a critical heterotic string compactified to six dimensions. 

For a more quantitative analysis, we focus on a $U(1)$ gauge symmetry and use a chain of dualities and mirror symmetry to determine the elliptic genus of the nearly tensionless string, which is given in terms
of certain meromorphic weak Jacobi forms.
 Their modular properties in turn allow us to determine the charge-to-mass ratios of certain string excitations near the tensionless limit. We then provide evidence that the tower of asymptotically massless charged states satisfies the (sub-)Lattice Weak Gravity 
Conjecture, the Completeness Conjecture, and the Swampland Distance Conjecture.
Quite remarkably, we find that the number theoretic properties of the elliptic genus conspire with the balance of gravitational and scalar forces of extremal black holes, such as to produce a narrowly tuned charge spectrum of superextremal states.
As a byproduct, we show how to compute elliptic genera of both critical and non-critical strings, when refined by Mordell-Weil $U(1)$ symmetries in F-theory.
\end{abstract}

\vfill

\thispagestyle{empty}
\setcounter{page}{0}
\newpage

\tableofcontents

%%%%%

\vfill\eject

%----------------------------------------------------------------------------------------------------------
\section{Introduction and Summary}

Quantum gravitational effects are deeply woven into the fabric of string theory. Effective field theories that derive from string theory should reflect these, and in particular must be consistent with any consistency constraints that quantum gravity may impose. 
String theory is therefore an ideal framework to investigate and exemplify general ideas about quantum gravity in a reasonably well-controlled and computable 
setting. The dichotomy between theories which can or which cannot be coupled to gravity manifests itself in the distinction between the so-called landscape of consistent vacua, and the swampland of effective theories without a UV completion  \cite{Vafa:2005ui}.
Various, mostly conjectured criteria characterize the boundary between both regions in theory space \cite{Brennan:2017rbf}. Since this boundary is still under considerable debate, it is important to use the opportunity of confronting such conjectures with as quantitative an analysis as possible in the context of string compactifications.
 
In this work we address the interplay between gauge and gravitational effects in what one may call the open string sector, by analyzing gauge symmetries under which the matter fields are charged. 
Realizing such a setup in F-theory allows us to relate general arguments about constraints of quantum gravity  to questions in algebraic and enumerative geometry,  number theoretic properties of modular forms, and conformal field theory.
This includes applying the theory of weak Jacobi forms to state counting via the elliptic genus, as well as computations  in topological strings and mirror symmetry.

Among the earliest conjectures on general quantum gravity properties is that  continuous global symmetries should not exist in the presence of gravity, as pointed out in the context of perturbative string theory in \cite{Banks:1988yz}. Rather, what appears as a global symmetry at low energies should derive from a gauge theory in the UV. This is motivated e.g.~by  the physics of quantum black holes: Global continuous symmetries would lead to charged stable remnants that would exceed certain covariant entropy bounds \cite{Banks:2010zn}.

More generally, according to the Weak Gravity Conjecture \cite{ArkaniHamed:2006dz}, gravity should be the weakest force at any point in moduli space. Starting from a gauge theory coupled to gravity and taking a limit in the stringy moduli space where the gauge symmetry becomes a global symmetry  should therefore be prohibited within the landscape of consistent theories.  
At a macroscopic level, this means that the global symmetry limit must lie at an infinite distance in the moduli space. 
Microscopically, the effective field theory should break down as the problematic limit is approached.

This  in turn ties in beautifully with another general expectation about the structure of quantum gravity, which can be summarized in the Swampland Distance Conjecture \cite{Ooguri:2006in} as follows: The effective field theory formulated at a given point in moduli space is  well-defined only within a finite distance from this reference point. As we travel an infinite distance in moduli space, a tower of infinitely many states should become massless at an exponentially fast rate. This leads to a breakdown of the effective field theory and, in turn, explains why such points at infinite distance cannot be attained within the effective field theory \cite{Heidenreich:2017sim, Grimm:2018ohb}. In combination with the rationale that no points with global symmetries can exist  in the presence of gravity,  this suggests the appearance of infinitely many {\it charged} states that become massless in the limit where a gauge symmetry becomes global.

This reasoning has been recently verified in the context of Type IIB compactifications to four dimensions with eight supercharges \cite{Grimm:2018ohb}: As infinite distance points in complex structure moduli space are approached, a tower of BPS states from D3-branes wrapping vanishing 3-cycles becomes massless. These states are charged under abelian Ramond-Ramond gauge symmetries which indeed become global symmetries at infinite distance.

If true, the Weak Gravity and Swampland Distance Conjectures would have far-reaching  consequences for cosmology and phenomenology, which, together with various refinements, have been the subject of intensive investigations in the recent literature, including \cite{Brown:2015iha,Brown:2015lia,Bachlechner:2015qja,Hebecker:2015rya,Hebecker:2015zss,Ibanez:2015fcv,Rudelius:2015xta,Heidenreich:2015wga,Heidenreich:2015nta,Baume:2016psm,Montero:2016tif,Heidenreich:2016jrl,Klaewer:2016kiy,Cottrell:2016bty,Hebecker:2017uix,Hamada:2017yji,Palti:2017elp,Heidenreich:2017sim,Lust:2017wrl,Valenzuela:2017bvg,Blumenhagen:2017cxt,Ibanez:2017vfl,Andriolo:2018lvp,Heidenreich:2018kpg,Aldazabal:2018nsj,Gonzalo:2018dxi,Blumenhagen:2018nts,Blumenhagen:2018hsh}.

In this article we initiate a systematic investigation of quantum gravity constraints for compactifications where the gauge symmetries are localised on branes,
in the framework of 6-dimensional F-theory \cite{Vafa:1996xn,Morrison:1996na,Morrison:1996pp}. 
We will address the Swampland Distance Conjecture, the Weak Gravity Conjecture and its refinement as a (Sub)Lattice Weak Gravity Conjecture \cite{Heidenreich:2015nta,Heidenreich:2016aqi} as well as  the Completeness Hypothesis of \cite{Polchinski:2003bq}.

Our starting point is the question whether it is possible to go to a point in moduli space where an (`open string') gauge symmetry becomes a global symmetry while gravity is kept dynamical, or rather what prevents us from realising such a limit.
This question turns out to have an answer directly from the compactification geometry. Recall that the 6-dimensional Planck mass is set by the volume of the F-theory compactification space.  This is a complex K\"ahler surface, $B_2$, which serves as the base of an elliptically fibered Calabi-Yau 3-fold, $Y_3$.
The open string gauge symmetry, on the other hand, arises from 7-branes that wrap holomorphic curves $C$ on $B_2$. Thus we have
\be
M^4_{\rm Pl} \sim {\rm vol}_J(B_2) \,,\quad \quad \quad \frac{1}{g^2_{\rm YM}} \sim {\rm vol}_J(C) \,,
\ee
where $J$ denotes a chosen K\"ahler form with respect to which the volumes are measured. 
In the special case of abelian gauge groups that we will sometimes restrict our discussion to, $C$ has a description in terms of the so-called height-pairing of a rational section of the elliptic fibration over $B_2$. 
The critical limit we are interested in thus corresponds to the limit
\be \label{limitintro1}
{\rm vol}_J(C) \to \infty  \quad \text{ with} \quad  {\rm vol}_J(B_2) \,\text{ finite}. 
\ee
Our first result is that this necessarily implies that $B_2$ must contain a rational curve $C_0$ whose volume goes to zero as ${\rm vol}_J(C) \to \infty$. The proof we provide makes use of various properties of K\"ahler geometry including in particular Mori's cone theorem. As we will see, the limit (\ref{limitintro1}) manifestly lies at infinite distance in the K\"ahler moduli space, which realizes one of the general expectations spelled out above. Technically, this means that the curve $C_0$ is not contractible in the sense of algebraic geometry, which would require the self-intersection of a contractible curve to be strictly negative. By contrast, the curve $C_0$ whose volume vanishes asymptotically is shown to have zero self-intersection. 

A D3-brane wrapping the curve $C_0$ gives rise to a solitonic string in the uncompactified six dimensions. Asymptotically, 
its tension $T$ is controlled by
\be
T \sim {\rm vol}_J(C_0)  \to\  0 \,. 
\ee
As it turns out, $C_0$ necessarily intersects the curve $C$ and hence the strings associated to the $C_0$  are charged under the 7-brane gauge group. A tower of light charged states thus appear in the effective theory, whose masses vanish exponentially fast as we approach the limit  $g_{\rm YM} \to 0$. 
This signals the breakdown of the effective theory,  which in a sense provides a microscopic censor that forbids the appearance of a global symmetry.

We can be considerably more quantitative. Since the curve $C_0$ is always a rational curve of vanishing self-intersection, its zero-mode structure \cite{Haghighat:2015ega,Lawrie:2016axq} coincides with the zero mode structure of a heterotic string.
Near the tensionless limit, the string in fact becomes identical to the familiar, critical heterotic string compactified to six dimensions.   Note that the physics of the nearly tensionless heterotic string is fundamentally different from the non-critical strings that arise from  curves with negative self-intersection.  These become tensionless at superconformal points of  6d $N=(1,0)$ gauge/tensor theories in the absence of gravity \cite{Seiberg:1996vs,Witten:1996qb} (see the recent review \cite{Heckman:2018jxk} for more details). The critical heterotic strings  we consider include gravity, however, and may also be weakly coupled.

The appearance of a weakly coupled heterotic string is evident when the base, $B_2$, of the elliptically fibered Calabi-Yau 3-fold, $Y_3$, is a Hirzebruch surface. In this case the 6d F-theory vacuum is known \cite{Morrison:1996na,Morrison:1996pp} to be dual to the perturbative heterotic string compactified on a $K3$ surface. The rational fiber of the Hirzebruch surface plays the role of the curve $C_0$, and the tensionless limit coincides with the heterotic weak coupling limit \cite{Morrison:1996na}.

More generally, we will see that in order for the limit (\ref{limitintro1}) in K\"ahler moduli space to exist, the elliptic fibration $Y_3$ must  always 
admit a $K3$ fibration, whose $K3$ fiber class $\widehat C_0$ is exactly the restriction of the elliptic fibration to $C_0$.
The D3-brane that wraps $C_0$ maps to an M5-brane along $\widehat C_0$ in the dual M-theory. Upon $S^1$-reduction, which takes us to the dual IIA string on $Y_3$,  the M5-brane turns into an NS5-brane wrapped on the $K3$ surface $\widehat C_0$. As an NS5-brane on $K3$ is known to be dual to the heterotic string in 6d \cite{Harvey:1995rn}, adiabatically fibering it over the base of the $K3$ fibration again identifies the string in question with the heterotic one, albeit compactified in a less standard way (see the recent work \cite{Braun:2016sks} for an account of that more general Type IIA/heterotic duality). 

The appearance of a potentially weakly coupled, critical heterotic string from a D3-brane wrapped on a shrinking curve $C_0$  allows us to determine the charge and mass spectrum in some detail, at least near that limit.  However, even for configurations where
the heterotic string is non-perturbative and not necessarily weakly coupled,  
estimates for the spectrum can be made by considering its elliptic genus. Such reasoning is quite analogous to using the elliptic genus to determine properties of
non-critical strings in 6 dimensions. This has been a rich and beautiful topic, pioneered in \cite{Klemm:1996hh} and with dramatic progress also in the recent literature, including \cite{Klemm:2012sx,Haghighat:2013gba,Haghighat:2014pva,Haghighat:2014vxa,Huang:2015sta,Haghighat:2015ega,Kim:2016foj,DelZotto:2016pvm,DelZotto:2017mee,Gu:2017ccq,Kim:2018gak,DelZotto:2018tcj} (see e.g. the review \cite{Heckman:2018jxk} for more references). 

The elliptic genus is defined as a weighted trace over the BPS spectrum of winding and Kaluza-Klein momentum modes of the string wrapped on $S^1$, and counts the index of BPS particles in the dual 5d M-theory.
In the limit of infinite radius, it reduces to a trace over the left-moving excitations of the 6d string. As a consequence, their partition function is meromorphic and thus
highly constrained by modular invariance.  This fact remains true in the presence of background gauge fields whose field strengths enter 
the partition function as `fugacities'. The corresponding modular forms are weak Jacobi forms, which are characterized by a ``fugacity index''
besides their modular weight. 
The fugacity index, in turn, is related to the anomaly polynomial of the string \cite{DelZotto:2016pvm,DelZotto:2017mee}.  The interplay of modular invariance and anomalies traces back to the earliest applications \cite{Schellekens:1986yi,Schellekens:1986xh,Lerche:1987qk} of the elliptic genus in string theory.

A new result that we will find is that for abelian flavour groups, the fugacity index is given by the intersection of the shrinking curve $C_0$ with the height-pairing of the rational section that is responsible for the appearance of the abelian symmetry. With this input one can then make, for any concrete model, an ansatz for the elliptic genus with a few free parameters. These can in turn be fixed by using a non-trivial duality \cite{Haghighat:2013gba,Haghighat:2014vxa} which relates the
elliptic genus to the topological string theory on $Y_3$.  

For given examples, this allows us to determine a characteristic part of the charge/mass spectrum of the nearly tensionless heterotic string explicitly, and to address questions related to the various Quantum Gravity conjectures mentioned above.
However, even without explicitly determining the elliptic genus for concrete examples, we can draw some important conclusions just from its general properties. These include the following:
\begin{enumerate}
\item
The excitation spectrum of the string contains states for every allowed value of the charge under the abelian gauge group. In this sense the full charge lattice is populated, in agreement with the Completeness Hypothesis of \cite{Polchinski:2003bq}. While this is a rather trivial fact for perturbative heterotic strings, 
it seems less so for non-perturbative variants for which no partition function is known. For example, in F-theory
the spectrum of open string excitations in the 7-brane sector alone would spectacularly fail to populate the full charge lattice. Indeed, at the massless level only a few charges are realized explicitly. Their stringy or Kaluza-Klein excitations do not carry higher charges either because the charge is set by the Chan-Paton factors of the open string. 
\item
At least in the limit of vanishing tension, the states of maximal charge per string excitation level satisfy the Sublattice Weak Gravity Conjecture bound of \cite{Heidenreich:2016aqi}, i.e. the charge-to-mass ratio exceeds that of an extremal (non-BPS) Reissner-Nordstr\"om black hole in six dimensions. This condition is required in order for non-BPS extremal black holes to decay, which in turn was argued \cite{ArkaniHamed:2006dz} to be a necessary property of any theory of quantum gravity  to evade various entropy bounds.
\end{enumerate}

The conclusions we draw concerning the censorship against global symmetries or concerning the Weak Gravity Conjecture can be read either as statements about six-dimensional F-theory vacua, or as analogous properties of the six-dimensional heterotic string with $N=(1,0)$ supersymmetry as such, even without any reference to an F-theory dual.

There are, however, a few caveats to the above reasoning and we therefore need to spell out our working assumptions concerning them.
Foremost is the absence of a BPS property of particle excitations in $(1,0)$ supersymmetric 6d theories. Rather what is BPS protected are strings. Their BPS nature allows one to infer the appearance of asymptotically tensionless strings in certain limits of the moduli space. It is natural to expect that their
particle excitations are gapped in terms of the tension, but strictly speaking there is little control of the mass spectrum.  Fortunately some of the geometries lead to
weakly coupled, perturbative heterotic strings, for which one can make quantitative statements at least in the asymptotic limit of vanishing coupling. For strongly coupled versions the only tool is, as mentioned above, the elliptic genus, but since it does not involve BPS protected states in 6 dimensions, it is not entirely clear what conclusions can be rigorously drawn. We assume, as our working hypothesis, that the statements about charge-to-mass relationships inferred from the elliptic genus remain at least qualitatively valid. The same applies also to the extremal black hole solutions with which masses and charges of the string states are to be compared, as these are not protected by a BPS property either.

This article is organized as follows:
In Section~\ref{sec_FonY3} we begin with a brief recapitulation of 6d F-theory vacua, mainly in order to set up our notation.
In Section~\ref{sec_Globalasalimit} we characterize the geometric limit (\ref{limitintro1}) in which a generic 7-brane-supported gauge symmetry becomes a global symmetry in the presence of gravity.
The technical details of the derivation are relegated to Appendix \ref{app_Jproof}. As we will see in Section~\ref{sec_infinitedistance}, the limit lies, quite as expected,
 at infinite distance in K\"ahler moduli space. In Section~\ref{sec_tensionlessstrings} 
we explain the appearance of a critical 6d heterotic string that becomes tensionless as the 7-brane gauge group becomes global.
The main properties  of the charge spectrum of this heterotic string are summarized in Section~\ref{sec_chargesummary}. The results of this section will be derived in the technical Section~\ref{sec_Ellgenus}, but we present the main points already at this stage in order to give the casual reader a quick overview. In Section~\ref{sec_ReltoQG} we put these findings into perspective  
with various conjectured quantum gravity constraints. 
Apart from verifying at least asymptotically the Sublattice Weak Gravity Conjecture, we give evidence that integrating out the tower of charged states reproduces the vanishing of the gauge coupling in the tensionless limit \cite{Heidenreich:2017sim,Grimm:2018ohb}.

The technical Section \ref{sec_Ellgenus} contains a detailed analysis of the tensionless heterotic string with regard to its elliptic genus.
In Section~\ref{sec_ellgenus1} we explain how to infer from the elliptic genus of the perturbative heterotic string the appearance of (a characteristic subset of the) physical charged states that become light in the tensionless limit.
We then review some basic properties of the elliptic genus, its relation to weak Jacobi forms and to the topological string in Section~\ref{sec_ellgenus}. The upshot of this discussion is to 
relate the elliptic genus of the heterotic string to the genus-zero free energy of the topological string on the F-theory 3-fold $Y_3$.
This allows us to obtain results also for
  situations where $Y_3$ is not a fibration over a Hirzebruch surface, in which case the nearly tensionless heterotic string is not necessarily weakly coupled.

In Section~\ref{subsec_GlobalMWU1} we provide a new result, namely the computation of the $U(1)$ fugacity index via the height pairing of a rational section in F-theory. Subsequently,
the general strategy to determine the charge-to-mass ratio of excited states via the elliptic genus is summarized in Section~\ref{sec_genstratmaxcharge}; a proof of some underlying mathematical facts is presented in Appendix~\ref{app_EJmaxcharge}. 

In Section~\ref{Ellgenusexamples} we apply the general machinery of Section~\ref{sec_Ellgenus} to various explicit 
examples of elliptic fibrations with a non-trivial Mordell-Weil group; these correspond to F-theory compactifications with a single unbroken $U(1)$ gauge symmetry.  Apart from being interesting by themselves, these computations serve as a check of the predicted charge-to-mass ratio of the string excitations as encoded in the elliptic genus. More specifically, 
in Section~\ref{subsec_F1base} we present models on  Hirzebruch bases and recover a $U(1)$-refined version of the familiar, conformal field theoretic heterotic elliptic genus on $K3$ \cite{Schellekens:1986yi,Schellekens:1986xh,Eguchi:1988vra,Kawai:1993jk,Kawai:1996te,Gritsenko:1999fk}. In Section~\ref{sec_dP2example} we exemplify the more general situation where the dual heterotic string theory involves additional tensor fields and thus is non-perturbative. The elliptic genus then differs from the familiar perturbative version, but is consistent with a tensor transition to a model on a more conventional Hirzebruch base. Details of the computation are relegated to Appendix \ref{sec_HirzebruchF2}.
 
In Section~\ref{sec_outlook} we conclude with an outlook and  some speculations.

\vfill\eject
%--------------------------------------------------------
\section{Global Symmetries and Their Geometric Realizations} \label{sec_Globalsymm}

In this section we will analyze the ``ungauging'' of a local gauge symmetry in 6d F-theory compactifications, while keeping gravity dynamical.
We begin with an analysis of the geometric realisation of this process as a very special limit in K\"ahler moduli space, and then study its physical implications.

\goodbreak
\subsection{F-theory on elliptic Calabi-Yau 3-folds} \label{sec_FonY3}

We will be working in the framework of F-theory \cite{Vafa:1996xn,Morrison:1996na,Morrison:1996pp}
 compactified on a Calabi-Yau 3-fold $Y_3$ which is elliptically fibered over a compact K\"ahler surface $B_2$,
\bea \label{ellfibrationpi}
\pi :\quad \mathbb{E}_\tau \ \rightarrow & \  \ Y_{3} \cr 
& \ \ \downarrow \cr 
& \ \  B_2 
\eea
The effective theory in six dimensions is a 6d $N=(1,0)$ supergravity theory.  Background on F-theory in general and its compactifications to six dimensions in particular can be found e.g.~in the reviews \cite{Denef:2008wq,Taylor:2011wt,Weigand:2018rez}.  In units where the string length is $\ell_s = 1$, the Planck mass in the 6d Einstein frame is determined by the volume of the base surface $B_2$ as follows:
\beq \label{Mpl}
M_{\rm Pl}^{4}  = 4 \pi\, {\rm vol}_J (B_2)\,.
\eeq
Throughout this article the value of $M_{\rm Pl}$ will be kept fixed
by normalizing the volume of the base $B_2$ as 
\beq\label{norm}
{\rm vol}_J(B_2) = 1\,.
\eeq
The volume is defined with respect to a K\"ahler form $J$. A summary of our conventions for the effective action in 6d is provided in Appendix \ref{App_DimRed}.

Non-trivial gauge symmetries arise essentially along the world-volume of 7-branes that wrap holomorphic divisors of the base $B_2$.
A non-abelian gauge symmetry algebra $\mathfrak{g}_I$ is localised on a stack of 7-branes that wrap a divisor $\Sigma_I$, which is in itself a component of the discriminant divisor of the elliptic fibration. More specifically, suppose we describe the elliptic fibration $Y_3$ by a Weierstrass model
\be
y^2 =  x^3 + f x z^4 + g z^6 \,,
\ee
where $[x : y : z]$ denote the homogeneous coordinates on the fiber ambient space, $\mathbb P_{231}$. 
If the elliptic fibration is to be Calabi-Yau, we must require that 
\be \label{fgclasses}
f \in H^0(B_2, \cO_{B_2}(4\bar K))\,, \qquad \quad g \in H^0(B_2, \cO_{B_2}(6 \bar K))
\ee
in terms of the anti-canonical divisor $\bar K$ on the base. In order for these sections to exist, $\bar K$ must be an effective class on $B_2$.
The vanishing locus of the discriminant 
\be
\Delta = 4 f^3 + 27 g^2
\ee
of the elliptic fibration is a divisor 
\be \label{Sigmadiv}
\Sigma = \Sigma_0  \, \cup \,  \bigcup_I  \Sigma_I \,.
\ee
Then to each component $\Sigma_I$ we associate a non-abelian gauge algebra $\mathfrak{g}_I$, while the residual discriminant divisor $\Sigma_0$
carries a trivial gauge algebra. For later purposes note that as a consequence of (\ref{fgclasses}) the divisor class of $\Sigma$ is
\be \label{Sigmaclass}
[\Sigma] = 12 \, \bar K \,.
\ee
On the other hand, {\it abelian} gauge symmetries are generated by non-trivial rational sections $S$ of the fibration $Y_3$ \cite{Morrison:1996na}, more precisely by the image of such a section under the Shioda homomorphism  \cite{Grimm:2010ez,Park:2011ji,Morrison:2012ei}
\be \label{Shioda-homo}
\sigma(S) = S - S_0 - \frak{D} \,.
\ee
Here $S_0$ is the zero-section of the fibration and $\frak{D} \in \pi^{*}H^2(B_2)$ is a base divisor class which is chosen such that $\sigma(S)$ satisfies certain transversality conditions.
In absence of additional non-abelian gauge algebra factors, the divisor class $\mathfrak{D}$ is given as
\be\label{DinSigma}
\frak{D} = \pi^{-1} (\pi_\ast((S -S_0)\cdot S_0)) \,,
\ee
which we will use later in Section~\ref{Ellgenusexamples} when analyzing explicit models. 
More generally, in presence of non-abelian algebras, $\frak{D}$ contains in addition a linear combination of the blow-up divisors that resolve the singularities in the fiber. For more information we refer to the summary in Appendix A of \cite{Lee:2018ihr} and references therein.

What is important for us is the fact that the gauge coupling $g_{\rm YM}$ associated  with both a non-abelian or an abelian gauge symmetry is determined by the K\"ahler volume of a certain divisor on $B_2$,
\be \label{gaugecoupling}
\frac{1}{g_{\rm YM}^2}  = \frac{1}{2\pi} \,  {\rm vol}_J(C) \,,
\ee
where\footnote{We are normalizing the gauge coupling as spelled out in Appendix \ref{App_DimRed}.   
}
\bea \label{Cdefcases}
C =  \begin{cases}  \,   \Sigma_I   \qquad & \text{if} \, \, \,   \mathfrak{g} = \mathfrak{g}_I               \\
                               \,   b: = - \pi_\ast(\sigma(S) \cdot \sigma(S))         \qquad  &   \text{if} \, \,  \,   \mathfrak{g} = \mathfrak{u}(1)    \end{cases}
\eea
The quantity $b$ is called the height pairing of the section $S$, and in absence of additional non-abelian gauge group factors it is given by
\be \label{bAexpression}
b = 2 \bar K + 2 \pi_\ast(S \cdot S_0) \,.
\ee
In the presence of several abelian gauge group factors, mixing can occur between the kinetic terms of the individual field strengths, which makes the definition of the gauge coupling more complicated. In such a situation, $b$ is the diagonal part of the gauge kinetic function. We will restrict ourselves, for simplicity, to situations with a single abelian gauge group factor to avoid the need to diagonalise the gauge kinetic function. 

Massless charged matter fields in the 6d $N=(1,0)$ supergravity theory arise from strings stretched between the 7-branes. Note that these matter fields transform under a finite number of irreducible representations, or have a finite number of $U(1)$ charges, respectively.

%-----------------------------------------------------------
\subsection{Global symmetries as a limit in K\"ahler moduli space} \label{sec_Globalasalimit}

Let  us suppose that the  coupling of a certain gauge group factor, be it a $U(1)$  or a simple non-abelian one, is parametrically smaller than the gravitational coupling, that is,
\beq
g_{\rm YM} \ll M_{\rm Pl}^{-1}\,.
\eeq
In view of (\ref{Mpl}) and (\ref{gaugecoupling}), this can be rephrased in geometric terms as
\beq\label{lim}
{\rm vol}_J (C) \gg ({\rm vol}_J(B_2))^{1/2} \,.
\eeq
As we will show, in such a geometric setup, there must arise an infinite tower of charged particles which become asymptotically massless. 
This is because the limit (\ref{lim}) drives the base $B_2$ to a very degenerate regime in K\"ahler moduli space, where the K\"ahler form $J$ approaches the boundary of the K\"ahler cone.
More precisely, we are interested in the regime where 
\be \label{limit-form1}
{\rm vol}_J (C) \sim t \to \infty \,, \qquad    {\rm vol}_J(B_2) \,\,  \text{finite} \,.
\ee
This forces the base $B_2$ to become very anisotropic in the sense that it grows `long and thin', as specified in technical terms below. 
In order for the base volume to remain finite while the volume of the curve $C$ goes to  infinity, there must exist another curve $C_0$ which intersects $C$ and whose volume goes to zero as $t \to \infty$.
This happens in such a way that the product of the volume of $C$ and $C_0$ must lie below a finite upper limit.
The infinitely many particles which become massless as $t \to \infty$ are the excitations of an effective string obtained by wrapping a D3-brane along $C_0$.  As we will see, the properties of this string depend crucially on the fact that the curve $C_0$ has vanishing self-intersection $C_0 \cdot C_0  =0$. This implies that the elliptic fibration whose base allows for a limit (\ref{limit-form1}) necessarily exhibits a $K3$-fibration.

While the properties of the string associated with $C_0$ are the subject of the subsequent sections, we will first analyze this degenerate limit of the base $B_2$ in more detail. 
To this end, let us formulate our above findings concerning the limit in the K\"ahler cone in more precise terms:
\begin{enumerate}
\item
Our first key result is that in the limit (\ref{limit-form1}), with the finite volume of $B_2$ normalised to~$1$, the K\"ahler form $J$ of the base $B_2$ must take the universal form 
\bea \label{baseK}
J = t J_0 +  \sum_{\nu} s_\nu I_\nu  \,, \ \ \   {\rm as}\ \   t \to \infty\,.
\eea
Here $J_0 \in H^{1,1}(B_2, \IZ)$ is a K\"ahler cone generator satisfying
\be \label{J0J0zero}
J_0 \cdot J_0 = 0 \,, \qquad 2m:= \int_C J_0 \geq 1 \,.
\ee
The remaining K\"ahler cone generators  $I_\nu \in H^{1,1}(B_2, \IZ)$ have  the property that
\be \label{nnusbound}
 \sum_\nu n_{ \nu} s_\nu   \to \sl{\frac{1}{t}}  \qquad  \text{as} \quad  t\to \infty    \,,
 \ee
where 
\be
n_{ \nu} = J_0 \cdot I_\nu 
\ee
and at least one $n_{ \nu} \neq 0$. 
The non-negative expansion parameters $s_\nu$ stay finite as $t \to \infty$ and are chosen such that ${\rm vol}_J(B_2) =1$.

To summarize, the K\"ahler form asymptotes to the direction of one K\"ahler cone generator, $J_0$, and the fact that $\int_C J_0 \geq 1$ guarantees that the volume of $C$ goes to infinity as $t \to \infty$.
This happens in a very special way such that the total volume of $B_2$ remains finite.

\item
The second key result is that there exists a rational curve $C_0$ which intersects $C$ and whose volume goes to zero in the limit (\ref{limit-form1})  as 
\be\label{volC0<1/2t}
{\rm vol}_J (C_0) \to \sl{\frac{1}{t}} \,.
\ee
It is of crucial importance that its self-intersection vanishes, $C_0 \cdot C_0=0$.
This  leads to the behaviour 
\beq\label{geom-bound}
{\rm vol}_J(C) \,{\rm vol}_J(C_0) \to \sl{2m} + \sl{\frac{s}{t}} 
\eeq
for the product of the volumes of the two curves, in the limit where $C$ becomes infinitely large while $B_2$ is kept finite. 
Here 
\beq
s:=\sum_\nu s_\nu \int_{C} I_\nu 
\eeq 
is a non-negative real number which remains finite in the limit.
\item
The curve $C_0$ is the only curve whose volume vanishes in the limit  with the property that $C_0 \cdot C_0 = 0$. All other vanishing curves have negative self-intersection.
\item
Whenever the base admits a limit of the form (\ref{lim}), the elliptic fibration over it admits the structure of a $K3$ fibration,
\bea \label{K3fibrrho_sec2.2}
\rho :\quad K3 \ \rightarrow & \  \ Y_{3} \cr 
& \ \ \downarrow \cr 
& \ \  C'
\eea
over some curve, $C'$. The class of the K3-fiber is 
\be\label{K3-fiber_sec2.2}
\widehat C_0 := \pi^{-1}(C_0) \in H_4(Y_3) \,,
\ee
i.e. the restriction of the elliptic fibration (\ref{ellfibrationpi}) to $C_0$.
In general, this $K3$ fibration is not compatible with the elliptic fibration (\ref{ellfibrationpi}).
\end{enumerate}
In Appendix \ref{app_Jproof} we derive these four results (Assertions 1 -- 4). The busy reader not interested in the technical details is invited to skip the derivation on a first reading.

\subsection{The global limit as a point at infinite distance}\label{sec_infinitedistance}

As explained in the previous section, we are interested in taking a limit in K\"ahler moduli space in which a self-intersection zero curve $C_0$ shrinks to zero volume. 
Before analyzing in more detail the physical consequences of this limit in the next section, note that on a surface $B_2$ a curve of non-negative self-intersection cannot be contractible in the sense of algebraic geometry. 
In fact, the limit in K\"ahler moduli space in which $C_0$
 shrinks to zero volume is strikingly different from the limit where a {\it contractible} curve assumes vanishing volume. In the latter case, we can take the curve volume to zero without any other curve on $B_2$ acquiring infinite volume. The result of such a contraction is a canonical singularity on $B_2$, which, for F-theory bases, is of the local form \cite{Heckman:2013pva} $\mathbb C^2/G$ with $G$ a discrete subgroup of $U(2)$. 
If we view the degeneration leading to a canonical singularity in complex structure moduli space, then these occur at finite distance with respect to the Weil-Petersson metric \cite{Wang97}. 
 The degeneration of $B_2$ in the present context, however, is much more severe. In particular, $C_0$ cannot be considered in isolation because we have shown that there must exist another curve $C$ with $C \cdot C_0 \neq 0$ with the property
\be
{\rm vol}_J(C) \to \infty  \qquad {\rm as} \quad {\rm vol}_J(C_0) \to 0 \,. 
\ee 
 In fact, as we show now, the point $t \rightarrow \infty$ in (\ref{baseK}) lies at infinite distance in the K\"ahler moduli space.\footnote{The following computation was worked out in collaboration with Diego Regalado.}  
 
 To see this, let us consider a basis $\omega_\alpha$, $\alpha = 0, \ldots, n_T$ of   $H^{1,1}(B_2)$. We recall first that the intersection form $\Omega_{\alpha \beta} = \int_{B_2} \omega_\alpha \wedge \omega_\beta$ on the compact K\"ahler surface has signature $SO(1,n_T)$ and can be assumed to take the form 
 \be
 \Omega_{\alpha\beta}={\rm diag}(+1,-1,\dots,-1) \,,
 \ee
 upon an appropriate diagonalization. 
 In physics terms, $n_T = h^{1,1}(B_2)-1$ is the number of tensor multiplets in the 6d $N=(1,0)$ F-theory vacuum. Any K\"ahler form $J$ normalised  such that {$\frac{1}{2}\int_{B_2} J^2  =1$} can therefore be expanded in the diagonal basis as
 \be\label{Jtwo}
 J = \sl{\sqrt{2}}j^\alpha \omega_\alpha \,.
 \ee 
 Thus, we may parameterise the coordinates on the hypersurface $\mathcal M_T$ of constant six-dimensional Planck mass, namely,
\be
j\cdot j = j^\alpha\Omega_{\alpha\beta}j^\beta = (j^0)^2 - \sum_{i=1}^{n_T}(j^i)^2 =1\,,
\ee
as
\be\label{emb}
j^0 = {\rm cosh}x \, \qquad j^i =  {\rm sinh}x \,  u^i(\phi^1, \ldots, \phi^{n_T-1})\,.
\ee 
Here $x$ takes any real value and the functions $u^i(\phi^1, \ldots, \phi^{n_T-1})$ are chosen such that $\sum_i  (u^i)^2 = 1$. Hence the coordinates $\phi^A$ ($A=1, \ldots, n_T-1$) parameterise the unit $(n_T-1)$-sphere $\mathbb S$.

Now, the metric on the $(n_T+1)$-dimensional K\"ahler moduli space spanned by $j^\alpha$ is known to take the form
\be
g_{\alpha\beta} = 2\Omega_{\alpha\rho}\Omega_{\beta\kappa}j^\rho j^\kappa - \Omega_{\alpha \beta}\,.
\ee
Given \eqref{emb} we may compute the induced metric on the hypersurface $\mathcal M_T$ of unit volume surfaces, which reads
\be
ds^2 = dx^2 + \sinh^2 x\,d\Omega_{\mathbb S} \,,
\ee
where $d\Omega_{\mathbb S}=h_{AB}(\phi) \,d\phi^A d\phi^B$ denotes the standard metric on the unit sphere $\mathbb S$. The distance between any two points $P, Q\in \mathcal M_T$ is given by
\be\label{LPQ}
L(P, Q) = \int _{x_P}^{x_Q} dx\,\sqrt{1 + \sinh^2 x\, h_{AB}(\phi(x)) \frac{d\phi^{A}}{dx} \frac{d\phi^{B}}{dx}}
\ee
where we used $x$ itself to parameterise the path between the two points. The important point is that for any path connecting $P, Q$, including the geodesic path, we have that
\be\label{dist}
L(P, Q) \ge |x_P - x_Q|
\ee
since $\sinh^2 x\, h_{AB}(\phi(x))\frac{d\phi^{A}}{dx} \frac{d\phi^{B}}{dx} \ge 0$ for any path $\phi^A(x)$.

It is interesting to compare the parameterisation of the K\"ahler form \eqref{Jtwo} with the one in \eqref{baseK}.
Since the behaviour of the moduli parameterising the finite volume sphere $\mathbb S$ is irrelevant for our purposes, it suffices to take 
 $h^{1,1}(B_2)= 2$ for simplicity. In this case \eqref{Jtwo} reads
\be
J = {\sqrt{2} (\cosh x \,\omega_0 + \sinh x \,\omega_1)} \,.
\ee
On the other hand, the ansatz (\ref{baseK}) for $h^{1,1}(B_2)= 2$ can be written explicitly as
\bea \label{Jh112}
J = \frac{1}{\sl{\sqrt{\frac{n_1}{2} + \frac{n_{11}}{8 \tilde t^2}}}}  \left( \tilde t J_0 + \frac{1}{2\tilde t} I_1 \right) \, , \qquad I_1 \cdot I_1 =:n_{11} \,,    \quad J_0 \cdot I_1  =:n_1 \,,
\eea
where $J_0$ and $I_1$ are the two generators of the K\"ahler cone. Then the parameter $t$ in (\ref{baseK}) controlling the asymptotic volume of $C$ is given by
\be
t =   (\sl{\frac{n_1}{2} + \frac{n_{11}}{8 \tilde t^2}})^{-\frac{1}{2}} \,  \tilde t 
\ee
and asymptotically \sl{proportional to} $\tilde t$.
We can now make an ansatz for $J_0$ and $I_1$ in terms of $\omega_0$ and $\omega_1$ and fix the coefficients by comparing the intersection forms. 
For large $t$, we find  that
\be \label{xtrel2}
t \simeq \frac{1}{2}e^{x}\,.
\ee

In particular, we are interested in measuring the geodesic distance $d$ between two points with $t(x_P)$ finite and $t(x_Q)  = \infty$. In view of (\ref{dist}) and (\ref{xtrel2}), this means that $x_Q = \infty$, and therefore the distance between an arbitrary point in $\mathcal M_T$ and the boundary $x\rightarrow \infty$ is infinite.
In general, due to the term proportional to ${\rm sinh}^2 x$ inside the square root in~\eqref{LPQ}, even the geodesic distance $d$ acquires a ``correction'' term from the lower bound~\eqref{dist}. However, such a correction is finite and we can still estimate $d$ to be the same as $x$ up to subleading terms for large $t$, so that 
\beq\label{t-d}
t \simeq \frac12 e^d \,.
\eeq

In conclusion this confirms the expectation that the limit in which a gauge symmetry becomes global in F-theory indeed lies at infinite distance in K\"ahler moduli space.  

\vfill\eject
%-----------------------------------------------------------------------------------------------------------------
\subsection{Asymptotically tensionless heterotic strings} \label{sec_tensionlessstrings}

The appearance of a genus-zero curve $C_0$ in $B_2$ with zero 
or almost zero volume has important physical consequences: A D3-brane wrapped on $C_0$ 
gives rise to a string propagating in 6 dimensions, which becomes tensionless as~${\rm vol}_J(C_0)\to0$.  
Evidently this tensionless string is quite different from 
the familiar non-critical strings that have been intensively investigated after the initial works of \cite{Klemm:1996hh,Minahan:1998vr}  in the context of $N=(1,0)$ superconformal field theories (SCFTs)
(see \cite{Heckman:2018jxk} for a recent review).  

The special properties of the string associated with $C_0$ stem from the fact that $C_0 \cdot C_0 = 0$, and, consequently,
 that  the restriction of the fibration to $C_0$ describes a $K3$ surface, $\widehat C_0$.
The appearance of this $K3$ surface has already been stressed
 at the end of Section~\ref{sec_Globalasalimit}.
By contrast, the familiar non-critical strings which appear in the context of $N=(1,0)$ SCFTs are due to D3-branes wrapping vanishing curves $\Gamma$  of  {\it negative} self-intersection. Correspondingly the surface $\widehat \Gamma = \pi^{-1}({\Gamma})$ is never a K3 surface because\footnote{The first equality follows because a contractible curve on an F-theory base is always rational \cite{Morrison:2012np}.}
\be
-2 = 2g(\Gamma) - 2 = \Gamma \cdot (\Gamma - \bar K)   \quad   \Rightarrow \quad \Gamma \cdot \bar K \leq 1 \quad  {\rm if}  \quad   \Gamma \cdot \Gamma < 0\,.
\ee
This means that the surface cannot contain 24 $I_1$ fibers which would have been required for an elliptic K3. This is because these are located at the intersection points of $\Gamma$ with the discriminant divisor $\Sigma$ in class $12 \bar K$, and by the above $\Gamma \cdot \Sigma \leq 12$.

The 2d supersymmetric field theory on the world-volume %$\mathbb R^{1,1}$
of the effective string can be deduced \cite{Haghighat:2015ega,Lawrie:2016axq} with the help of a certain topological duality twist \cite{Martucci:2014ema} along $C_0$, and turns out to have $(0,4)$ world-sheet supersymmetry. 
Given a general curve $C_\beta$ in the base $B_2$, the complex structure of the elliptic fiber of $Y_3$ is identified with the complexified gauge coupling of the ${N}=4$ SYM theory on the D3-brane along $C_\beta$, which therefore varies holomorphically over $C_\beta$. 
Using the topological duality twist \cite{Martucci:2014ema,Haghighat:2015ega,Lawrie:2016axq} the effective world-sheet theory can be obtained by dimensional reduction of the underlying ${  N}=4$
SYM. The spectrum of massless fields as determined by the twist is shown in table \ref{table_2dfields}. 
More specifically, the space $\mathbb R^4 \subset \mathbb R^{1,5}$ normal to the string world-sheet 
comes with a transverse rotation group 
\be \label{SO4T}
SO(4)_T = SU(2)_+ \times SU(2)_- \,,
\ee
 and the spectrum organizes into $(0,4)$ supermultiplets distinguished by their $SU(2)_+ \times SU(2)_-$ quantum numbers,
  as well as by their spin with respect to the $SO(1,1)$ Lorentz algebra along the string world-sheet. 
The fields shown in the first three rows of table \ref{table_2dfields} arise from dimensional reduction of the underlying $N=4$ supermultiplet, while the half-Fermi multiplets in the last line localize at the intersection of $C_\beta$ with the discriminant divisor $\Sigma$. While the naive number of such half-Fermi multiplets would be $12 \bar K \cdot C_\beta$ given that $\Sigma$ is in the class of $12 \bar K$, not all of these modes are independent due to $SL(2,\mathbb Z)$ monodromies. The correct counting   is $8 \bar K \cdot C_\beta$ and can be deduced \cite{Haghighat:2015ega,Lawrie:2016axq} e.g. from the requirement of gravitational anomaly cancellation or by duality with MSW strings \cite{Maldacena:1997de}.  These half-Fermi multiplets carry net charge with respect to the gauge group of the 6d $N=(1,0)$ field theory. As a result, also the excitations of the effective string will be charged.

Applied to the rational curve $C_0$ with $C_0 \cdot C_0 = 0$ and $C_0 \cdot \bar K =2$, we recover $16$ left-moving chiral fermions ($\lambda_-$) along with $4+4$ left-moving scalars. The latter describe
  the center-of-mass motion in the four extended normal
directions ($\varphi$) as well as the internal degrees of freedom of the gauge field ($a, \bar a$) and the normal bundle modes within $B_2$ ($\sigma, \bar \sigma$). The right-moving sector consists of 
the right-moving counterparts of these scalars along with their 
 their fermionic superpartners.

\begin{table}
  \centering
  \begin{tabular}{|c|c|c|c|c|c|}
    \hline
      \multicolumn{2}{|c|}{Fermions} &
    \multicolumn{2}{c|}{Bosons}     & $(0,4)$ &  Multiplicity \cr\hline\hline
     ${\bf (2,1)}_1$  &$\psi_{+}$& ${\bf (1,1)}_0$, ${\bf (1,1)}_0$
    & $ \bar{a}, \bar{\sigma}$ &\multirow{2}{*}{Hyper} & \multirow{2}{*}{ $g - 1 + \bar K \cdot C_\beta$ }      \cr
    ${\bf (2,1)}_1$ &$\tilde{\psi}_{+}$& ${\bf (1,1)}_0$, ${\bf
    (1,1)}_0$ & $a,\,\sigma$ & &   \cr\hline
      ${\bf (1,2)}_1$ & $\mu_+$& \multirow{2}{*}{${\bf (2,2)}_{0}$} & \multirow{2}{*}{$\varphi$} &  Twisted  & \multirow{2}{*}{$1$} \cr
      ${\bf (1,2)}_1$ &   $\tilde{\mu}_+$&  &  & Hyper  &  \cr\hline
   ${\bf (1,2)}_{-1}$ &$\tilde{\rho}_{-}$& & &\multirow{2}{*}{Fermi}  &\multirow{2}{*}{$g(C_\beta)$}\cr
    ${\bf (1,2)}_{-1}$ & ${\rho}_{-}$& &&&   \cr\hline\hline 
    ${\bf (1,1)}_{-1}$ & $\lambda_{-}$& &&{\rm half-Fermi} & $8 \bar K \cdot C_\beta$    \cr\hline\hline 
    %  ${\bf (2,1)}_{-1}$ & $\lambda_-$& ${\bf (1,1)}_{2}$& $v_+$ &\multirow{2}{*}{Vector} &\multirow{2}{*}{$h^1(C,K_C \otimes {\cal L}_D) = 0$}\cr
    %  ${\bf (2,1)}_{-1}$ & $\tilde{\lambda}_- $&${\bf (1,1)}_{-2}$ & $v_-$ &&\cr\hline
  \end{tabular}
  \caption{Spectrum of massless 2d $N=(0,4)$ multiplets of the effective world-sheet theory of the string that arises from a D3-brane that wraps $C_\beta \subset B_2$ (not contained in the discriminant). %The first column indicates the $SU(2)_+ \times SU(2)_-$ representation and the $SO(1,1)$ quantum number of the fields. 
The representations are under $SU(2)_+ \times SU(2)_- \times SO(1,1)$.   
The table is a modification of Table 3 in \cite{Lawrie:2016axq}.  \label{table_2dfields}}
  \end{table}

This spectrum identifies the effective string with the heterotic string. This can be equivalently understood by going to the M-theory picture and recalling that the M$5$-brane, when wrapped on a $K3$ surface, is dual to the heterotic string \cite{Harvey:1995rn,Cherkis:1997bx}.  More precisely, in our situation, where the $K3$ is embedded in an elliptic threefold, we have less supersymmetries than in the usually discussed situation, where the heterotic string is compactified on $T^4$. Rather we will obtain a chiral heterotic string 
 with $(1,0)$ space-time supersymmetry in $d=6$, which means that it is, morally speaking, compactified on some $K3$ 
 (generically with gauge bundles on top, potentially dressed by pointlike instantons, and not necessarily weakly coupled). We will denote this generalized $K3$ compactification geometry by $\otherK3$. Note that there is no reason why $\otherK3$ should coincide with the vanishing $K3$ surface, $\widehat C_0$, in our $F$-theory setup.
 
Hence the strings of our present interest are orthogonal to the familiar non-critical
strings that appear in 6d SCFTs.  This is of course as expected, because the latter arise in the limit of decoupling gravity at finite distance in the moduli space, while the reason for the appearance of a zero-volume curve $C_0$ in our context was the requirement of preserving gravity, and keeping the Planck mass fixed. Thus it makes sense that we obtain a critical string theory (albeit compactified), rather than a non-critical one.

Since this is an important point, let us investigate the relation to the heterotic string in more detail.
Before discussing the general case, assume for a moment that the base $B_2$ is given by one of the Hirzebruch surfaces, $\mathbb F_a$. While we will review the pertinent properties of these surfaces in more detail in Section~\ref{subsec_F1base},
what is important for us here is that they have the structure of a $\mathbb P^1$-fibration
\bea \label{HZBpfibr}
p :\quad {\mathbb P}^1_f \ \rightarrow & \  \ {\mathbb F}_a \cr 
& \ \ \downarrow \cr 
& \ \  {\mathbb P}^1_b
\eea
As we will see in Section~\ref{subsec_F1base}, the limit (\ref{baseK}) is compatible with this structure if and only if we identify 
\be
\mathbb P^1_f = C_0 \,.
\ee
This means that the elliptic $K3$ surface, $\widehat C_0 = \pi^{-1}(C_0)$, which we advertised on general grounds at the end of Section~\ref{sec_Globalasalimit}, turns out to be in itself fibered over the base, $C'=\mathbb P^1_b$, of the Hirzebruch surface. Thus, in this case, the elliptic fibration of $Y_3$ is  compatible with the $K3$ fibration.
F-theory on this $K3$ fibration $Y_3$ is then dual to the heterotic string on a $K3$ surface~$\otherK3$ which is an elliptic fibration by itself:
\bea \label{heteroticfibrationr}
r :\quad  T^2 \ \rightarrow & \  \    \otherK3 \cr 
& \ \ \downarrow \cr 
& \ \  {\mathbb P}^1_b
\eea
To avoid confusion, let us recapitulate and  spell out again that we generally deal with two $K3$ surfaces, which are given
by $\widehat C_0$ on the F-theory side and  $\otherK3$ on the heterotic side, respectively. That is,
the nearly tensionless heterotic string that is obtained by wrapping a D3-brane on the shrinking base $C_0$ of $\widehat C_0$ coincides with the heterotic string compactified on $\otherK3$. 
In the situation where the base $B_2$ of $Y_3$ is a Hirzebruch surface, $\widehat C_0$ is an elliptic $K3$ with base 
 $C_0=\mathbb P^1_f$, which by itself is fibered over $\mathbb P^1_b$. Moreover on the heterotic side, $\otherK3$
 is an elliptic fibration over $\mathbb P^1_b$.

Having straightened out any possible confusion, we now continue with an analysis of the various couplings.
The coupling of the heterotic string (measured in the heterotic string frame) is determined by the duality as \cite{Morrison:1996na}
\be
(g_s^h)^2 = \frac{{\rm vol}_J(\mathbb P^1_f )}{{\rm vol}_J(\mathbb P^1_b )} \,.
\ee
In our limit (\ref{baseK}), ${\rm vol}_J(\mathbb P^1_f ) = {\rm vol}_J(C_0 ) \to 0$ while ${\rm vol}_J(\mathbb P^1_b ) \to \infty$. Therefore the string under consideration indeed  asymptotically describes the weakly coupled heterotic string. The fact that the heterotic string is weakly coupled is a priori true only in the specific limit (\ref{baseK}), and away from this limit quantitative statements are harder to make. Fortunately, this is sufficient for our purposes as we are interested 
just in the asymptotic behaviour. 

We can be more precise about the dual heterotic theory in the weak coupling limit, where we continue to work for now in the string frame when discussing the heterotic side of the duality,
and in the Einstein frame when referring to the Type IIB side.
First note that the tension of the string associated with $C_0$ in the Type IIB Einstein frame takes the form 
\be
T = \frac{2 \pi}{\ell_s^4} {\rm vol}(C_0) \,,
\ee
where we have reinstated the Type IIB string length $\ell_s$ (outside of this section, we will generally set $\ell_s \equiv 1$.). The precise normalization is computed in Appendix \ref{App_DimRed}, which also states our conventions.
Equating this with the heterotic string tension 
determines the heterotic  string length $\ell_h$ (measured in heterotic string frame) as
\be \label{ellhexpre1}
\frac{2\pi}{\ell_h^2} = \frac{2\pi}{\ell^4_s} {\rm vol}_J(C_0) \,.
\ee
Furthermore, the unwrapped D3-brane corresponds to a heterotic $5$-brane on the fiber $T^2$ of (\ref{heteroticfibrationr}), and comparing the tensions leads to
\be
\frac{2\pi}{\ell_h^6 (g_s^h)^2} {\rm vol}(T^2)  = \frac{2\pi}{\ell^4_s} \,.
\ee
This gives an alternative expression for the heterotic coupling,
\be \label{gshexpre2}
(g_s^h)^2 = \frac{{\rm vol}(T^2) {\rm vol}(\mathbb P^1_f)}{\ell_h^4} \,.
\ee
Combining (\ref{gshexpre2}) and (\ref{ellhexpre1}) we find for the Planck scale of the heterotic theory
\be
(M_{\rm Pl}^h)^4 = \frac{4 \pi}{\ell_s^8 }   {\rm vol}(\mathbb P^1_f) \,  {\rm vol}(\mathbb P^1_b) \,.  
\ee
If we apply the limit (\ref{baseK}) to the special case of a base space with $h^{1,1}(B_2)=2$, the product of $C_0$ and the remaining curve class indeed remains constant, see (\ref{geom-bound}), and therefore the Planck scale of the heterotic theory remains fixed.
On the other hand, the heterotic gauge coupling is related to the heterotic Planck scale as
\be
\frac{1}{(g^h_{\rm YM})^2} =  (M_{\rm Pl}^h)^4 \frac{\ell_h^2}{16 \pi^2} \,.
\ee
 In summary, in the limit (\ref{baseK}), the heterotic string becomes tensionless, $\ell_h \to \infty$, and weakly coupled, $g_s^h \to 0$. Gravity remains dynamical as $M_{\rm Pl}^h$ stays constant, while the gauge coupling $g^h_{\rm YM}$ scales to zero. 
 
We recall that the above discussion applies to geometries where $Y_3$ is elliptically fibered over a  Hirzebruch surface.
Let us now discuss the situation for general base spaces, $B_2$, that are compatible with the limit (\ref{baseK}).
What we have established is the existence of a $K3$ fibration of the form (\ref{K3fibrrho_sec2.2}), which, however, is in general not compatible with the elliptic fibration of $Y_3$. Nevertheless, we will argue that in the tensionless limit we will again obtain the  critical heterotic string. 

To see this, we use a chain of dualities as follows.
First, we use F/M-theory duality to relate F-theory on the elliptic fibration $Y_3$ times a circle $S^1$
 to M-theory on the same 3-fold, $Y_3$. This duality makes use of the elliptic fibration (\ref{ellfibrationpi}) in the usual manner. The D3-brane wrapping $C_0$ but not wrapping the circle $S^1$ dualises to an M5-brane wrapped on $\widehat C_0$. 
We are therefore interested in the effective theory on the string world-sheet associated with this M5-brane. M-theory on the 3-fold $Y_3$ is dual to strongly coupled Type IIA theory on $Y_3$, with the M5-brane dualising to an NS5-brane on $\widehat C_0$.

Finally, we make use of the $K3$ fibration (\ref{K3fibrrho_sec2.2}):
Type IIA string theory compactified on the $K3$ fibration (\ref{K3fibrrho_sec2.2}) is dual to a 4d ${\cal N}=2$ compactification of heterotic string theory.
This is a fibered version of the 6d IIA - heterotic duality, which equates the  (strongly coupled) Type IIA string on K3 with the weakly coupled heterotic string on $T^4$ \cite{Witten:1995ex}. Famously, under this duality the NS5-brane wrapped along K3 turns into the heterotic string \cite{Harvey:1995rn,Cherkis:1997bx}. Adiabatically fibering the 6d duality over a rational curve $C'$ leads to the duality with the heterotic string in four dimensions \cite{Kachru:1995wm}.
If the $K3$ fibration (\ref{K3fibrrho_sec2.2}) were compatible with the elliptic fibration of the $K3$ fiber, then the dual heterotic string theory would be compactified on $T^2 \times \otherK3$. Here $\otherK3$ is the $K3$ surface (\ref{heteroticfibrationr}) that appears also in the duality between F-theory on $Y_3$ and the heterotic string in six dimensions. When in addition the base of the $K3$ fibration of $Y_3$ becomes large, the dual heterotic string is weakly coupled, as discussed above. 

More generally, however, the { K3-surface $\otherK3$   of the dual heterotic compactification space $\otherK3 \times T^2$ is not elliptically fibered {\it globally}.}
Type IIA - heterotic duality under such more general conditions has been revisited in the recent work~\cite{Braun:2016sks}. 
More specifically, degenerations of the $K3$ fiber in the fibration  (\ref{K3fibrrho_sec2.2}) of $Y_3$ can lead to heterotic 5-brane defects in the dual theory. Even in the limit of large base $C'$, such localised defects on the heterotic side can spoil a fully perturbative CFT description. Nonetheless, as long as the base $C'$ is large, it is reasonable to think of the resulting theory as being `weakly coupled away from the defects', while these provide important further degrees of freedom that render the full theory consistent.

Despite these complications, we will find evidence in the course of this paper that the string that arises from wrapping a D3 brane on 
$C_0$, shares essential properties of the heterotic string, in particular its elliptic genus.
As we will see, this includes a generalization which describes not just single wrapped, but also multiply wrapped strings, which makes it evident
that we really deal with heterotic strings that arise from wrapping D3 branes on $C_0$.

%-------------------------------------------------------------------------------------------------------
\subsection{$U(1)$-charge spectrum of the heterotic string}\label{sec_chargesummary}

To summarise the discussion so far, we have seen that in the asymptotic limit in K\"ahler moduli space, 
the string from a D3-brane wrapped on a shrinking curve, 
$C_0$, becomes tensionless and can be identified with a critical heterotic string on some $K3$ surface, $\otherK3$.
The excitation spectrum of this string takes the usual form
\be \label{MassTension}
M_n^2  = \alpha \, 2\pi \,  T \,   (n -1) \,,        \qquad \quad T  = 2 \pi \,  {\rm vol}_J(C_0) \,,
\ee
where we have set $\ell_s =1$ in the expression for the string tension, and $n$ refers to the (left-moving) excitation level. 
This will be discussed in more detail in Section~\ref{sec_ellgenus}. 
Furthermore for the critical heterotic string the overall normalization is 
\be
\alpha = 4 \,.
\ee
Note that, strictly speaking, we can argue for the specific relation (\ref{MassTension}) only as we approach  the tensionless limit, which coincides with the weak coupling limit (modulo complications for general $B_2$, as discussed above).
In this limit we are essentially changing the duality frame from F-theory/Type IIA to the heterotic one.
In particular, 
 we will find later that the massless sector of the heterotic string
 correctly reproduces the massless F-theory spectrum, at least at the level of the elliptic genus that we will be computing. This fact holds true both if the base is a Hirzebruch surface and in more general situations.

Before we discuss the role of the (nearly tensionless) heterotic string in the context of the Weak Gravity Conjecture in the next section,
let us first analyze the charges of the string excitations in some more detail.
Indeed the excitations necessarily carry charge under the gauge symmetries that become global in the tensionless string limit. 
The reason is that, as we have shown, the D3-brane wrapped around $C_0$ intersects the stack of 7-branes on $C$ (see Point 2 made in Section~\ref{sec_Globalasalimit}). 
The intersection locus between $C_0$ and $C$ hosts massless 3-7 strings between the D3-brane on $C_0$ and the 7-brane on $C$.
The resulting  fermionic zero-modes  have already been discussed in Section \ref{sec_tensionlessstrings} in terms of the 
 chiral (half-)Fermi supermultiplets charged under 
the 7-brane gauge group. 
From the space-time point of view, these charges translate into global ``flavor" charges of the string excitations. 

In order to be more quantitative, we will  focus on a single $U(1)$ gauge symmetry.
The detailed technical arguments will be developed in Section \ref{sec_Ellgenus}, while for now we present the following key results:

%-------------------------------------------------------------------------------------------
\begin{figure}[t!]
\centering
\includegraphics[width=12cm] {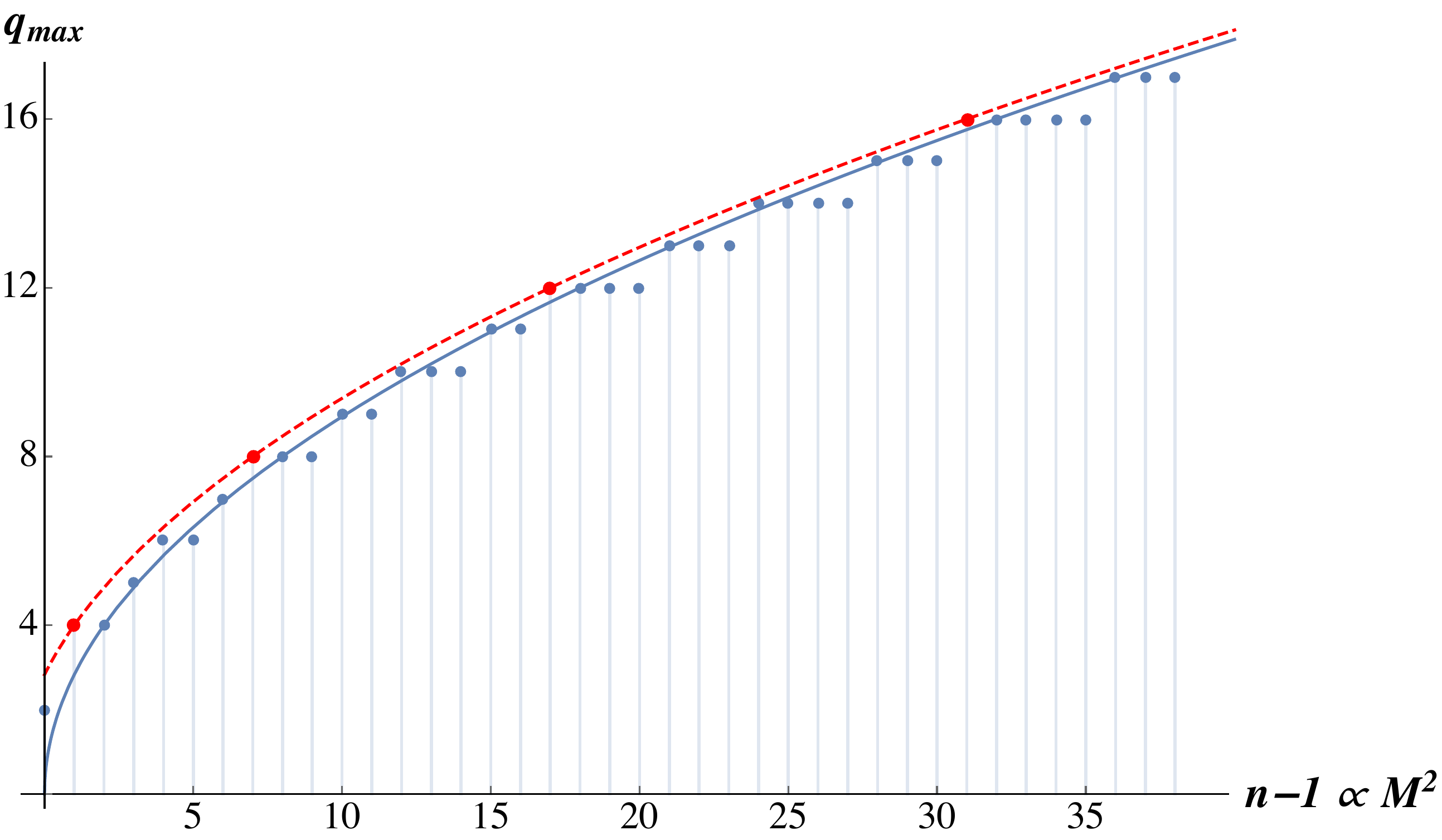}%{newF144.jpg}
\caption{ Maximal charge $\frak{q}_{\rm max}(n)$ per excitation level $n$ for a 6d F-theory compactification on $Y_3$ with base $B_2 = \mathbb F_1$. The model refers to the values $({\rm x},{\rm y}) = (4,4)$ in the notation of Section~\ref{sec_dP1}, and  has charge index $m=\frac12 C\!\cdot\! C_0=2$.
The solid blue curve is given by $\frak{q}(n)=\sqrt{4m(n-1)}$, which
corresponds to the modified, scalar weak gravity bound derived in ref.~\cite{toappear} for the relevant extremal black holes.
\\
Observe that  the charges for some excitations 
lie just barely above this curve, as a consequence of  
the plateau pattern in conjunction with the offset of the vacuum energy by~$-1$.
The maximally superextremal states, marked in red,  lie on the dashed curve given by $\frak{q}_{\rm max}(n)=\sqrt{4mn}$ and 
populate a charge sublattice with spacing given by $\Delta\frak{q}=2m=4$. Together with the additional superextremal states lying between the red and the blue
curves they
populate the full charge lattice. This is a feature of this particular example.
%This is all what is  needed for the Sublattice Weak Gravity Conjecture to hold. Physically, this means that extremal black holes would be able to decay into these states.\\
%The dashed red line corresponds to the value of $\beta_\epsilon$ in (\ref{betansharper}) with $\epsilon = \frac{5}{8}$.\\
%The green-dashed curve will be discussed in the text.  xxx??
 }

\label{f:plot_1B_F1_100}
\end{figure}
%-------------------------------------------------------------------------------------------

\begin{enumerate}
\item 
For each value of $U(1)$  charge $\frak{q} \in \mathbb Z$ there exists at least one state in the excitation spectrum of a 
single  string along $C_0$.
All of these states become massless in the limit ${\rm vol}_J(C_0) \to 0$, which coincides with the limit $g_{\rm YM} \to 0$. Oftentimes, we will only discuss non-negative charges $\frak q \in \mathbb Z_{\geq 0}$ given that the states in the theory come in pairs of $U(1)$ charges $\frak{q}$ and $-\frak{q}$.
\item
At each excitation level $n$, the spectrum of these string excitations contains states of maximal charge
\bea \label{chargemass}
\frak{q}^2_{\rm max}(n) \geq \beta(n)   \,  n   \,   ,  
\eea
for some numerical prefactor $\beta$ that is fixed by the specifics of the abelian gauge group and its realisation in F-theory. 
For large $n$, this prefactor can be taken as a positive constant with an upper bound given by $2 C \cdot C_0$, in the following  sense: 
For each value of $\epsilon\in(0,1)$, we can find some $N(\epsilon)$ such that 
{
\be \label{betansharper}
\beta(n) \geq   (1 - \epsilon) \,  4 m =: \beta_\epsilon \qquad \quad \forall \, n \geq N(\epsilon)  \,,
\ee
with 
\be
m = \frac{1}{2} C \cdot C_0
\ee
as introduced in (\ref{J0J0zero}).
}
This implies, in particular, that 
\be \label{betasharper2}
\frak{q}^2_{\rm max}(n) \geq  \beta_\epsilon \,  (n-1) \qquad    \forall \, n \geq N(\epsilon)   \,.       %      \beta(n)   \,  n   \,  
\ee
As we will explain in Section~\ref{sec_genstratmaxcharge} and Appendix \ref{app_EJmaxcharge}, these properties of the excitation spectrum can be deduced on general grounds without reference to a concrete model.
We can reduce the bound $N(\epsilon)$ at the cost of increasing $\epsilon$ and therefore decreasing $\beta_\epsilon$.
In concrete examples we will give an even lower value for $\beta_\epsilon$ such that the bound holds for all values of $n$. 

\item
{
Conversely, there exists a {\it sublattice} of charges of index $2m$ such that for each $\frak{q}_{k}$ in the {sublattice}, i.e. for each
\be \label{sublatticedef}
\mathfrak{q}_{k} = 2m \, k \,, \qquad     k \in \mathbb Z \,,
\ee
there exists a state at excitation level $n(k) = \frac{\mathfrak{q}_{k}^2}{4m} = m \, k^2$.
Put differently, for each charge in the sublattice (\ref{sublatticedef}), there exists a state with excitation level $n$ such that 
\be \label{sublatticerel}
\mathfrak{q}_{k}^2 = 4m \, n(k) \,.
\ee
A derivation will be given in Section~\ref{sec_genstratmaxcharge}. 
The importance of these states is that they are superextremal with respect to a (dilatonic) charged black hole and hence an extremal such black hole can decay into them, see Section \ref{sec_ReltoQG} for a detailed discussion.
In general, there may be further superextremal states (possibly even corresponding to a smaller index sublattice), but already the states associated with the sublattice (\ref{sublatticedef})
suffice for the Sublattice Weak Gravity Conjecture to hold.

% $n(k)$
% some $n(\frak{q})$ with $\frak{q}^2 \geq \beta_\epsilon (n(\frak{q})-1)$.
% The coarseness of the sublattice depends on the bound $N(\epsilon)$, as it determines the finite set of charges that we have to exclude for the statement to hold. 
% Again, by lowering the universal prefactor $\beta_\epsilon$ the bound $\frak{q}^2 \geq \beta_\epsilon (n(\frak{q})-1)$ can be made to hold for a state with every charge $\frak{q} \in \mathbb Z$ in the full lattice. 
}
\end{enumerate}

This general behaviour is illustrated in Figure \ref{f:plot_1B_F1_100}, which refers to a particular example that will be discussed in more detail in Section \ref{sec_dP1}. Depicted is the maximal $U(1)$ charge per excitation level, $\frak{q}_{\rm max}(n)$, for a D3-brane wrapping a curve $C_0$ in the base $B_2=\mathbb F_1$ with ${C_0}^2=0$ and $ m =\frac12 C_0 \cdot C = 2$.
Note that at given $n$, each value of the charges with $|\frak{q}| \leq \frak{q}_{\rm max}(n)$ is populated by string states. We can see that the 
maximally superextremal excitations indeed obey eq.~(\ref{sublatticerel}) and form the sublattice defined by (\ref{sublatticedef}).
In this particular example, the full set of superextremal states  in fact happens to populate the {\it full} charge lattice.

In Section~\ref{sec_Ellgenus} of this paper we will substantiate the claims (\ref{chargemass}), (\ref{betansharper}) {as well as (\ref{sublatticedef}), (\ref{sublatticerel})} by computing the elliptic genus for the nearly tensionless heterotic strings we are encountering.
Our main focus will be on the massive states that are charged under a $U(1)$ gauge symmetry that becomes global in the limit $t \rightarrow \infty$.
A key role will be played by the charge refinement of the elliptic genus in terms of the height pairing associated with the rational section underlying $U(1)$.

%--------------------------------------------------------------------------------------
\subsection{Relation to Quantum Gravity Conjectures} \label{sec_ReltoQG}

As ${\rm vol}_J(C_0) \to 0$, the tension of the heterotic string discussed in the previous section tends to zero, and so do the masses of its excitations.
{A priori we need to distinguish between the appearance of the asymptotically tensionless (BPS protected) heterotic {\it string} as such and the (non-BPS) {\it particle} type excitations of the string.
 The latter result in a tower of nearly massless, charged (particle) states.
In this section we discuss the implications of the tower of particle states in the context of four interrelated Quantum Gravity conjectures, reserving some speculations about the role of the string  for Section \ref{sec_outlook}.}

Before entering the discussion let us note that the results of this section can be read in two ways:
From the outset we view them as statements about the 6d F-theory compactification and its relation to the Quantum Gravity Conjectures.
Equivalently, however, we can interpret them directly in the duality frame of the heterotic string introduced in Section \ref{sec_tensionlessstrings}, and as such they make sense for 6d heterotic string compactifications {\it per se},  without any reference to an F-theory dual.

\subsubsection* {\bf 1) Censorship against global symmetries in Quantum Gravity}

The appearance of infinitely many massless {particle-like} states in the tensionless limit is responsible for the breakdown of the 6d effective field theory. Such a breakdown is expected and, in fact, required by quantum gravity arguments: In the limit $t \rightarrow \infty$, a global symmetry emerges while at the same time gravity remains dynamical as the 6d Planck mass remains constant; this would be in stark contrast with general reasoning (see \cite{ArkaniHamed:2006dz,Banks:2010zn} and references therein) that continuous global symmetries are incompatible with quantum gravity. 
In this sense, the appearance of a tensionless string and its associated infinite tower of massless modes acts as a censor that prevents the theory from an inconsistency with general quantum gravity arguments: The field theory breaks down before reaching a catastrophic global symmetry limit \cite{ArkaniHamed:2006dz,Banks:2010zn}.

%-----------------------------------------------------------------------------------------------------------------
\subsubsection* {\bf 2) The Sublattice Weak Gravity Conjecture}

The particle excitations of the nearly tensionless string play an important role also in the context of the so-called Sublattice Weak Gravity Conjecture (SWGC) \cite{Heidenreich:2015nta,Heidenreich:2016aqi}.
In the sequel, we will show that,  under certain assumptions, the string excitation spectrum contains charged particle-like states whose charge-to-mass ratio, 
near the tensionless limit, is in accordance with the SWGC bounds.

Applied to a single {1-form} $U(1)$ gauge symmetry in $d$-dimensional spacetime, the SWGC posits that there must exist some finite index sublattice of the charge lattice such that for each value of charges $\frak{q}$ in that sublattice, there exists an elementary particle-like state of mass $M$ with the property that its charge-to-mass ratio exceeds that of a charged (non-BPS) extremal black hole.
{This gives rise to a bound of the form
\be \label{SLWGCa}
\frak{q}^2  g_{\rm YM}^2 \stackrel{!}{\geq} \mu  \frac{M^2}{M^4_{\rm Pl}}  \,. % \qquad \mu|_{d=6} = \frac{d-3}{d-2}|_{d=6} = \frac{3}{4} \,,
\ee
The precise value of the numerical constant $\mu$ depends on the type of extremal black hole with respect to which the particles must be super-extremal.
}

Certainly in six dimensions many kinds of black objects are known to exist, in particular black strings\footnote{Unlike black holes, these can be BPS in six dimensions with $N=(1,0)$ supersymmetry; however, BPS objects are not of interest here in the context of the SWGC.}
(see, for instance, \cite{Cvetic:1996dt,Cvetic:1996zq,Emparan:2008eg,Emparan:2009vd} and references therein for a sample of the vast possibilities in higher dimensions).
Given the enormous richness of possible black objects in six dimensions, we adopt the working assumption
that the relevant objects with respect to which super-extremal states are to exist are certain { extremal %, non-dilatonic\footnote{Note that in our context we do not have to consider dilatonic black holes because the dilaton does not couple to the gauge theory obtained from 7-branes in the Einstein frame, as shown in Appendix \ref{App_DimRed}.} 
Reissner-Nordstr\"om  (RN) black holes};
in this we follow ref.~\cite{Heidenreich:2015nta}. 
{If we could completely ignore the influence of massless scalar field fluctuations, the corresponding black objects to consider would be {\it non-dilatonic} extremal RN black holes \cite{Duff:1996hp} in $d$-dimensional Einstein-Maxwell theory.} With respect to these, the {constant $\mu$ in (\ref{SLWGCa})   takes the value % following  SWGC bound was derived
\cite{Heidenreich:2015nta}
\be \label{SLWGCmu1}
%\frak{q}^2  g_{\rm YM}^2 \stackrel{!}{\geq} \mu  \frac{M^2}{M^4_{\rm Pl}}  \,,  \qquad 
\mu|_{d=6, {\rm non-dil.}} = \frac{d-3}{d-2}|_{d=6} = \frac{3}{4} \,.%= \frac{k}{8 \pi} \frac{M^2}{M^4_{\rm Pl}} \,,
\ee
}
%{Below we will investigate to what extent the particle excitations of the nearly tensionless heterotic string satisfy this bound. Before we enter this analysis, however, a few comments are in order.}
{However, in a supergravity context, the effect of massless scalar field fluctuations cannot be neglected \cite{Heidenreich:2015nta,Palti:2017elp}.
In the forthcoming work \cite{toappear} we will derive in detail that including the massless scalar field fluctuations in the tensor multiplets  increases the numerical value of the bound $\mu$ in (\ref{SLWGCa}) to
\be \label{SLWGCmu2}
\mu|_{d=6, {\rm dil.}} =  \frac{d-3}{d-2}|_{d=6}  + \frac{1}{4} = 1 \,.
\ee
}
% First of all, we expect that there will be contributions to $\mu$ from the scalar fields, and the gauge theory  couples in particular to the scalars in the tensor multiplets, which set the value of the gauge coupling. 
% The results of \cite{Palti:2017elp} (in the context of 4d compactifications) suggest  that a proper inclusion of all scalar fields increases the numerical value of the bound $\mu$ in (\ref{SLWGCa}). 
% Including all scalar fields in our six-dimensional framework is beyond the scope of this work
% and we will hence adopt the bound (\ref{SLWGCa}) as a simplifying, conservative working assumption.

We proceed by 
evaluating whether the charge spectrum  (\ref{chargemass}) obeys the 
requirement (\ref{SLWGCa}) {with this value  $\mu = \mu|_{d=6, {\rm dil.}} =1$.} Such a comparison is meaningful  a priori only  in the tensionless limit $t \to\infty$, where we have perturbative control of the heterotic string spectrum and trust
in particular the mass relation (\ref{MassTension}). With this in mind,
first deduce from~\eqref{MassTension} and~\eqref{chargemass}  that
\be
{ \frak{q}^2_{\rm max}(n) \geq  \beta(n) \left(\frac{1}{4 \pi^2 \alpha} \frac{M_n^2}{{\rm vol}(C_0)}  +1\right)   }
\ee
near the tensionless limit.
Next we multiply both sides with $g^2_{\rm YM}$, given in (\ref{gaugecoupling}), and furthermore express the result in terms of the 6d Planck mass (\ref{Mpl}), using that we have fixed  ${\rm vol}_J(B_2) =1$.
As a result we obtain
\bea \label{q2maxbeforelimit}
{ \frak{q}^2_{\rm max}(n)   \, g^2_{\rm YM} \geq \beta(n)     \left( \frac{ 2 }{\alpha}   \frac{1}{{\rm vol}(C) \, {\rm vol}(C_0)}   \frac{M_n^2}{M^4_{\rm Pl}}  +   g^2_{\rm YM}  \right)\,. }
\eea
In the limit where  ${\rm vol}_J(C) \sim t \to \infty$, which coincides with the tensionless limit in which we are working, we have established the behaviour (\ref{geom-bound}) for the product ${\rm vol}_J(C) \, {\rm vol}_J(C_0)$.
Hence in this limit we can estimate
\bea \label{equality1}
{ \frak{q}^2_{\rm max}(n)   \, g^2_{\rm YM}    \geq  \beta(n) \left(      \frac{1}{ \alpha \, m}   \frac{M_n^2}{M^4_{\rm Pl}}  + g^2_{\rm YM}   \right)   }       \qquad \text{as} \quad t \to \infty \,.
\eea
At this stage we { recall that there exists a sublattice of charges $\frak{q}_{k}^2$, see (\ref{sublatticedef}), with the property (\ref{sublatticerel}). These charges hence saturate the inequality (\ref{equality1}) with $\beta(n)$ replaced by $4 m$.
Using $\alpha=4$ in (\ref{chargemass}), as is appropriate for the perturbative heterotic string, we find for this charge sublattice the relation  %and dropping the additive $+1$ on the righthand side leads to
\bea \label{boundqmax2-a}
{ \frak{q}^2_{k}   \, g^2_{\rm YM}    =    \frac{M_n^2}{M^4_{\rm Pl}}  + 4m g_{\rm YM}^2   >   \frac{M_n^2}{M^4_{\rm Pl}}  }       \qquad \text{as} \quad t \to \infty \,,
\eea
where $g_{\rm YM}^2 \to 0$ as $t \to \infty$. This beautifully matches the Weak Gravity bound (\ref{SLWGCa}) for the value $\mu =1$ as in (\ref{SLWGCmu2}).
Specifically, after dividing by $g^2_{\rm YM}$ the two sides of this inequality correspond to the two curves shown in Figure~\ref{f:plot_1B_F1_100}, in the context
of an explicit example.  We see that
 that the plateau-like structure of the maximal charges, in combination with the offset of the vacuum energy by $-1$, is crucially
 responsible for the existence of superextremal string exitations that lie just minimally above the Sublattice Weak Gravity bound.
 This is close in spirit to, and generalizes,  the original analysis of the Weak Gravity Conjecture for heterotic strings
 compactified on $T^6$  \cite{ArkaniHamed:2006dz}.
 
 We find it quite remarkable that the purely number theoretic relation (\ref{sublatticerel}) implies that the corresponding states in this sublattice just narrowly satisfy the (Scalar) Weak Gravity Conjecture bound, which after all has been derived in a completely independent manner that 
also takes the effects of scalar fluctuations \cite{Heidenreich:2015nta,Palti:2017elp,toappear} into account.
}

Before claiming victory, however, we hasten to point out a few caveats.
The inequality (\ref{boundqmax2-a}) has been established only asymptotically for  $t \to \infty$. Specifically,
eq.~(\ref{boundqmax2-a}) makes use of the behaviour (\ref{geom-bound}), which applies only for $t\to\infty$ in K\"ahler moduli space. 
Moreover, only in this limit can we rely on the perturbative mass relation (\ref{MassTension}) of the non-BPS string excitations, for geometries where the base space is  rationally fibered.
For more general base spaces, we have argued that in the tensionless limit the heterotic string can be trusted too. Away from the limit, however, it is expected
that the precise value of the mass per excitation level is subject to renormalisation, in particular since the states are not BPS protected.
A related point is that the Weak Gravity relation (\ref{SLWGCa}) rests on the formula for the mass of an extremal Reissner-Nordstr\"om black hole in 6d. A priori one { could expect it to receive quantum corrections due to absence of the BPS property. It would be interesting to understand these effects and their potential interplay. In particular it is tempting to conjecture that the various corrections away from the limit $t \to \infty$ conspire such that the 
string states continue to satisfy the Sublattice Weak Gravity Conjecture, but this is pure speculation at this point. }

% Finally, we have already noted that neglecting the effect of the scalar fields is an oversimplification \cite{Palti:2017elp}.
%  From the perspective of the elliptic genus in relation to geometry, the most canonical bound 
%  on the charge-to-mass ratio is the extremal value in (\ref{boundqmax2-a}) for $\epsilon =0$. We may call it the geometric bound, since
% it is entirely determined by geometry.\footnote{In terms of heterotic strings, the quantity $m=\frac12 C \cdot C_0$ corresponds to the level of the underlying $U(1)$ Kac-Moody algebra.} 
% That is,  via (\ref{betasharper2}) it corresponds to  $\frak{q}^2_{\rm max}(n)=2C\cdot C_0(n-1)$.

% Evidently, the geometric bound  $\epsilon =0$ lies considerably above the estimated, conservative bound (\ref{SLWGCa}) deduced from the extremal black hole solution that we used for our analysis. It would be very interesting to understand whether the geometric bound with $\epsilon =0$ has a physical realisation; 
% ideally as  an exact version of the Weak Gravity Conjecture which takes the effect of quantum corrections and massless scalar fields on the extremal black hole solutions  \cite{Palti:2017elp} fully into account, such as to yield $\mu=2$ in eq.~(\ref{SLWGCa}).

\vspace{2mm}

%-----------------------------------------------------------------------------------------------
\subsubsection* {\bf 3) The Completeness Conjecture }

The Sublattice Weak Gravity Conjecture is a  variant of the Completeness Conjecture \cite{Polchinski:2003bq,Banks:2010zn}, according to which each point  of the charge lattice should be populated by a state.
As stressed in section \ref{sec_chargesummary}, for the excitations of a  string from a single D3-brane  wrapping $C_0$,  we find that indeed each charge $\frak{q} \in \mathbb Z$ is realized for some mass in the spectrum. 

While this fact is pretty self-evident in terms of the weakly coupled heterotic dual string, it
can be contrasted with the subsector of charged matter states in F-theory which are due to 7-7 strings. For these states the SWGC and in fact even the completeness hypothesis appears to be badly violated, at least at the level of stable one-particle states. To see this,
consider again an abelian $U(1)$ gauge symmetry and suppose that the massless string excitations contain states of certain charge. Then the tower of massive 7-7 string states above the massless ground states does not contain any charges higher than that. From a perturbative Type IIB perspective, the $U(1)$
charges are a consequence of the Chan-Paton factors of open strings stretched between 7-branes, and these do not change at higher string excitation levels.
One might wonder if this situation is remedied in F-theory due to the appearance of multi-pronged strings, but this is not the case: Under F/M-theory duality the spectrum of $(p,q)$ string states maps to M2-branes wrapping exclusively curves in the fiber in the dual M-theory.
Computation of the BPS invariants for these curves shows that no charges appear beyond the ones present already at the massless level in the F-theory spectrum. 
By contrast, M2-branes wrapping linear combinations of charged fibral curves, the full fiber and curves in the base do give to rise to arbitrarily highly charged states. In fact, these are precisely encoded in the Gromov-Witten invariants which will be computed in Section~\ref{Ellgenusexamples}. But the associated states in M-theory are the 5d BPS states obtained from wrapped 6d strings \cite{Klemm:1996hh}.
So again the string is needed to render the spectrum complete.

Let us now come back to the fate of the heterotic string excitations away from the tensionless, perturbative limit $t \to \infty$.
It is natural to speculate that these states continue to be realized as physical states even though we cannot trust the weak coupling expression (\ref{MassTension})  for their masses.
 In fact, we can turn tables around and use the Completeness Conjecture to argue for the existence of these states,
  as otherwise the states of high charge would appear to be missing in the F-theory spectrum. 
  
The situation seems reminiscent of the duality between Type I string theory and the $Spin(32)/\mathbb Z_2$ heterotic string in 10 dimensions:
The D1-string of Type I theory becomes light in the limit of large Type I string coupling. In this limit its physical excitation spectrum coincides with the spectrum of the weakly coupled dual heterotic string. Unlike the fundamental Type I string, this includes excitations in spin representations of $Spin(32)/\mathbb Z_2$. As the Type I string coupling becomes smaller, the description of the D1-string becomes non-perturbative in heterotic language, but it is reasonable to expect that the spinorial representations stay in the spectrum of the Type I theory, albeit at high masses.

%-----------------------------------------------------------------------------------------------------------------
\subsubsection* {\bf 4) Swampland Conjectures and emergence of gauge symmetries }

As shown in Section~\ref{sec_infinitedistance}, the limit $t \to \infty$ occurs at infinite distance in K\"ahler moduli space. 
According to the general Swampland Conjecture of \cite{Ooguri:2006in} and its refinement in \cite{Klaewer:2016kiy}, as one approaches a point at infinite distance in moduli space, a tower of infinitely many states becomes massless, and the mass scale of the tower of states which become massless must be at least exponentially suppressed by the distance in moduli space.
This fits perfectly with the behaviour in the concrete situation studied in this paper.
Indeed, from (\ref{volC0<1/2t}) and (\ref{t-d}) we 
notice that as we take the limit $t \rightarrow \infty$, the volume of the curve $C_0$ goes to zero parametrically as 
\be
{\rm vol}_J(C_0) \sim \frac{1}{t}  \sim e^{-d}\,,
\ee
 where $d$ is the (geodesic) distance in moduli space as we approach the point $t \rightarrow \infty$. At the same time, from (\ref{MassTension}) we see that the masses of the particles that arise from excitations of the D3-brane wrapped on $C_0$ are suppressed by $e^{-d/2}$. %go to zero as 
 Hence we confirm the expectations of  \cite{Ooguri:2006in,Klaewer:2016kiy} that at infinite distance in moduli space a tower of infinitely many states becomes exponentially light.

However, as we have discussed, the tower of massless states is also charged under the $U(1)$ 
 gauge symmetry that becomes a global symmetry at infinite distance in moduli space.
A very detailed analysis of a similar phenomenon in the context of 4d ${\cal N}=2$ compactifications of Type IIB string theory on a Calabi-Yau 3-fold %$X_3$
has been presented in \cite{Grimm:2018ohb}; the relevant points lie at infinite distance in complex structure moduli space, and a tower of infinitely many  BPS states has been identified that become massless in the vicinity of these points. The points at infinite distance considered in  \cite{Grimm:2018ohb} indeed coincide with the limit where the couplings of the RR gauge fields vanish.
The appearance of infinitely many charged massless particles is in perfect agreement with field theoretic expectations: Such massless states can be thought of as inducing a running of the gauge coupling in such a way as to enforce its vanishing in the extreme limit where all of them become massless \cite{Harlow:2015lma,Klaewer:2016kiy,Heidenreich:2017sim,Grimm:2018ohb,Heidenreich:2018kpg}. 
Whereas the appearance of a finite number of charged massless states leads to a divergence of $1/g^2_{\rm YM}$ which is logarithmic in the mass, as in the prototypical example of the conifold transition \cite{Strominger:1995cz,Vafa:1995ta}, an infinite number of charged massless states gives rise to a polynomial divergence of $1/g^2_{\rm YM}$ with the particle mass \cite{Grimm:2018ohb}.

In the six-dimensional F-theory context that we are considering,
the gauge coupling indeed vanishes polynomially in the K\"ahler moduli as
\be
\frac{1}{g^2_{\rm YM}}   = \frac{1}{2\pi} {\rm vol}_J(C)   \sim t  \qquad {\rm as} \quad t \to \infty \,.  
\ee
By a very crude estimate, we can relate this behaviour to the appearance of the tower of massless charged string excitations in the spirit of the above discussion, as follows.

We are considering the effective theory at an IR scale $\Lambda_{\rm IR}$, which is identified with the scale of the charged particle with lowest mass.
Above this scale, a tower of massive charged states arises. The field theory description breaks down at the UV cutoff $\Lambda_{\rm UV}$, where gravitational effects are non-negligible. 
If we denote by $N$ the number of massive states below $\Lambda_{\rm UV}$, then in a theory in $d$ spacetime dimensions, the UV cutoff is to be identified \cite{Heidenreich:2017sim,Grimm:2018ohb}   with the so-called species scale $\Lambda_{\rm UV}$ determined by \cite{Dimopoulos:2005ac,Dvali:2007hz,Dvali:2007wp}  $\Lambda^{D-2}_{\rm UV} = \frac{M^{D-2}_{\rm Pl}}{{N}}$, which for $d=6$ gives
\be
\Lambda^{4}_{\rm UV} = \frac{M^{4}_{\rm Pl}}{{N}} \,.
\ee
The fact that, in the presence of a tower of light states, the correct cutoff scale is indeed the species scale was stressed in \cite{Heidenreich:2017sim,Grimm:2018ohb}  and is an important ingredient in the following analysis. 
In our context, we identify $N$ with the excitation level of the highest mass state which lies below the UV cutoff. This is, a priori, a significant oversimplification because we know that the excitation numbers already from the single string sector  at a given excitation level $n$ are certainly far from uniform and, in fact, considerably larger than one.
Nonetheless, as far as the parametric behaviour is concerned, this simplification leads to the desired result.
Indeed, with the above interpretation, and using that  the mass $M_n$ at level $n$ scales with $\sqrt{n}$, see (\ref{MassTension}), we can alternatively express $\Lambda_{\rm UV}$ 
as
\be
\Lambda_{\rm UV}  = \sqrt{(4 \pi^2 \alpha)}  \, \sqrt{ N \,  {\rm vol}_J (C_0)} \,,
\ee
where we neglect the constant offset in (\ref{MassTension}) for large $N$. 
Combining both relations and ignoring unimportant numerical prefactors, we arrive at the relation  
\be
N^3  \sim  \frac{M^4_{\rm Pl}}{{{\rm vol}^2_J (C_0)}}
\ee
for the maximal excitation level of states below $\Lambda_{\rm UV}$.  

The 1-loop running of the inverse gauge coupling from the UV to the IR cutoff  is  - up to potential subleading contributions - 
\be
\frac{1}{g^2_{\rm YM}} \biggr\rvert_{\rm IR} = \frac{1}{g^2_{\rm YM}} \biggr\rvert_{\rm UV} -   x  \sum_{n=1}^N   \frak{q}_n^2 \,M_n^2  \,, %  {\rm log} \frac{\Lambda_{\rm UV}}{M_n^2} \,,
\ee
where the numerical prefactor $x$ depends on the specific type of state considered.
To obtain the dominant contribution to the running one can hence very crudely estimate, with $\frak{q}_n^2 \sim n$ from (\ref{chargemass}) and $M_n^2 \sim n \,  {\rm vol}_J(C_0)$ from (\ref{MassTension}), that
 \be
\sum_{n=1}^N   \frak{q}_n^2 \,   M_n^2  \sim \sum_{n=1}^N   n^2  {\rm vol}_J(C_0)   \sim N^3  \,  {\rm vol}_J(C_0)  \sim \frac{1}{ {\rm vol}_J(C_0)} \sim t \,.
\ee
The underlying approximation is that at each excitation level $n$, the dominant contribution to the running comes from the states of the highest charge $\frak{q}_n$.
This indeed reproduces the  vanishing of the gauge coupling as the infinite distance point $t \rightarrow \infty$ is attained.

\section{$U(1)$-Charge Spectrum from Elliptic Genus} \label{sec_Ellgenus}

In this section we address the counting of charged states of the (nearly) tensionless heterotic string,
by evaluating its elliptic genus. The main result of our analysis is an underpinning of our claim (\ref{chargemass})
for the maximal charge per excitation level. In particular, we will see that the counting of string states charged under an abelian group $U(1)$ is determined geometrically via the intersection product between the curve $C_0$ wrapped by the D3-brane and the height pairing associated with the section $S$. This result may be of interest also to other applications.

We begin by reviewing the overall strategy how to determine the charge spectrum, or at least a characteristic portion thereof,  of a nearly tensionless heterotic string that arises from a given elliptically fibered threefold, $Y_3$. 
Following the pioneering work of refs.~\cite{Klemm:1996hh,Minahan:1998vr} (as reviewed in \cite{Heckman:2018jxk}),
  the procedure consists of several logical steps which are connected as follows:
\be
%5{\rm Type}\  IIB{\rm\ on\ } Y_3 
%\,\longrightarrow\,
%{\rm mirror\ map}
%\,\longrightarrow\,
{\rm topological~string~on~} Y_3
\,\longrightarrow\,
%{\rm mirror\ map}
%\,\longrightarrow\,
{\rm elliptic~genus\ of\ }\otherK3
\,\longrightarrow\,
{\rm charge\ spectrum}
\ee
In the next sections we will discuss each step in detail, while roughly following the arrows in reverse order.
In Section \ref{sec_ellgenus1}, we give some overview and comments on the charge spectrum. Then in Section \ref{sec_ellgenus} we will proceed with a technical review of the elliptic genus in terms of weak Jacobi forms.   We will also review the relation between the elliptic genus and the topological string partition function, which will later be used in actual computations for concrete examples. 
In Section \ref{subsec_GlobalMWU1} we  will derive the abovementioned result about the relation between the ``fugacity index $m$" of the elliptic parameter related to  the $U(1)$ gauge symmetry in the 6d theory, and the height pairing of its underlying section.
Finally in Section \ref{sec_genstratmaxcharge} we will summarize the preceding technical sections, and the casual reader is invited to directly jump there.

%----------------------------------------------------------------------------------------------------------------------------------------
\subsection{Perturbative and non-perturbative elliptic genera} \label{sec_ellgenus1}

Our aim is to determine the charge spectrum (or rather a characteristic part of it) of the nearly tensionless heterotic string in six dimensions with $N=(1,0)$ space-time supersymmetry, with respect to a nearly global $U(1)$ symmetry whose specific embedding is determined by the given elliptic threefold, $Y_3$. 

A familiar method is to use the elliptic genus associated with the 6-dimensional heterotic string theory.  Computations of elliptic genera of the various non-critical strings have been a long-standing theme over the years, 
following the works of refs.~\cite{Klemm:1996hh,Minahan:1998vr}. 
There one starts with the compactification of such strings on $S^1$  to 5 dimensions, making use of the fact that the wrapped strings can be straightforwardly counted as BPS particles in one dimension lower.  Indeed,  ref. \cite{Klemm:1996hh} considers the non-critical $E$-string, obtained in F-theory by wrapping a  D3-brane on a rational curve $\Gamma$ with $\Gamma \cdot \Gamma = -1$.
The authors then analyze  the level-matching condition for the 6d non-critical $E$-string wrapped on an $S^1$ with winding number $\kappa$ and Kaluza-Klein level $n$. 
The total excitation number $N$ of the string turns out to be related to the winding number and KK level as
\be
\kappa \, n = N \,.
\ee
In particular for a single wrapped string, $\kappa=1$, one can identify the KK momentum $n$ with the excitation number $N$.  In this way one can write down a counting function for the excited string states, which defines,  in the limit of large $S^1$, what is meant by the elliptic genus of the 6d non-critical string. Concretely, for the non-critical $E$-string one obtains \cite{Klemm:1996hh,Minahan:1998vr} the expression
\be
Z_{E_8}(\tau)\ =\  {\Theta_{E_8}(\tau)\over \eta^{12}(\tau)}  \ ,
\ee
where $\Theta_{E_8}(\tau)$ is the partition function of the root lattice of $E_8$. 

As is well-known, a crucial property of the elliptic genus is that it is a meromorphic modular form (for early works, see \cite{Schellekens:1986yi,Schellekens:1986xh,Witten:1986bf,Alvarez:1987wg,Kawai:1993jk}), and this often allows us to determine it with very little physical input; for instance, for the $E$-string it is given in terms of an Eisenstein series,  $\Theta_{E_8}(\tau) =E_4(\tau)$. When adding background fields, the elliptic genus still maintains good modular properties and turns into a meromorphic, in general Weyl-invariant Jacobi form \cite{Kawai:1997em,Kawai:1998md,Gritsenko:1999fk}.    

We now turn to the nearly tensionless heterotic strings under consideration.
The simplest situation occurs when the 
heterotic string is  perturbative  and weakly coupled. As we have discussed in Section \ref{sec_tensionlessstrings}, it arises by wrapping a  D3-brane along the  fiber $C_0$  of a Hirzebruch surface in the tensionless limit (\ref{baseK}). Being weakly coupled, the string can be formulated in terms of a 2d superconformal world-sheet
theory,  which guarantees a well-defined partition function.
From modular invariance and the mere fact that we have chiral $N=(1,0)$ space-time supersymmetry in $d=6$, 
it immediately follows \cite{Schellekens:1986xh} that, in the absence of background fields,
the elliptic genus of this perturbative heterotic string must be given by
\be
\label{oldellgen}
Z_{K3}(\tau) \ \equiv \ {\rm Tr}_R\left[(-1)^F {F^2} q^{H_L} \bar q^{H_R}\right] \ =\ 2{E_4(\tau)E_6(\tau)\over \eta^{24}(\tau)}\ = \frac2q-480-282888 q+\dots.
\ee
Here, as usual  $q=e^{2\pi i\tau}$, and $F$ denotes the right-moving fermion number.\footnote{The factor of $F^2$ is required in order to obtain a non-vanishing result, where the trace is taken over the entire Hilbert space. Equivalently, one can factor out the spin content of a half-hypermultiplet and only trace over the remaining quantum numbers of the states in the Hilbert space, but without insertion of $F^2$.}
This expression incorporates besides the conformal field theoretical, ${\cal N}=(0,4)$ world-sheet supersymmetric elliptic genus pertaining to any $K3$ surface \cite{Eguchi:1988vra,Kawai:1997em,Kawai:1998md}, also the extra left-moving degrees of freedom of the perturbative heterotic string.
In the following we will always implicitly assume this when we 
refer to the elliptic genus of $K3$, and in particular this applies to the elliptic genus related to the specific,
 not necessarily perturbative compactification geometry
$\otherK3$ that was introduced in Section \ref{sec_tensionlessstrings}.  

We will be interested in a refinement induced by the $U(1)$ symmetry that is determined by the elliptic fibration $Y_3$. 
In terms of $\otherK3$ this refinement translates into 
certain bundles on top such that the only unbroken gauge symmetry is just this $U(1)$. For reasons of modularity, the elliptic genus must then take the more general form \cite{Kawai:1993jk,Kawai:1998md}
\be\label{Kindex}
Z_{\otherK3}(\tau,z) = {\Phi_{10,m}(\tau,z)\over \eta^{24}(\tau)}  \ ,
\ee
where $\Phi_{10,m}(\tau,z)$ is a weak Jacobi form of weight 10. {The concept of weak Jacobi forms and their behaviour under modular transformations is reviewed in Appendix \ref{app_Jacring}. The $U(1)$ { `fugacity index'} $m$ } depends on the  specific $U(1)$ embedding and will be determined in Section~\ref{subsec_GlobalMWU1}
($z$ denotes the field strength of the $U(1)$ background field, or its `fugacity'). As is familiar in this context, and will be re-addressed later in Section~\ref{sec_genstratmaxcharge}, the ring of such modular forms is finitely generated and this is why one can easily determine $Z_{\otherK3}(\tau,z)$  for concrete examples.  

Note that the elliptic genus (\ref{Kindex})  is not a partition function for physical states, at least not directly. This is because the left-moving tower of states lacks level-matching; most notably, the tachyon is not physical. Rather, it needs to be properly combined with right-moving excitations. This means that these states cannot be BPS protected (and indeed there are no BPS particles for 6d $(1,0)$ supersymmetry in the first place). 

We can put a slightly different perspective on this trivial fact by employing space-time supersymmetry, which can be implemented by spectral flow of  the elliptic genus \cite{Lerche:1987ca,Kawai:1993jk,Harvey:1995fq}  from the R to the NS sector. This transports the tower of left-moving states paired with the right-moving Ramond ground  states to a partition function summed over all fermionic periodicities.\footnote{
One can see this more directly by making use of certain generalized Riemann theta function identities that map the parity-odd sector to a sum over
all sectors \cite{Lerche:1988zy,Lerche:1988zz}. } 
Since the left sector will not be touched by this operation, the total sum will still be holomorphic, now as a consequence of space-time supersymmetry rather than of world-sheet supersymmetry. 

\begin{figure}[t!]
\centering
\includegraphics[width=7cm] {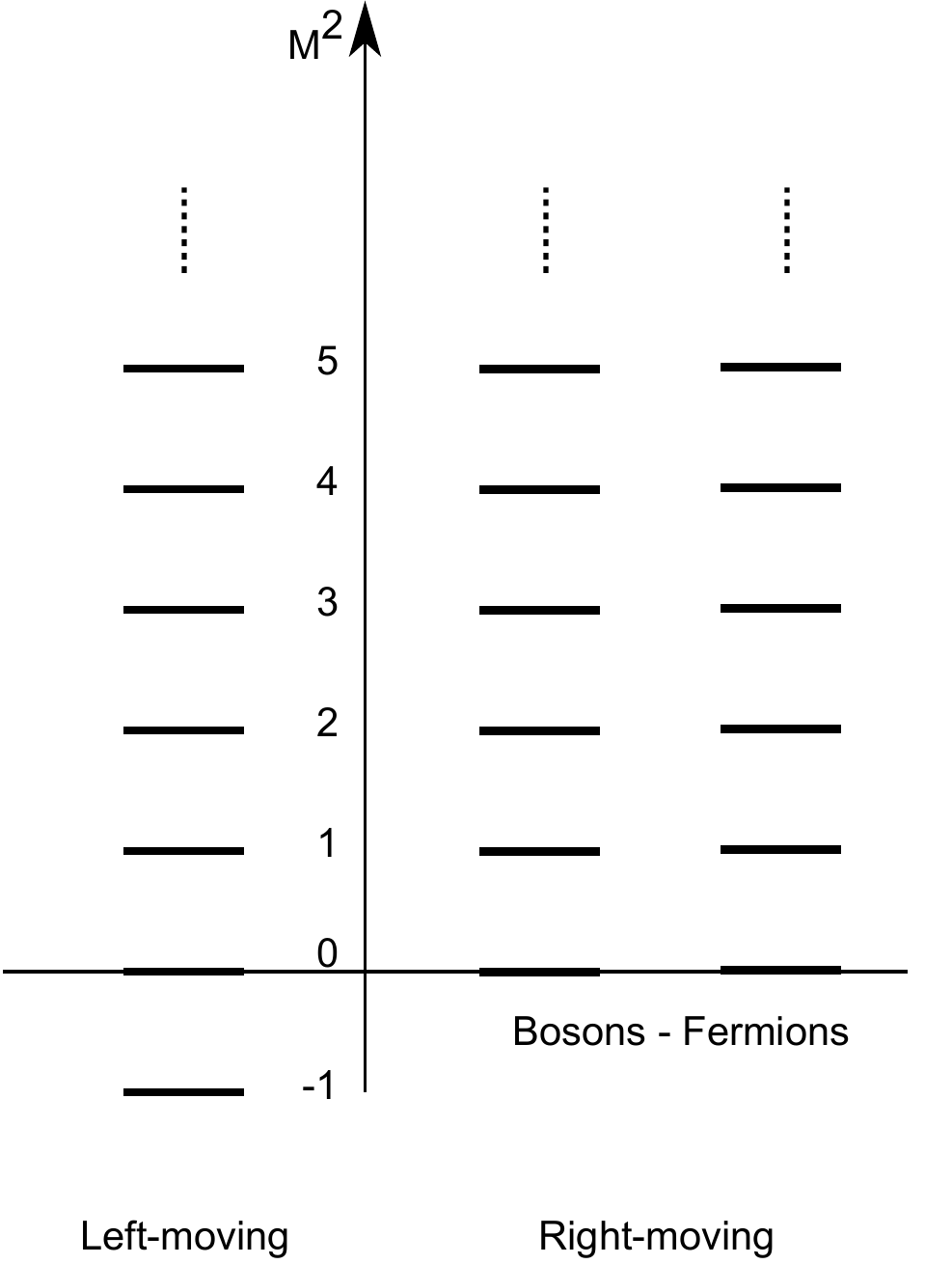}
\caption{ 
Shown are left- and right-moving spectra relevant for the elliptic
genus of a perturbative heterotic string with $(1,0)$ supersymmetry in $d=6$. 
In the right-moving periodic Ramond sector, 
the massive towers cancel due to world-sheet supersymmetry, but they can also be
mapped into towers that cancel between bosons and fermions due to space-time supersymmetry.
In this way we see that the left-moving, $U(1)$-charged excitations can be level-matched to excitations in the vacuum modules
of the right-moving sector, to form physical supermultiplets with fixed masses.
}
\label{f:ellgenus}
\end{figure}

The partition function can then be disentangled  into bosonic and fermionic space-time sectors, and one can infer the existence of physical, level-matched states coming from the super-Virasoro modules over the bosonic and fermionic ground states (corresponding to massless representations of the $\cN=(0,4)$ world-sheet supersymmetry \cite{Kawai:1993jk}).  Certainly this does not give, by far, the full partition function, rather it gives a piece of it that is holomorphically factorized.  
This subsector is protected, and in particular moduli independent, by an interplay between modularity and world-sheet/space-time supersymmetry; see Fig.~\ref{f:ellgenus} for an illustration.  
While this is indeed only a very small subsector of the theory,
it suffices to demonstrate the existence of physical states with certain charge-to-mass ratios, and this is all we need for discussing weak gravity conjectures as in Section \ref{sec_ReltoQG}.

%----------------------------------------------------------------------------
%----------------------------------------------------------------------------

Note that these arguments apply primarily when we have a well-defined world-sheet formulation in terms of conformal field theory,
i.e. to perturbative heterotic strings in the limit of vanishing coupling.  As mentioned in Section~\ref{sec_tensionlessstrings}, these arise when we consider threefolds $Y_3$ that are elliptic fibrations over 
Hirzebruch bases, $\mathbb F_a$, in the limit $ {\rm vol}_J(C_0) \to0$.  
However, tensionless  heterotic strings can also arise  from 
shrinking zero-intersection curves $C_0$ on non-Hirzebruch bases. These will in general  contain more than one massless tensor field and will not be easily describable in terms of conformal field theory. 

For such strongly coupled heterotic strings the physical particle spectrum is less clear, since the massive excitations are not BPS protected and a world-sheet formulation is also not at hand. 
 For such cases, one may adopt as working assumption that it suffices to consider the elliptic genus at face value, even though it may not directly count physical states. 
  
As we will show below, an elliptic genus can indeed be defined also for strongly coupled
heterotic strings via dualities, analogously as for the non-critical $E$-strings.  For such heterotic strings the elliptic genus is still of the form (\ref{Kindex}),  where however $\Phi_{10,m}(\tau,z)$ is in general only quasi-modular (which means that may involve the quasi-modular Eisenstein series, $E_2(\tau)$).
The deviation from the familiar conformal field theoretic expression (\ref{oldellgen}), for which $\Phi_{10,m}(\tau,0)=E_4(\tau)E_6(\tau)$, is 
due to the modified anomaly associated with the extra massless tensor fields. The mild non-modularity reflects that we do not have a standard world-sheet description in this case.
We will continue to refer to this more general object as  $Z_{\otherK3}(\tau,z)$, even though it may not necessarily
 coincide with the familiar, conformal field theoretical elliptic genus associated with a $K3$ surface. In the following we explain how to compute  $Z_{\otherK3}(\tau,z)$ also in these cases.

%-----------------------------------------------------------------------------------------------------------------
\subsection{From elliptic genus to topological string partition function}\label{sec_ellgenus}

We now review and discuss some technical aspects of the elliptic genus that are relevant to us. 
The discussion will be structured according to the following  four different aspects: 
\begin{enumerate}
\item 
For a string arising from a D3-brane wrapping a curve $C_\beta$ on the base of an F-theory model,  it is convenient to define a refined notion of an elliptic genus, $Z_{C_\beta}(\tau,\lambda_s,z)$, as in (\ref{Zbetadefinition}) below, by introducing an extra fugacity parameter $\lambda_s$ with respect to the group of rotations transverse to the string. We will sometimes refer to $Z_{C_\beta}(\tau,\lambda_s,z)$ as the `spin-refined elliptic genus'.
The elliptic genus we are ultimately interested in, namely $Z_{\otherK3}(\tau,z)$ in (\ref{Kindex}),   is contained in this more general notion of the elliptic genus as we will explain.
\item
The spin-refined elliptic genus $Z_{C_\beta}(\tau,\lambda_s,z)$ enjoys good modular properties and can hence be expanded in terms of suitable weak Jacobi forms.
\item 
Duality with M-theory relates $Z_{C_\beta}(\tau,\lambda_s,z)$ to the topological string partition function on the elliptic 3-fold $Y_3$ as in (\ref{Ztopstringdef}) and (\ref{Ztopdual}) below. 
\item
The topological string genus-zero free energy
determines the elliptic genus $Z_{\otherK3}(\tau,z)$ of the heterotic string, as  shown in (\ref{ZKgenus0}). 
For perturbative heterotic strings, this relation reproduces the conformal field theoretical elliptic genus, and we will take it as a definition 
for the non-perturbative generalisation with extra massless tensor fields. 
\end{enumerate}
Most of the material corresponding to the first three points is well-known in the literature (see for example \cite{Klemm:2012sx,Haghighat:2013gba,Haghighat:2014vxa,Haghighat:2015ega,Huang:2015sta,Kim:2016foj,DelZotto:2016pvm,DelZotto:2017mee,Gu:2017ccq,Kim:2018gak,DelZotto:2018tcj}  and the review \cite{Heckman:2018jxk} for further references), 
and we will review it below to set the notation and to make the article self-contained. The informed reader (or those not interested primarily in the technical details) may skip directly to point 4.

%-----------------------------------------------------------------------------------------
\subsubsection*{1) The spin-refined elliptic genus of wrapped strings}

We consider strings that arise from a D3-brane wrapping some curve, $C_\beta \in H_2(B_2,\mathbb Z)$, in the base $B_2$ of an elliptically fibered Calabi-Yau, $Y_3$.  We assume that the curve is at general position, i.e. that it is not a component of the discriminant. 
As pointed out in Section \ref{sec_tensionlessstrings}, the restriction of the elliptic fibration to $C_\beta$ describes an in general non-trivially fibered elliptic surface, $\widehat{C}_\beta \in H_4(Y_3,\mathbb Z)$. For the nearly tensionless heterotic string that we consider, this surface turns out to be a $K3$ surface $\widehat{C}_0$. This is not to be confused with $\otherK3$, which represents the
$K3$ surface on which the dual heterotic string is compactified, whose elliptic genus  we are up to determine.

Following \cite{Klemm:1996hh,Minahan:1998vr} we consider a compactification of the 6d string on $S^1$, while also making the time direction periodic. Thus we can think of space-time formally to be given by
\be
\mathbb R^4 \times T^2.
\ee
An elliptic genus-like partition function, $Z_{C_\beta}(\tau, \lambda_s, {z})$,  can then be defined \cite{Klemm:1996hh,Minahan:1998vr} as a trace over the spectrum of 5d BPS momentum and winding states of the wrapped string, with periodic boundary conditions:
\be \label{Zbetadefinition}
Z_{C_\beta}(\tau, \lambda_s, { z})  = {\rm Tr}_R\left[ (-1)^F  F^2 q^{H_L} \bar q^{H_R}   u^{2 J_-}   \xi^J\right] \,.
\ee
Here $u = e^{2 \pi i \lambda_s}$, where the parameter $\lambda_s$ implements a refinement
by the universally present global $SU(2)_-$ symmetry with generator $J_-$, which is part of the transverse Lorentz group (\ref{SO4T}).
Thus, the insertion of $u^{2 J_-}$ in the trace weighs the BPS states depending on their $SU(2)_-$ spin.
Moreover
\be
\xi^J = e^{2 \pi i zJ}
\ee
induces a grading with respect to the almost global $U(1)$ symmetry, with generator $J$.
It is here that the dependence of the BPS spectrum on the gauge symmetry of the specific F-theory background enters.
Moreover, we have defined
\be
q = e^{2 \pi i \tau} \,,
\ee
where $\tau$ refers to the torus $T^2$ and does, a priori, {\it not} refer as usual to the string world-sheet. 
Rather, $Z_{C_\beta}(\tau, \lambda_s, { z})$  counts the index of winding and Kaluza-Klein 5d BPS states arising from the wrapped string on $S^1$. 

The spin-refined elliptic genus enjoys an expansion of the form
\be \label{Zbetaexpansion}
Z_{C_\beta} (\tau, \lambda_s, { z})  = q^{E_0} \left( f^{(0)}(\lambda_s, z) + \sum_{n=1}^\infty q^n  f^{(n)}(\lambda_s, z) \right)\,,
\ee
where $n$ corresponds to the Kaluza-Klein level of the string. The shift in vacuum energy is due to the Casimir energy 
of the ${\cal N}=(0,4)$ worldsheet theory \cite{DelZotto:2016pvm,Kim:2018gak},
\be \label{E0formular}
E_0\ = -  \frac{1}{2} C_\beta \cdot \bar K\ ,
\ee
where $\bar K$ is the anti-canonical bundle of $B_2$.

The important point \cite{Klemm:1996hh,Minahan:1998vr} is that the functions $f^{(n)}(\lambda_s, z)$ for $n \geq 1$ appearing in (\ref{Zbetaexpansion}) have an equivalent interpretation directly in terms of the string propagating in $\mathbb R^{1,5}$, namely as 
counting left-moving string excitations at level $N=n$. This makes contact with the conformal field theoretical
elliptic genus that we mentioned in the previous section, for which $q$ refers to the modular parameter of a stringy world-sheet.

\subsubsection*{2) Modular properties of the elliptic genus}

As is well-known, key to determining the elliptic genus is to make use of the specific behaviour under modular transformations of $\tau$ and the elliptic parameters $\lambda_s$ and $z$.  That is, it must transform as a Jacobi form
of given weight $w$ and index $m$ as defined in eqs.~(\ref{modulartrafo1}) and (\ref{quasiperiod}) of the appendix.
Modular invariance implies that the total weight $w$ of $Z_{C_\beta}(\tau,\lambda_s, {z})$ must vanish, $w=0$. The indices of $Z_{C_\beta}(\tau,\lambda_s, {z})$ with respect to the elliptic parameters $\lambda_s$ and $z$ are encoded in the modular anomaly equation\footnote{The factor of $(2 \pi)^2$ reflects our conventions that $\lambda_s$ and $z$ are quasi-periodic as in (\ref{quasiperiod}) or (\ref{quasiperiod-2}).} \cite{Klemm:2012sx,Alim:2012ss,Haghighat:2014vxa,Huang:2015sta,DelZotto:2016pvm,Gu:2017ccq,DelZotto:2017mee}
\be \label{holanomaly}
\left(\frac{\partial}{  \partial E_2(\tau)} + \frac{(2 \pi)^2}{12}  f_{C_\beta}(\lambda_s, {z})  \right)  Z_{C_\beta}(\tau,\lambda_s, {z}) = 0 \,.
\ee
Indeed, as reviewed in Appendix \ref{app_Jacring}, the index of a weak Jacobi form with respect to the elliptic parameters can be read off from the quadratic terms   in  $f_{C_{\beta}}(\lambda_s, {z})$, see eq.~(\ref{partialE2equ}). 
%  a weak Jacobi form $\varphi_{k,m}(\tau,\xi)$ satisfies the equation
% \be
%  \left(  \frac{\partial}{\partial E_2}  + \frac{(2\pi)^{2}}{12} m \xi^2  \right) \varphi_{k,m}(\tau, \xi) = 0 \,,
% \ee
% which allows us to read off the elliptic index $m$ with respect to $\xi$. 
A general strategy to determine the function $f_{C_\beta}(\lambda_s, {z})$  is via the t' Hooft anomaly 
 polynomial of the $(0,4)$ theory on the string \cite{DelZotto:2016pvm,DelZotto:2017mee,DelZotto:2018tcj}. 
 In Section \ref{subsec_GlobalMWU1} we will show how to determine the index $m$ related to the $U(1)$ symmetry 
in terms of the height pairing of a rational section on the elliptic fibration $Y_3$. 

An analysis of the pole structure of the elliptic genus suggests that it can be written in terms of a ratio of weak Jacobi forms,
with an explicit expression for the denominator. Specifically, for the special case of a single string associated to a curve $C_\beta$
 corresponding to one of the generators of the Mori cone,
the elliptic genus takes the form \cite{Huang:2015sta,Haghighat:2015ega}
\be \label{ZCbetaansatz}
Z_{C_\beta}(\tau,\lambda_s, {z}) = \left( \frac{1}{\eta^2(\tau)}  \right)^{6 C_{\beta} \cdot \bar K}    \frac{\Phi_{W, L, {m}}(\tau,\lambda_s,{z})}{   \varphi_{-2,1}(\tau, \lambda_s)} \,.
\ee
Here $\varphi_{-2,1}(\tau, \lambda_s)$
is the unique weak Jacobi form of weight $-2$ and index $1$ with respect to $\lambda_s$, see eq.~(\ref{phim21}). The numerator is then a  Jacobi form whose weight $W$ is determined by the requirement that $Z_{C_\beta}(\tau,\lambda_s, {z})$ has total weight zero, where we recall that $\eta^2(\tau)$ transforms like a modular form of weight $1$.
The total index $m_-$ of $Z_{C_\beta}(\tau,\lambda_s, {z})$ with respect to the elliptic parameter $\lambda_s$ 
is determined by the modular anomaly equation (\ref{holanomaly}) to be 
\be \label{m-}
m_- = \frac{1}{2} \, C_\beta \cdot (C_\beta +  K). 
\ee
Altogether this leads to 
\be \label{WLindices}
W = 6 C_{\beta} \cdot \bar K - 2, \qquad  \quad L = \frac{1}{2} \, C_\beta \cdot (C_\beta +  K) + 1 = g(C_\beta) \,,
\ee
where $g(C_\beta)$ is the genus of $C_\beta$, and $K$ the canonical bundle of $B_2$.
The index ${m}$ of $\Phi$ with respect to the $U(1)$ fugacity agrees with that of $Z_{C_\beta}(\tau,\lambda_s, {z})$, because the denominator does not involve $z$. The reason is that the only modes charged under the global symmetries are the half-Fermi multiplets in the last line of Table \ref{table_2dfields}, which are purely fermionic and whose contribution therefore appears only in the numerator of the elliptic genus.  

More generally, suppose that the curve $C_\beta$ is a linear combination
\be
C_\beta = \sum_{i=1}^k  \beta^i  C_i
\ee
of some Mori cone generators, $C_i$. Then the pole structure is slightly more complicated and the total elliptic genus takes the form (see e.g. \cite{Haghighat:2015ega})
\be \label{ZCbetaansatz-general}
Z_{C_\beta}(\tau,\lambda_s, {z}) = \left( \frac{1}{\eta^2(\tau)}  \right)^{6 C_{\beta} \cdot \bar K}    \frac{\Phi_{W, L, {m}}(\tau,\lambda_s,{z})}{  \prod_{i=1}^k  \prod_{s_i=1}^{\beta^i} \varphi_{-2,1}(\tau, s_i \lambda_s)} \,.
\ee
The weight $W$ and index $L$ of the numerator now  change accordingly,
 such that $Z_{C_\beta}(\tau,\lambda_s, {z})$ is still of weight zero and of $SU(2)_-$ fugacity index as given in~(\ref{m-}).

Once modular indices and weight are given, the elliptic genus $Z_{C_\beta} (\tau, \lambda_s, { z})$ is fixed up to a finite number of parameters.  This well-known
feature is a consequence of the fact that the ring of weak Jacobi forms is finitely generated  \cite{EichlerZagier}. 
So all one needs to do is to fix these parameters
by matching the general ansatz for $Z_{C_\beta} (\tau, \lambda_s, { z})$, with given modular indices and weight, to some extra physical input.

\subsubsection*{3) Relation to the topological string on $Y_3$}

The elliptic genus $Z_{C_\beta}(\tau,\lambda_s, {z})$, as given in (\ref{Zbetadefinition}), is closely related to the topological string partition function $Z_{\rm top}$ on the elliptically fibered Calabi-Yau $Y_3$, via a chain of dualities \cite{Haghighat:2013gba,Haghighat:2014vxa}. 
This was already used in the original work \cite{Klemm:1996hh} in the context of non-critical $E$-strings. More generally, 
the topological string partition function on elliptic Calabi-Yau 3-folds has been studied in detail in \cite{Klemm:2012sx,Alim:2012ss,Huang:2015sta}.

Specifically, it is known by F/M-theory duality that the 6d effective action of F-theory on $Y_3$ reduced on $S^1$ coincides 
with the 5d effective action of M-theory on $Y_3$. 
A critical or non-critical string from a D3-brane on $C_\beta$ can either wrap the $S^1$ or not. 
If it does, it produces a BPS particle in 5d. In the dual M-theory description, the 5d BPS particle arises from
 an M2-brane wrapping some curve on $Y_3$. 
If we consider for simplicity a generic Weierstrass model $Y_3$, then the string with wrapping number $\kappa=1$ along 
$S^1$ and Kaluza-Klein momentum level $n$ maps to a 5d BPS particle  
that arises from the M2-brane wrapping the curve
\be
C_\beta + n C_{\mathcal E} \,.
\ee
Here $C_{\mathcal E}$ is the class of the elliptic fiber while $C_\beta$ is the curve in the base.
Indeed, the Kaluza-Klein $U(1)$ symmetry of the $S^1$ reduction is generated by the divisor \cite{Park:2011ji}
\be
\tilde S_0 = S_0 + \frac{1}{2}\pi^{-1} \bar K \,, 
\ee 
where $S_0$ is the zero-section of the elliptic fibration $Y_3$ and $\bar K$ the anti-canonical divisor of the base.
Hence an M2-brane along $C_\beta + n C_{\mathcal E} $ has KK charge 
\be
\tilde S_0 \cdot_{Y_3} (C_\beta + n C_{\mathcal E} ) = - \frac{1}{2} \bar K \cdot_{B_2} C_\beta + n = E_0 + n \,,
\ee
where we used the relation $S_0 \cdot S_0 = -  S_0 \cdot \pi^{-1}(\bar K)$ for the zero-section of an elliptic fibration.\footnote{More precisely, for a curve $C_\beta \in H_2(B_2)$ in the base, the corresponding base curve $\frak{C}_\beta \in H_2(Y_3)$ we consider in $Y_3$ is given as $\frak{C}_\beta := S_0 \cdot_{Y_3} \pi^{-1}(C_\beta)$. By abuse of notation, we often denote this also by $C_\beta$. Then, $S_0 \cdot_{Y_3} \frak{C}_\beta  = S_0 \cdot_{Y_3}  S_0 \cdot_{Y_3} \pi^{-1}(C_\beta) = - S_0 \cdot_{Y_3}  \pi^{-1}(\bar K) \cdot_{Y_3} \pi^{-1}(C_\beta) = - \bar K \cdot_{B_2} C_\beta$. Furthermore $S_0 \cdot_{Y_3} C_{\cal E} =1$ because the fiber intersects the zero-section once. } Note that the shift by $- \frac{1}{2} \bar K \cdot C_\beta$ beautifully
coincides with the shift of the energy of the wrapped string in (\ref{Zbetaexpansion}), which is due to the Casimir energy $E_0$.

To describe the precise relation between the elliptic genus and $Z_{\rm top}$, we introduce the following notation.
A basis of $H_2(Y_3,\mathbb Z)$ can be split into a basis $\{C^{\rm b}_i\}$ of base curve classes and a basis $\{C_{\mathcal E}, C^{\rm f}_a\}$ of fibral curves, where in the latter we have singled out $C_{\mathcal E}$ as the class of the generic elliptic fiber. To these basis elements we assign the K\"ahler parameters ${\bf t} = \{ t_i \}$, $\tau $ and ${\bf z}=\{z_a\}$, respectively.
Given a base curve  $C_\beta \in H_2(B_2,\mathbb Z)$ with expansion $C_\beta = \beta^i C^{\rm b}_i$, its complexified K\"ahler volume can then be expanded as $t_\beta = \beta^i t_i$, and likewise for a fibral curve $n C_{\mathcal E} + r^a C^{\rm f}_a$, the volume is given as $n \tau + r^a z_a$. Hereafter, we will restrict ourselves to the geometries with exactly one independent non-zero section, i.e., to those effective theories with a single $U(1)$, and the subscripts in $C^{\rm f}_a$ and $z_a$, etc. will thus be removed. Furthermore, we will assume w.l.o.g. that an appropriate basis for the fibral curves $\{C_{\mathcal E}, C^{\rm f}\}$ has been chosen such that the M2-brane wrapped on $C^{\rm f}$ has $U(1)$ charge $\frak{q}=1$ and KK charge $0$. 

In this notation, the topological string partition function on $Y_3$ takes the form
\be \label{Ztopstringdef}
Z_{\rm top} = {\rm exp} \left(  {\cal F}( \lambda_s, \tau,  {\bf  t}, { z} )  \right)   =  {\rm exp} \left( \sum_{g \geq 0}   {\cal F}^{(g)}( \tau,  {\bf t}, {  z} )   \lambda_s^{2g -2} \right) \,.
\ee
Here ${\cal F}( \lambda_s, \tau,  { \bf t}, { z} ) $ denotes the topological string free energy, which enjoys a perturbation series in
contributions of genus $g$ maps, where the expansion parameter $\lambda_s$ plays the role of the string coupling.
Each genus $g$ free energy,  ${\cal F}^{(g)}(\tau,  { \bf t}, { z} )$, can in turn be expanded in terms of the exponentiated K\"ahler parameters of~$Y_3$,
\be
q = e^{ 2 \pi i \tau} \,, \qquad Q_{\beta} = e^{2 \pi i t_{\beta }} \,, \qquad \xi = e^{2 \pi i z} \,,
\ee
as
%,  $Q_C = {\rm exp}(2 \pi i  t_C)$,  and the genus $g$ Gromov-Witten invariants $N^{g}_C \in \mathbb Q$ as 
\be \label{Fgexp1}
% {\cal F}^{(g)}(\tau,  { \bf t}, {  \bf z}) = \sum_{C_\beta \in H_2(B, \mathbb Z)}  {\cal F}_{C_\beta}^{(g)}(\tau, {\bf z})   \,  \widehat Q_{\beta} \,,
 {\cal F}^{(g)}(\tau,  { \bf t}, {  z}) = \sum_{C_\beta \in H_2(B_2, \mathbb Z)}  {\cal F}_{C_\beta}^{(g)}(\tau, {z})   \,  Q_{\beta} \,.
\ee
Here, the object 
\be \label{Fgexp2}
  {\cal F}_{C_\beta}^{(g)}(\tau, {z})   =   \sum_{n, \gamma}  N^{(g)}_{C_\beta}(n, r)  \,     q^n   \,  \xi^{r}
\ee
gives an expansion in terms of the genus $g$ Gromov-Witten invariants $N^{(g)}_{C_\beta}(n, r)$, which are a priori rational numbers.
However, for a single wrapped curve classe these agree with the integral Gopakumar-Vafa BPS invariants. In the present context, single wrapping refers to the class $C_\beta$ not being a multiple of some other curve class.

Now we are ready to write the relation between the  topological partition function on $Y_3$, $Z_{\rm top}$, and  the elliptic genus,  $Z_{C_\beta}(\tau, \lambda_s, {z})$, of the  6d, $(1,0)$ supersymmetric string theory obtained by compactification of F-theory on $Y_3$ \cite{Haghighat:2013gba,Haghighat:2014vxa}:
\be \label{Ztopdual}
Z_{\rm top} = Z_0(\tau, \lambda_s, z)  \left( 1 + \sum_{ C_\beta}  Z_{C_\beta}(\tau, \lambda_s, {z}) \,  \widehat{Q}_\beta  \right) \,.
\ee
The overall factor $Z_0(\tau, \lambda_s, z)$ is the contribution to the 5d index from 5d BPS states due to the 6d supergravity modes and their Kaluza-Klein tower. 
The expansion parameter in  (\ref{Fgexp1}) is shifted as \cite{Huang:2015sta,DelZotto:2017mee}
\be
\widehat Q_{\beta} = \sigma(\tau,{z}) \, Q_{\beta} \,.
\ee
This shift has been motivated in \cite{Huang:2015sta} by requiring that  ${\cal F}_{C_\beta}^{(g)}(\tau, { z})$ enjoys good modular properties.
Specifically, if the base curve $C_\beta$ is not contained in the discriminant, the shift is given as $\sigma(\tau, z) = e^{\pi i (C_\beta \cdot \bar K) \tau}$, that is, 
\be \label{widehatQbeta}
\widehat Q_{\beta} = q^{\frac{1}{2} C_\beta \cdot \bar K} \, Q_{\beta}  = e^{2 \pi i  (t_\beta + \frac{1}{2} (C_{\beta} \cdot \bar K)  \tau ) } \,.
\ee

Observe, importantly, that by virtue of the duality, $\tau$ can be viewed as either the modular parameter of the elliptic genus, or as the K\"ahler parameter of the elliptic fiber of $Y_3$.  Similarly, $\lambda_s$ can be interpreted as either the elliptic parameter with respect to $SU(2)_-$, or as the topological  string coupling; moreover $z$ as either the elliptic parameter with respect to the global $U(1)$ symmetry of the string, or as the fibral K\"ahler parameter of $Y_3$.

We can elucidate the relation between the free energy of the topological string and the elliptic genus by  
taking the logarithm of (\ref{Ztopstringdef}) and expanding
\bea \label{calFZexpansion}
\sum_{g \geq 0} {\cal F}^{(g)}( \tau,  {\bf t}, {  z} )   \lambda_s^{2g -2} &=& {\rm log} Z_0(\tau, \lambda_s, z) +  {\rm log} \left(1 + \sum_{C_\beta} Z_{C_\beta}(\tau,\lambda_s,{z}) \, \widehat Q_\beta \right) \\
 &=& {\rm log} Z_0(\tau, \lambda_s, z) + \sum_{C_\beta} Z_{C_\beta} (\tau,\lambda_s,{z})   \, \widehat Q_\beta  - \frac{1}{2}   \sum_{C_{\beta'}, C_{\beta''}} Z_{C_{\beta'}} Z_{C_{\beta''}}\widehat Q_{\beta'} \widehat Q_{\beta''} \nonumber \\ 
 && + 
 \nonumber \ldots \,.%\ldots \nonumber \,.
\eea
A considerable simplification occurs if $C_\beta$ corresponds to one of the generators of the Mori cone and therefore cannot be written as the 
 sum of two effective curves $C_{\beta'}$ and $C_{\beta''}$. 
 In this case, the elliptic genus obeys the simpler form (\ref{ZCbetaansatz}). 
Furthermore, in view of   (\ref{Fgexp1}) we can identify 
\be \label{ZCbetagen1}
 \sum_{g \geq 0}   {\cal F}^{(g)}_{C_\beta}( \tau, {  z} )  \, \lambda_s^{2g -2}      = q^{\frac{1}{2}  C_\beta \cdot \bar K}  \, Z_{C_{\beta}} ( \tau, \lambda_s, {  z} )  =
\left( \frac{q^{\frac{1}{2}C_\beta \cdot \bar K} }{\eta(\tau)^{12C_{\beta} \cdot \bar K} }  \right)    \frac{\Phi_{W, L, { m}}(\tau,\lambda_s,{z})}{   \varphi_{-2,1}(\tau, \lambda_s)} \,.
\ee
The important factor of $q^{\frac{1}{2} C_\beta \cdot \bar K}$ is due to the shift in $\widehat Q_\beta$.
In view of (\ref{Zbetaexpansion}) and (\ref{E0formular}) it precisely cancels the overall factor $q^{E_0}$ in  $Z_{C_{\beta}} ( \tau, \lambda_s, {  z} )$, as is necessary for the match with the topological string prepotential to work.
Due to this extra factor, while  $Z_{C_{\beta}} ( \tau, \lambda_s, {  z} )$ is a weak Jacobi form of weight zero, the LHS and RHS of (\ref{ZCbetagen1})  do not have good modular properties.

One may wonder what this implies for the modular {behaviour} of
the  $ {\cal F}^{(g)}(\tau,{z})$, viewed as expansion coefficients of  the free energy
with regard to the elliptic parameter $\lambda_s$.
For this, note that a weak Jacobi form of weight $w$ and index $m$ has the  expansion
\be \label{varphijquasimodularexp}
\varphi_{w,m}(\tau, z) = \sum_{j=0}^\infty  \phi_j(\tau)  z^j \,,
\ee
where $\phi_j(\tau)$ are in general only {\it quasi-modular}\footnote{See, e.g., page 16 of \cite{Gu:2017ccq} for a review of this fact. The same conclusion has been obtained in \cite{Klemm:2012sx,Alim:2012ss,Huang:2015sta} without making use of the duality, solely based on properties of the topological string partition function for elliptic fibrations.} forms of weight $j+w$. {As explained at the end of Appendix \ref{app_Jacring}, this} means that in general the $\phi_j(\tau)$ must lie in the ring generated by {the modular Eisenstein functions $E_4(\tau)$ and $E_6(\tau)$, as well as the quasi-modular Eisenstein function $E_2(\tau)$; the quasi-modular transformation behaviour of $E_2(\tau)$ can be found in (\ref{E2quasimodularity}).}
In the same spirit, we may expand  the numerator on the RHS of eq.~(\ref{ZCbetagen1}), 
\be \label{PhiWLMexpinlambda}
 \Phi_{W, L, { m}}(\tau,\lambda_s,{z}) =     \Phi_{W, { m}}(\tau,{z}) + {\cal O}(\lambda_s) \,,
 \ee
as well as use the expansion (\ref{varphim21exp1}) of  the denominator,
\be \label{lsexpofphi}
\varphi_{-2,1}(\tau, \lambda_s)  = - \lambda_s^2  \, (1 + {\cal O}(\lambda_s^2)) \,.
\ee
Comparing the terms of order $\lambda_s^{-2}$ of both sides of (\ref{ZCbetagen1}) finally identifies the genus-zero prepotential with the elliptic genus as
\be \label{F0inPhiexpansion}
{\cal F}^{(0)}_{C_\beta}( \tau, {  z} )  =  - \frac{q^{\frac{1}{2} C_\beta \cdot \bar K   }}{ \eta(\tau)^{12 C_{\beta} \cdot \bar K}  }   \Phi_{W, { m}}(\tau,{z})  \,.
\ee

Note that the  expansion (\ref{PhiWLMexpinlambda}) has higher order terms 
only if the index $L$ (as given in (\ref{WLindices}))  is non-zero. In this case, the function  $\Phi_{W, { m}}(\tau,{z})$ will be generically only
a quasi-modular form, which follows from the remarks above. This fits with the fact that the genus-zero prepotential satisfies an in general non-trivial modular anomaly equation \cite{Klemm:2012sx,Alim:2012ss,Huang:2015sta}. 
On the other hand, if $C_\beta$ is a genus-zero curve {corresponding to a Mori cone generator}, then $L=0$ and hence  $\Phi_{W, { m}}(\tau,{z})$ gives the complete numerator 
of eq.~(\ref{ZCbetagen1}), because
$ \Phi_{W, L=0,{ m}}(\tau,\lambda_s,{z})$ is independent of $\lambda_s$. In this case   $ q^{-\frac{1}{2} C_\beta \cdot \bar K} {\cal F}^{(0)}_{C_\beta}( \tau, {  z} ) $ is by itself a good modular Jacobi form and fully determines the elliptic genus.
Recall, however, that to arrive at this result, we have had to make the assumption that $C_\beta$ corresponds to a Mori cone generator as otherwise (\ref{ZCbetaansatz}) would not hold. 
In this case, the relation between $Z_{C_\beta}(\tau,\lambda_s,{z})$ and the genus-zero prepotential is more complicated, as will be explained below; what is still correct, even in this more general situation, is the ansatz (\ref{F0inPhiexpansion}) for ${\cal F}^{(0)}_{C_\beta}( \tau, {  z} )$, with $\Phi_{W, { m}}(\tau,{z})$ now being only quasi-modular.

%----------------------------------------------------------------------------------------------------------
\subsubsection*{4) Relation to ${\cal N}=(0,4)$ elliptic genus $Z_\otherK3$ }

After these preparations we are  ready to apply the formalism reviewed above to our  object of interest, namely the elliptic genus $Z_\otherK3(\tau,z)$
of the heterotic sting on the $K3$ geometry $\otherK3$.
There are two cases to distinguish.
Recall from Section~\ref{sec_tensionlessstrings} that if the self-intersection-zero rational curve $C_0$ is the fiber of a Hirzebruch surface $\mathbb F_a$, there exists a perturbative heterotic dual, and in the tensionless  limit under consideration the heterotic string becomes weakly coupled. 
As will be explained in more detail in Section~\ref{subsec_F1base}, the fiber of a Hirzebruch surface is indeed a generator of the Mori cone and therefore the spin-refined elliptic genus $Z_{C_0}(\tau,\lambda_s,z)$ takes the simple form (\ref{ZCbetaansatz}).
Furthermore, the elliptic genus without the spin refinement is the leading term in a $\lambda_s$ expansion of $Z_{C_0}(\tau,\lambda_s,z)$ around $\lambda_s = 0$.
According to the discussion around (\ref{F0inPhiexpansion}) this coincides up to a sign and a factor of $q$ with the genus-zero prepotential of the topological string on $Y_3$. That is, we can identify:\be \label{ZKgenus0}
Z_\otherK3(\tau,z) = - q^{-1}  {\cal F}^{(0)}_{C_0}( \tau, {  z} )  = -  q^{-1}  \sum N^{(0)}_{C_0}(n,r) \,   q^n \, \xi^r  \,.
\ee
 This fits perfectly with the fact that, up to the factor of $q$, the genus-zero prepotential is by itself a modular form of precisely the correct weight to furnish the elliptic genus (\ref{Kindex}) for the perturbative heterotic string.\footnote{Note that for this be true, the GW invariants $N^{(0)}_{C_0}(n,r)$ must be integers, which is the case for the expansion~(\ref{Fgexp1}) where we consider only
 terms linear in $Q_0$.}

More generally, suppose that $C_0$ lies in a non-Hirzebruch surface. In this case, it will generically not be one of the Mori cone generators. 
Assume for definiteness that $C_0 = C_1 + C_2$ for two Mori cone generators $C_1$ and $C_2$. 
The general relation (\ref{calFZexpansion}) now allows us to write
\be
Z_{C_0}(\tau,\lambda_s,z) = {\cal F}_{C_0}(\tau,\lambda_s,z) + \frac{1}{2} {\cal F}_{C_1}(\tau,\lambda_s,z)  {\cal F}_{C_2}(\tau,\lambda_s,z) \,,
\ee
where ${\cal F}_{C_i}(\tau,\lambda_s,z) = \sum_{g\geq 0} \lambda_s^{2g-2}{\cal F}^{(g)}_{C_i}(\tau,\lambda_s,z)$ is the all-genus prepotential that multiplies $Q_{C_i}$ in the expansion of the topological string free energy.
From (\ref{ZCbetaansatz-general}) we see that the denominator of $Z_{C_0}$ contains the product $\varphi^2_{-2,1}(\tau, \lambda_s)$. The leading term in the $\lambda_s$ expansion {(\ref{lsexpofphi})} hence goes as $1/\lambda_s^4$. This term must equal the $1/\lambda_s^4$ contribution from $\frac{1}{2} {\cal F}_{C_1}  {\cal F}_{C_2}$. The next term is of order $1/{\lambda_s^2}$ and receives contributions from both the single string prepotential  ${\cal F}_{C_0}$ and the 2-string prepotential ${\cal F}_{C_1}  {\cal F}_{C_2}$.
 The `1-string irreducible' contribution is only due to the leading piece in $ {\cal F}_{C_0}$, which again coincides with the genus-zero prepotential. 
 Hence it is natural to identify this with the non-spin-refined elliptic genus describing the dual, non-perturbative heterotic string. This reasoning leads to the same expression (\ref{ZKgenus0}).
 The genus-zero free energy has to have the correct modular properties for this to be possible: Up to the same factor of $q$, it must be quasi-modular 
 in accordance with the
 modular anomaly equation~\cite{Klemm:2012sx,Alim:2012ss,Huang:2015sta}, which under the present assumptions takes the form
\beq\label{modular-anomaly}
\frac{\partial \mathcal F^{(0)}_{C_0}}{\partial E_2} = \frac{1}{24} \sum_{C_1+C_2= C_0} (C_1 \cdot C_2) \,\mathcal F^{(0)}_{C_1}\mathcal F^{(0)}_{C_2} \,.
\eeq
This generically leads to a dependence on the quasi-modular Eisenstein series $E_2(\tau)$ for the non-spin-refined elliptic genus advertised in Section~\ref{sec_ellgenus}. 
 We will exemplify this behaviour for an F-theory base $dP_2$ in Section~\ref{sec_dP2example}. In particular we will verify that the genus-zero prepotential associated with $C_0$ correctly encodes the massless spectrum as expected from the geometry. 

%----------------------------------------------------------------------------------------------------------

\subsubsection*{5) Multi-wrappings}

As a side remark, it is also instructive to consider the situation where the D3-brane is wrapped not just once, but $k$ times on $C_0$.
 This can be investigated by computing via mirror symmetry the terms with higher powers of $Q_0$ of the prepotential
\be
\mathcal F^{(0)}(Q_0,q,\xi)\ =\ \mathcal F^{(0)}_{{\rm class}} +\sum n^{(0)}_{k\cdot C_0}(n,r)\,{\rm Li}_3({Q_0}^k q^n\xi^r)\,.
\ee
Here $n^{(0)}_{k\cdot C_0}(n,r)$ denotes the (integral) Gopakumar-Vafa invariants.
In fact, multi-wrapped heterotic strings are expected~\cite{Klemm:1996hh,Minahan:1998vr,Haghighat:2015ega} not to form bound states, so that the 
elliptic genus for multiple heterotic strings should be given \cite{Dijkgraaf:1996xw,Haghighat:2014pva}
in terms of the  elliptic genus of the $k$-fold symmetric product of $\otherK3$. Taking multi-coverings into account, this translates into
\be\label{multiwrap}
\mathcal F^{(0)}_{k\cdot C_0}(q,\xi) \ =\  - q^k Z_{Sym^k\otherK3}(q,\xi)  -q^k\mathcal F^{(0)}(Q_0,q,\xi) {\big \vert}_{{Q_0}^k} \,,\ \ k>1\,,
\ee
where for simplicity we assumed $k$ to be prime. Above, 
\be
Z_{Sym^k\otherK3}(q,\xi)\ =\   T_k^{(-2)}(Z_\otherK3(\tau,z))\,,
\ee
where the Hecke transform of a Jacobi form $\varphi$  of modular weight $w$  is defined by \cite{EichlerZagier}
\be
T_k^{(w)}\varphi(\tau,z)\ =\ k^{w-1} \sum_{{ad=k}\atop{b\ {\rm mod}\ d}}d^{-w}
\varphi\left({a\tau+b\over d},a\,z\right).
\ee
We have checked for the examples we will consider later that (\ref{multiwrap}) is indeed satisfied, to some low orders of
expansions in $Q_0,q,\xi$.  This  implies for the relevant Gopakumar-Vafa BPS invariants that $n^{(0)}_{k\cdot C_0}=0$ for $k>1$, which in turn  
means that the heterotic strings do not form bound states, as expected.
While this fact has been known since long for perturbative heterotic strings \cite{Klemm:1996hh,Minahan:1998vr}, it may appear
slightly less trivial for non-perturbative heterotic strings. This is why we will re-address this point for the example  in Section~\ref{sec_dP2example}.

%-------------------------------------------------------------------------------------------------------------------------------------

\subsection{Global Mordell-Weil $U(1)$ symmetries } \label{subsec_GlobalMWU1}

It is well-known \cite{Morrison:1996na,Grimm:2010ez,Park:2011ji,Morrison:2012ei} that $U(1)$ gauge symmetries arise in F-theory when the elliptic 3-fold $Y_3$ exhibits a non-trivial Mordell-Weil group of rational sections. {The $U(1)$ gauge symmetries  } translate into global symmetries of the tensionless string. 
Although we will consider geometries with just  one independent non-zero section, and hence models with a single $U(1)$ factor, the generalization to multiple $U(1)$ factors is immediate. We now explain how the index $m$ of the elliptic genus $Z_{C_\beta} (\tau, \lambda_s, { z})$,{(\ref{ZCbetaansatz}) or (\ref{ZCbetaansatz-general})}, with respect to the background field $z$ is determined by the geometry of the Mordell-Weil group of the elliptic fibration.\footnote{Already in \cite{Klemm:1996hh,Huang:2013yta} BPS numbers for examples of elliptic fibrations with extra rational sections have been computed and a refinement with respect to the associated $U(1)$ charges has been observed. Our results hold very generally for any type of rational section generating an abelian gauge symmetry in F-theory.}

As already pointed out, this index is given by { the  terms of 
$f_{C_\beta}(\lambda_s, {z})$ in (\ref{holanomaly}) which are quadratic in $z$} as a consequence of the general relation (\ref{partialE2equ}). 
The function $f_{C_\beta}(\lambda_s, {z})$ in turn can be identified with the 't Hooft anomaly potential of the world-sheet theory of the string \cite{Benini:2013xpa,DelZotto:2016pvm,DelZotto:2017mee,DelZotto:2018tcj}. 

The anomaly polynomial of a single ${\cal N}=(0,4)$ world-sheet supersymmetric string has been already
computed in \cite{Berman:2004ew,Ohmori:2014kda,Shimizu:2016lbw,Kim:2016foj,Lawrie:2016axq},  however for situations in which the the F-theory gauge group only contains non-abelian factors. In our notation it takes the form
\be
\ba \label{I4}
{I}_4 &= - \frac{1}{4} ( \bar K \cdot C_\beta)\,   p_1(T) - \frac{1}{2} \sum_I {\rm Tr} F_I^2 \,  ([\Sigma_I] \cdot C_\beta)  + \frac{1}{2} \tr F_{\cal R}^2       \cr
&  + \frac{1}{2} \tr F_{+}^2     \left( \frac{1}{2} C_\beta \cdot C_\beta + \frac{1}{2} \bar K \cdot C_\beta  \right)    + 
\frac{1}{2} \tr F_{-}^2   \left( - \frac{1}{2} C_\beta \cdot C_\beta + \frac{1}{2} \bar K \cdot C_\beta  \right)  \,.
\ea
\ee
Here $F_-$ and $F_+$ refer to the field strengths associated with the global symmetries $SU(2)_-$   and $SU(2)_+$ in (\ref{SO4T}), and  $F_{\cal R}$ denotes the $SU(2)$ R-symmetry of the 6d $N=(1,0)$ theory that is inherited by the string.\footnote{In Table \ref{table_2dfields}, the fermions of the twisted hypermultiplet transform as a ${\bf 2}$ of this symmetry, while all other fermions are singlets. This explains why the  't Hooft anomaly coefficient is a universal constant. } 
Finally $F_I$ is the field strength associated with a non-abelian gauge algebra $\mathfrak{g}_I$ of a 7-brane wrapping the divisor component $\Sigma_I$. The trace is normalised with respect to the trace in the fundamental representation via ${\rm Tr} F^2_I = \frac{1}{\lambda_I} {\rm tr} F_I^2$,  where $\lambda_I$ is 
the Dynkin index (in particular, $\lambda_I =1$ for $\mathfrak{su}(N)$).

To read off the index with respect to the global symmetry $SU(2)_-$, one identifies $I_4$ with $f_{C_\beta}(\lambda_s, {z})$ upon replacing
\be \label{replacementforls}
\frac{1}{2} \tr F_{-}^2   \rightarrow  - \lambda_s^2, \qquad  p_1(T) \rightarrow 0, \qquad \frac{1}{2} \tr F_I^2 \rightarrow 0, \qquad  \frac{1}{2} \tr F_{+}^2 \rightarrow 0 \,.
\ee
This reproduces the result (\ref{m-}) for the value of the index with respect to $\lambda_s$. 
Similarly one can obtain the index with respect to non-abelian global symmetries of the string from the term $- \frac{1}{2} \sum_I {\rm Tr} F_I^2 \,  ([\Sigma_I] \cdot C_\beta)$ in the anomaly polynomial. 

We now come to the case of interest for us where we have a single  $U(1)$ gauge symmetry. 
The 't~Hooft anomaly polynomial of the stringy world-sheet contains an additional term
\be
I_4 \rvert_{U(1)} = - \frac{1}{2}  F^2 \,  (b \cdot C_\beta)\,,
\ee
where $b = - \pi_\ast (\sigma(S) \cdot \sigma(S) )$ is the { { height pairing} of the section $S$ as introduced in Section \ref{sec_FonY3}}. The precise form of this term including its normalization has been derived from first principles in \cite{Weigand:2017gwb}. The normalization is in conventions where the lowest charge with respect to $U(1)$ is given by $\frak{q}_{\rm min} =1$. 
Note that, on the other hand, the index $m$ of a weak Jacobi form is defined in a normalization where the weight lattice is $\mathbb Z/\sqrt{2}$, or equivalently, that the smallest co-weight has norm 2, see Appendix \ref{app_Jacring}.  
Since the charges are determined by intersection numbers of the Shioda map $\sigma(S)$ with suitable fibral curves of $Y_3$, we must rescale the section by $\frac{1}{\sqrt{2}}$ and hence  the height pairing $b$ by $\frac{1}{2}$.
Taking this into account, and making the replacement
\be
 \frac{1}{2}  F^2 \rightarrow -  z^2\/,
\ee
in analogy to (\ref{replacementforls}), we conclude that  the index with respect to the $U(1)$ fugacity $z$ is
\be
m = - \frac{1}{2} \pi_\ast (\sigma(S) \cdot \sigma(S)) \cdot C_\beta  = \frac{1}{2} b \cdot C_\beta\,.
\ee
Note that this quantity is guaranteed to be integer because $b$ is an even divisor as can be seen from (\ref{bAexpression}). 
%This ties together with its interpretation as the level of the underlying $U(1)$ Kac-Moody algebra in the language of heterotic strings. 

%-----------------------------------------------------------------------------------------------------------------
\subsection{Synopsis: determining the maximal charge per excitation level} \label{sec_genstratmaxcharge}

We now  combine the arguments of the foregoing sections and outline how to practically determine (a subset of) the $U(1)$ charge spectrum of the nearly tensionless heterotic string we consider. Recall that the string arises from a wrapped D3-brane on a shrinking curve $C_0$ 
in the base of an elliptic 3-fold, $Y_3$.
As a first step one makes   a general ansatz for the quasi-modular form $\Phi_{W, {m}}(\tau,{z})$ in eq.~(\ref{F0inPhiexpansion}) of weight
  $W = 6 C_{0} \cdot \bar K-2=10$ and $U(1)$ fugacity index ${ m} = \frac{1}{2} C_0 \cdot b$. The ansatz consists of a linear combination of all monomials in the generators of the ring of quasi-modular  Jacobi forms (see Appendix \ref{app_Jacring}), { i.e.},
  \be \label{E2E4E6varphiansatz}
 \Phi_{10,m}(\tau,z)\ = \sum c_{a_0,a_1,a_2,a_3,a_4}  E_2^{a_0} (\tau) \,  E_4^{a_1}(\tau) \, E_6^{a_2}(\tau)  \, \varphi^{a_3}_{0,1}(\tau, {z}) \, \varphi_{-2,1}^{a_4}(\tau, z) \, 
  \ee
with
  \be
  \qquad 2 a_0 + 4 a_1 + 6 a_2 -2 a_4 = 6 C_{0} \cdot \bar K-2\ =\ 10, \qquad 
  a_3 + a_4 = \frac{1}{2} C_0 \cdot b = m\,.
  \ee
As furthermore explained in Appendix \ref{app_Jacring}, 
  the only  source of quasi-modularity in $\Phi^{(0)}_{10, {m}}(\tau,{z})$ are possible factors of $E_2(\tau)$.
 { We have discussed in Section \ref{sec_ellgenus} that} if $C_0$ is a Mori cone generator and of genus zero, there cannot be an explicit dependence on $E_2(\tau)$, as in this case  the prepotential is modular, not quasi-modular.  This immediately implies that $\Phi_{10,m}(\tau,0)= E_4(\tau)E_6(\tau)$, which is a classic result for perturbative heterotic strings on $K3$ \cite{Schellekens:1986xh}. This is in particular the case where $Y_3$ is a fibration over a Hirzebruch surface, ${\mathbb F}_a$. For non-perturbative heterotic strings, however, $\Phi_{10,m}(\tau,0)$ may not reduce to $E_4(\tau) E_6(\tau)$ due to the possible appearance of extra terms involving $E_2(\tau)$; this reflects the fact that there can be a variable number of extra massless tensor fields which contribute to the gauge and gravitational anomalies in six dimensions.
  
In any concrete model, the ansatz (\ref{E2E4E6varphiansatz})  leaves us with a finite number of free parameters, which can be fixed by comparison with the
genus-zero prepotential via eq.~(\ref{F0inPhiexpansion}).  Specifically, for the geometry we consider we have
\bea
{\cal F}^{(0)}_{C_0}( \tau, {  z} )  &\equiv& \sum N^{(0)}_{C_0}(n,r) \,   q^n \, \xi^r \\
&=&  - \frac{q}{ \eta(\tau)^{24}  }  \,    \Phi_{10, {m}}(\tau,{z})  \ \equiv\ -q\,Z_\otherK3(\tau,{z}) \ .\nn
 \eea
It suffices to consider a small number of Gromow-Witten invariants $N^{(0)}_{C_0}(n,r)$ of $Y_3$, 
which can be straightforwardly computed by using mirror symmetry.
Having obtained  in this way an exact, analytic expression for the $U(1)$-graded elliptic genus $Z_\otherK3(\tau,{z})$, one can then go on and analyze the charge-versus-mass spectrum in order to investigate various weak gravity conjectures.  This will be done explicitly for a number of examples in the following Section \ref{Ellgenusexamples} and in Appendix~\ref{sec_HirzebruchF2}.
 
However, even without analyzing concrete models, 
one can deduce important general properties of the $U(1)$ charges of the string excitations that are encoded in the elliptic genus. 
These properties follow from the modular properties of Jacobi forms. One way
to make use of these is to note that the generators $\varphi_{0,1}(\tau, {z})$ and $\varphi_{-2,1}(\tau, z)$ that appear {in} the ansatz  (\ref{E2E4E6varphiansatz}), 
can be rewritten in terms of the Eisenstein-Jacobi series $E_{w,l}(\tau,z)$ via the identities (\ref{phi0102Eisen1}).
This property of the Eisenstein-Jacobi series allows us to establish a lower bound for the maximal charge of the states at any given mass level $n$, 
as discussed in detail in Appendix~\ref{app_EJmaxcharge}.  The result of this analysis is the general behaviour that we mentioned already in Section~\ref{sec_chargesummary}: namely, in a model with $U(1)$ index $m = \frac{1}{2} C_0 \cdot b$, the maximal charge per excitation level $n$ is bounded as follows:
\be \label{betasharper3}
\frak{q}^2_{\rm max}(n) \geq   (1 - \epsilon)  4 \,  m   \, n  =:  \beta_\epsilon \,  n \qquad    \forall \, n \geq N(\epsilon)   \,.       %      \beta(n)   \,  n   \,  
\ee
This part of the analysis is completely general and does not rely on any explicit evaluation of the elliptic genus for a given model.
On the other hand, in order to find the actual value of $N(\epsilon)$ for given~$\epsilon$, we must explicitly analyze the elliptic genus by following the steps outlined above.  In particular, this will lead to lower, universal bounds for $\frak{q}^2_{\rm max}(n)$ valid for all values of $n$ and not just for $n$ above a certain limit,
$N(\epsilon)$. 

In a similar vein one can make use of the fact that any weak Jacobi form,
and thus in particular the elliptic genus, can be expanded as \cite{EichlerZagier}\footnote{Correspondingly
 one may describe the $U(1)$ sector of the theory in terms of a free periodic boson $H$ that is associated with a Kac-Moody current 
of level $m$ of the form $J=i\sqrt m\partial H$. Its winding states are then counted by the theta functions below. This is analogous to the familiar  $U(1)$ current that is part of an $N=2$ superconformal algebra on the world-sheet.}
\be\label{thetadef}
 \varphi_{w, {m}}(\tau,{z})\ =\ \sum_{\ell\in\IZ\, {\rm mod}\,2m} h_\ell(\tau)\Theta_{m,\ell}(\tau,z)\ .
\ee
Here 
 \be
 \Theta_{m,\ell}(\tau,z)\ =\ \sum_{k\in\IZ}q^{(\ell+2mk)^2/4m}\xi^{\ell+2mk}
 \ee
is  the standard theta function of index $m$ and characteristic $\ell$, and the coefficient functions $h_\ell(\tau)$ are certain vector-valued modular forms of weight $w-1/2$. A particular distinguished subset of states is defined by $\ell=0$, for which we can read off
\be
n(k)=m\,k^2\,,\qquad \frak q_k=2\,m\,k  \,, \qquad  k\in\IZ\,.
\ee
These are extremal in the sense that they saturate the bound (\ref{betasharper3}) 
\be\label{exbound}
{\frak q_k}^2\ =\  4 \,  m   \, n(k) \,,
\ee
which amounts to a vanishing ``discriminant'', $D\equiv 4mn-{\frak q}^2=0$.
Note that they form a sublattice with spacing $\Delta \frak q=2m$. 
Obviously, all these charges strictly satisfy the weak gravity bound  (\ref{SLWGCa}),
which translates to 
\be\label{wgbound}
{\frak q}(n)^2\ \geq\  4 \,  m   \, (n-1) \,.
\ee
The other states with $\ell\not=0$ decompose into conjugacy classes of charge lattices shifted by $\ell$ and will generically not saturate the bound (\ref{exbound}). 
Which precise subset of these states satisfies the weak gravity bound (\ref{wgbound}) depends on the model and may be determined by a case-by-case analysis.  Recall, however, that given that the $\ell=0$ states already satisfy the gravity bound, the $\ell\not=0$ states are irrelevant for consistency of the theory with the Sublattice Weak Gravity Conjecture.

%-----------------------------------------------------------------------------------------------------------------
\section{Examples of $F$-theory Models with $U(1)$ Symmetry} \label{Ellgenusexamples}

In this section we explore various explicit 6-dimensional F-theory compactifications with a single  $U(1)$ gauge group. Our main purpose is to explicitly compute the elliptic genus of heterotic strings that arise from wrapped D3-branes on shrinking rational curves, $C_0$, of self-intersection zero on the base $B_2$ of an elliptic 3-fold $Y_3$. The results of this section serve as concrete examples for the advertised general properties of the massive, $U(1)$ charged string excitations, in the limit of vanishing tension and global $U(1)$ symmetry. 
  
We will start in Section~\ref{sec_dP1} by analyzing two explicit models on the Hirzebruch surface $B_2 =\mathbb F_1$. In this case the string associated with the vanishing curve $C_0$ is the perturbative, critical heterotic string. We will  see that indeed, starting from the geometry and applying mirror symmetry, the correct $U(1)$ refinement of the elliptic genus is obtained for the perturbative heterotic string theory on the $K3$-surface $\otherK3$, defined in (\ref{heteroticfibrationr}). Exactly the same method can systematically be applied to models for more general toric bases $B_2$, e.g., any other Hirzebruch surfaces $\mathbb F_a$ and the toric del Pezzo surfaces, $dP_r$ ($r \leq 3$). 
For more Hirzebruch examples, the results for $\mathbb F_2$ are collected in Appendix~\ref{sec_HirzebruchF2}. 

For a further illustration, a fibration over $B_2=dP_2$ will be analyzed in Section~\ref{sec_dP2example}.
On this non-Hirzebruch base the string associated with $C_0$ does not become a perturbative heterotic string, and correspondingly we will observe 
that the elliptic genus is different from the familiar elliptic genus of the perturbative heterotic string. 

In all these examples the type of $U(1)$ gauge symmetry is engineered by realising the fiber of the elliptic 3-folds $Y_3$
as a most general hypersurface of degree $4$  within  the space ${\rm Bl}_1\mathbb P^2_{112}$. Such an elliptic fibration has a Mordell-Weil group of rational sections of rank one \cite{Morrison:2012ei}. It is cut out by the hypersurface
\be \label{MP-hypers}
\hat P =  
 s w^2   + b_{0}  s^2 u^2 w  + b_1s u v w  + b_2 v^2 w 
 + c_{0}s^3 u^4   + c_{1} s^2  u^3 v  + c_{2}s u^2 v^2   + c_3 u v^3 \,,
\ee
where $[u : v : w :s]$ are the homogeneous coordinates of the toric fibral ambient space ${\rm Bl}_1\mathbb P^1_{112}$. The $b_i$ and $c_i$ are sections of certain bundles on $B_2$, the associated cohomology classes of which take the form
\be
\begin{aligned}
& [b_0] = \beta \,,  \qquad [b_1] = \bar K \,,  \qquad [b_2] = 2 \bar K - \beta \,, \qquad [c_0] = 2 \beta \,, \cr   \label{cibetadetf}
& [c_1] = \beta + \bar K \,, \qquad [c_2] = 2 \bar K \,, \qquad [c_3] = 3 \bar K - \beta \,, \qquad [c_4] = 4 \bar K -2 \beta \,.  
\end{aligned}
\ee
Apart from the concrete choice of base space $B_2$, the fibration is therefore specified by a class $\beta \in H^2(B_2,\mathbb Z)$. It  must satisfy the condition
\be \label{betaconstraint-a}
0 \leq \beta \leq 2 \bar K
\ee
in order for all $b_i$ and $c_i$ to be realisable as holomorphic polynomials on $B_2$. Here, an inequality between two divisor classes indicates that the difference of the divisors is effective.  
The class $\beta$ also defines the transformation behaviour of the fiber coordinates, which are  sections of the bundles
\bea
\label{fiberbundles}
&& u \in H^0(Y_3, {\cal L}_u) \,, \, \qquad v \in H^0(Y_3, {\cal L}_u \otimes {\cal L}_s \otimes {\cal O}(\beta - \bar K))  \,, \,\qquad w \in H^0(Y_3, {\cal L}_u^2 \otimes  {\cal L}_s \otimes {\cal O}(\beta) ) \,, \cr
&& s \in H^0(Y_3, {\cal L}_s) \,.
\eea
Let us denote by 
\be
S_0= \{u=0\} \,, \qquad \quad S= \{s = 0\}
\ee
the divisor classes associated with the zero-section and an extra independent rational section, respectively. The extra section $S$ is responsible for
the $U(1)$ gauge symmetry whose global limit we investigate. Since $\pi_\ast(S_0 \cdot S) = b_2$, the Shioda homomorphism (\ref{Shioda-homo}) takes the form\footnote{By slight abuse of notation, we denote by $b_2$ also the divisor associated with $b_2=0$ on $B_2$.}
\be
\sigma(S) =  S - S_0  - \pi^{-1}(\bar K + b_2) \,,
\ee
where the last term is computed via~\eqref{DinSigma} as
\beq
\frak{D}=\pi^{-1}(\pi_*(S-S_0)\cdot S_0) = \pi^{-1} (b_2+\bar K) \,.
\eeq
This divisor can be seen to generate the $U(1)$ gauge symmetry in the 6d F-theory effective action. Of special importance for us is the height-pairing 
\be
b = - \pi_\ast(\sigma(S) \cdot \sigma(S)) = 2 \bar K + 2 [b_2] = 6 \bar K  - 2 \beta   \,.
\ee
As reviewed in Section~\ref{sec_FonY3}, we can define a holomorphic curve $C$ with class $[C] = b$ such that its volume determines the gauge coupling of $U(1)$ as in (\ref{gaugecoupling}).

The fibration $Y_3$ over $B_2$ contains two types of curve classes:
First, as recalled in Section~\ref{sec_ellgenus}, the classes in $C_\beta \in H_{2}(B_2)$ can be used to define the base curve classes $S_0 \cdot \pi^{-1}(C_\beta )\in H_2(Y_3,\mathbb Z)$, both of which will be denoted by $C_\beta$ by abuse of notation. The fibral curve classes, on the other hand, are given by the full fiber class $C_{\mathcal E}$
and one more fibral curve class $C^{\rm f}$ with the respective properties
\bea
&S_0 \cdot C_{\mathcal E} = 1 \,, \qquad S \cdot C_{\mathcal E}  =1  & \qquad \Longrightarrow \quad  \sigma(S) \cdot  C_{\mathcal E} = 0 \label{intfibrala1} \\
&S_0 \cdot C^{\rm f} = 0 \,, \qquad S \cdot C^{\rm f}  =1  &\qquad \Longrightarrow \quad  \sigma(S) \cdot  C^{\rm f} = 1 \label{intfibrala2}\,.
\eea

There are two types of massless, charged hypermultiplets in the 6d F-theory compactification, which are localised in the fiber over the following two loci on $B_2$ \cite{Morrison:2012ei},
\be \label{CICIIDef}
\begin{aligned}
&C_I = \{b_2 = 0\} \cap \{c_3 = 0\}: \qquad &\frak{q}_I = 2 \cr
&C_{II} =    V(f_1, f_2)  \setminus C_I: \qquad &\frak{q}_{II} = 1  \,,
\end{aligned}
\ee
where $  V(f_1, f_2) $ denotes the vanishing ideal of the functions 
\begin{equation}
\begin{aligned}
&f_1 =  - c_1 b_2^4 +b_1b_2^3 c_2+b_0 b_2^3 c_3-b_1^2 b_2^2 c_3-2 b_2^2c_2 c_3+3 b_1 b_2 c_3^2-2c_3^3 \\
& f_2 = -c_3^2 b_0^2 +b_1
   b_2 b_0^2 c_3-b_1 b_2^2 b_0
   c_1+b_2^2 c_1^2+b_1^2 b_2^2
   c_0-4 b_1 b_2 c_0 c_3+4 c_0
   c_3^2 \,.
\end{aligned}
\end{equation}
Both $C_I$ and $C_{II}$ represent a set of isolated points on $B_2$.
Over each of these points, the elliptic fiber factorises into two components.
One of the two fibral components over the locus $C_I$ is a rational curve in class $2 C^{\rm f}$, and one of the components over $C_{II}$ is the rational curve in class $C^{\rm f}$.
M2-branes wrapping these isolated rational curves give rise to states of charge $\frak{q}_I =2$ and $\frak{q}_{II}=1$, respectively. These become massless in the F-theory limit and furnish the advertised charged hypermultiplets.
Note that for the special value $\beta = 2 \bar K$, the class $[b_2] = 0$ and hence the charge-two state locus $C_I$ is absent.

%------------------------------------------------------------------------------------------------

\subsection{Hirzebruch surface base $B_2  = \mathbb F_1$ with extra section} \label{sec_dP1} \label{subsec_F1base}

Let us first recall some topological properties of the Hirzebruch surfaces $\mathbb F_a$. A Hirzebruch surface can be viewed as a $\IP^1$ bundle over $\IP^1$ of the form (\ref{HZBpfibr}), i.e. as the total space of the projectivisation of the bundle $\cO_{\IP^1} \oplus \cO_{\IP^1}(-a)$ over a base curve of genus zero. The space $H^2(\mathbb F_a, \IZ)$ of its divisor classes is spanned by the class $h$ of the base $\mathbb P^1_b$ and the  class $f$ of the fiber $\mathbb P^1_f$ with  intersection numbers
\beq
h\cdot h = -a \,,\quad f \cdot f = 0\,,\quad h\cdot f = 1 \,,
\eeq
and the anti-canonical divisor has the class
\beq
\bar K= 2h + (2+a) f \,.
\eeq
The Mori cone $\Mcone$ and (the closure of) the K\"ahler cone $\Kcone$, both viewed as a cone in $H^2(\mathbb F_a, \IR)$, are then described, respectively, as
\bea
\Mcone(\mathbb F_a) &=& {\rm Span}\left<f, h\right> \,, \\
\overline{\Kcone(\mathbb F_a)} &=& {\rm Span}\left<f, h+a f \right> \,,
\eea
where ${\rm Span}\left<\cdot \right>$ denotes the set of non-negative linear combinations of the indicated divisor classes. 

Let us begin by specifying the toric data for the ${\rm Bl}_1\mathbb P^2_{112}[4]$  fibrations with base $B_2 = \mathbb F_a$ in terms of the abelian GLSM charges. 
The divisor class $\beta \in H^2(\mathbb F_a,\mathbb Z)$ introduced above is parametrized as 
\be
\beta({\rm x}, {\rm y}) = {\rm x} \, h + {\rm y} \, f \,.
\ee
The constraint (\ref{betaconstraint-a}) implies that 
 $\beta({\rm x}, {\rm y})$ has to take values within the finite range
\beq \label{xyrange}
0 \leq {\rm x} \leq 4\,,\quad\quad 0 \leq {\rm y} \leq 4+2a \,.
\eeq
If we introduce toric coordinates $\nu_{z_i}$ for the base space and $\nu_{u}$, $\nu_{v}$, $\nu_{w}$, $\nu_{s}$ for the fiber, then (\ref{fiberbundles}) translates into the GLSM charges shown in 
 Table~\ref{tb:F_a}. 

\begin{table}
\begin{center}
\begin{tabular}{c |cccc|cccc}
& $\nu_{z_1}$ & $\nu_{z_2}$ & $\nu_{z_3}$ & $\nu_{z_4}$ & $\nu_u$ & $\nu_v$ & $\nu_w$ & $\nu_s$ \\ \hline
$U(1)_h$ &0&1&0&1&0&${\rm x}-2$&x&0 \\ \hline
$U(1)_f$ &1&$a$&1&0&0&${\rm y}-(2+a)$&y&0\\ \hline
$U(1)_U$ &0&0&0&0&1&1&2&0\\ \hline
$U(1)_S$ &0&0&0&0&0&1&1&1\\ \hline
\end{tabular}\end{center}
\caption{GLSM charges of the toric coordinates of the ${\rm Bl}_1 \mathbb P^2_{112}$ fibration over $\mathbb F_a$.} 
\label{tb:F_a}
\end{table}

The inverse gauge coupling of the gauge group $U(1)$ is controlled by the volume of the curve
%the height pairing is given as
\beq
C = b= 6\bar K -2\beta = (12-2 {\rm x}) h + (12+6a - 2 {\rm y}) f \,.
\eeq
We are interested in the limit~\eqref{baseK} in K\"ahler moduli space, where the volume of $C$ goes to infinity.
For a base with $h^{1,1}(B_2)=2$ this limit is conveniently parametrized as in (\ref{Jh112}).
We therefore need to identify a boundary ray $J_0$ of $\Kcone(\mathbb F_a)$ such that $\int_{\mathbb F_a} J_0 \wedge J_0 = 0$ and $\int_C J_0 \neq 0$. Since the two boundary rays $f$ and $h+af$ have self-intersection numbers $0$ and $a$, respectively, the only possibility to satisfy the first requirement is to take $J_0 = f$ (when $a=0$ both $f$ and $h$ can be taken as $J_0$, but in this case the fibration is trivial so that $f$ and $h$ may be treated on an equal footing). This choice also satisfies the second constraint, that is,
\beq
\int_C J_0 = 12-2{\rm x} \geq 4  \,
\eeq
within the range~\eqref{xyrange} so that $\int_C J_0 \neq 0$. 
Altogether we are therefore interested in the limit
\be
J = \frac{1}{\sqrt{\sl{\frac12 + \frac{a}{8 \tilde t^2}}}} \left( \tilde t \, f + \frac{1}{2\tilde t} \, (h + a f)  \right) \,,
\ee
for  $t: =\tilde t (\sl{\frac{1}{2} + \frac{a}{8 \tilde t^2}})^{-1/2}   \to \infty$. 
In agreement with the general discussion of Section~\ref{sec_Globalasalimit}, the class $J_0 = f$ is the class of a holomorphic curve \
\be
C_0 = f \,.
\ee
This is, in fact, the only curve class (modulo integer multiples) whose volume  with respect to $J_0$  vanishes  as $t \to \infty$.

With this preparation, the elliptic genus $Z_{C_0}$ of the nearly tensionless heterotic string associated with $C_0$ can be 
determined  by following the procedure that was outlined in Section~\ref{sec_genstratmaxcharge}:

\begin{itemize}

\item 
For a Hirzebruch surface $\mathbb F_a$ an important simplification occurs, as we recall from Section~\ref{sec_ellgenus}: Since $C_0$ is a Mori cone generator, the full elliptic genus $Z_{C_0}$ is given by \eqref{ZCbetaansatz}. Since $L = g(C_0)=0$,  the numerator is a weak Jacobi form of weight
$W = 6 C_0 \cdot \bar K -2 =10$ and $U(1)$ fugacity index
\be
m=\frac12 C_0 \cdot b=6- {\rm x} \,,
\ee
so in total we have:
\be\label{ZC0-Step1}
Z_{C_0}(\tau, \lambda_s, z) =   \frac{1}{\eta^{24}(\tau)} \frac{\Phi_{10, 6- {\rm x}}(\tau,z)}{
\varphi_{-2,1}(\tau,\lambda_s)}  \,.
\ee

\item Since $C_0$ is a Mori cone generator and thus the numerator does not depend on $ \lambda_s$, 
the genus-zero topological string partition function $\mathcal F_{C_0}$ is given by
\bea\label{ansantz-F0}
\mathcal F_{C_0}^{(0)}(\tau,z) &=&  - \frac{q}{\eta^{24}(\tau)} \Phi_{10, 6- {\rm x}}(\tau,z) \\ \label{FC0-ZK}
&=& - q Z_\otherK3(\tau,z) \,,
\eea
where $Z_\otherK3(\tau,z) $ is the $U(1)$-refined heterotic elliptic genus associated to the elliptically fibered $K3$ surface ${\cal K}$ defined in eq.~(\ref{heteroticfibrationr}).
The minus sign in \eqref{ansantz-F0} originates from the minus sign of the leading term in the expansion $\varphi_{-2,1}(\tau, \lambda_s)  = - \lambda_s^2  \, (1 + {\cal O}(\lambda_s^2))$, which is needed for matching
the RHS of \eqref{ZC0-Step1} to the genus-zero free energy. 

\item As per (\ref{E2E4E6varphiansatz}), we can now make an ansatz for $\Phi_{10, 6- {\rm x}}(\tau,z)$ as a sum of products of $E_4(\tau)$, $E_6(\tau)$, $\varphi_{0,1}(\tau,z)$, $\varphi_{-2,1}(\tau,z)$ of total weight $10$ and fugacity index $6-{\rm x}$.
To fix the finite number of coefficients in this ansatz, we compare it with a finite number of 
genus-zero Gromov-Witten invariants of curve classes of the form
\beq
\Gamma_{C_0}(n,r) := C_0 + n C_\mathcal{E} + r C^{\rm f} \,, \quad n=0, 1 \ldots, 
\eeq
for a small number of choices of $n$.
The fibral classes $C_{\cal E}$ and $C^{\rm f}$ have the properties (\ref{intfibrala1}) and (\ref{intfibrala2}). 
The expansion  (\ref{Fgexp2})  of the generating function
\beq\label{F0-expansion}
\mathcal F^{(0)}_{C_0} (\tau, z)=\sum_{n \geq 0} \sum_{|r|\leq \frak{q}_{\rm max}(n)} N^{(0)}_{C_0}(n,r) q^n \xi^r \,\qquad \text{with}~ q=e^{2\pi i \tau},\, \, \, \xi=e^{2\pi i z} 
\eeq 
can then be obtained to the chosen finite order, where $\mathfrak{q}_{\rm max}(n)$ denotes the maximal power of $\xi$ appearing in the expansion at a given level $n$ in $q$.
By computing a sufficient number of low-degree Gromov-Witten invariants $N^{(0)}_{C_0}(n,r)$, one can proceed to fix the coefficients in the ansatz for $\Phi_{10, 6-{\rm x}}$. This provides an analytic expression for $\cF^{(0)}_{C_0}$ and thus for  $Z_\otherK3(\tau,z)$, as well as for its spin-refined cousin, $Z_{C_0}(\tau, \lambda_s, z)$.

\end{itemize}

In the remainder of this subsection we will consider $a=1$, that is $B_2 = \mathbb F_1$. Although one may in principle analyze all possible choices of $\beta$ within the range~\eqref{xyrange}, we will focus here on the two models with $({\rm x}, {\rm y})=(4,4)$ and $(4,6)$, respectively.
Some further models over $\mathbb F_2$ can be found in Appendix \ref{sec_HirzebruchF2}.

%-------------------------------------------------------------
\subsubsection*{\bf{Model 1: Hirzebruch base $\mathbb F_1$  with extra section, for $({\rm x}, {\rm y})=(4,4)$}}

Some results for the Hirzebruch base $\mathbb F_1$ with $({\rm x}, {\rm y})=(4,4)$ have already been presented in Section~\ref{sec_chargesummary} to exemplify the general behaviour of the charge spectrum of the nearly tensionless heterotic string.
%The detailed computation of the elliptic genus along the above steps can be found in Appendix \ref{}.
Using the toric data as summarized in Appendix \ref{app_44F1}, we can compute via mirror symmetry a collection of low order genus-zero Gromov-Witten invariants.  These assemble into the expansion (\ref{F0-expansion})  as follows:
\bea\label{F0-expansion-numbers-1B}
\mathcal F^{(0)}_{C_0}(\tau,z) &\equiv&   \sum N^{(0)}_{C_0}(n,r)\,q^n\xi^r\ \nn  \\
% &=&~    -2 + \left(\frac{4}{\xi^2} + \frac{128}{\xi}+216 + 128 \xi + 4 \xi^2\right) q \\ 
% &&+ \left(- \frac{2}{\xi^4} + \frac{128}{\xi^3} + \frac{9808}{\xi^2} + \frac{70528}{\xi} + 121964  + {70528} \xi + 9808 \xi^2 + 128 \xi^3 - 2 \xi^4\right) q^2 \nn\\
&=&~    -2 + \left(216 + 128 \xi^{\pm 1} + 4 \xi^{\pm 2} \right) q \\ 
&&+ \left( 121964  + {70528} \xi^{\pm 1} + 9808 \xi^{\pm 2} + 128 \xi^{\pm 3} - 2 \xi^{\pm 4}\right) q^2 \nn\\
&&+ \,  \cO(q^3) \,. \nn
\eea
Here and in the sequel we use the abbreviation
\be \label{zetanotation}
\xi^{\pm n} :=\xi^n + \xi^{-n} \,.
\ee

These Gromov-Witten invariants are more than enough to determine the unknown coefficients in the ansatz
(\ref{E2E4E6varphiansatz}), and this yields:
\be
\begin{split}
\label{F0C0-M1}
\mathcal F^{(0)}_{C_0}(\tau,z)  &= -q\, Z_\otherK3(\tau,z)     \cr
&= \frac{q}{\eta^{24}} \left( \frac{1}{54} E_4^3 \varphi_{-2,1} \varphi_{0,1} + \frac{1}{108} E_6^2 \varphi_{-2,1} \varphi_{0,1}    - \frac{1}{72} E_4^2 E_6 \varphi_{-2,1}^2  - \frac{1}{72} E_4 E_6 \varphi_{0,1}^2\! \right).
\end{split}
\ee
As a non-trivial consistency check, we have confirmed that the Gromov-Witten invariants $N^{(0)}_{C_0}(2,r)$ at order $n=2$ in $q$ are correctly reproduced. 
Moreover, in the un-refined limit $z=0$, for which $ \varphi_{-2,1}=0$ and  $\varphi_{0,1}=12$,  the well-known expression  (\ref{oldellgen}) for the conformal field theoretic elliptic genus $Z_{K3}(\tau)$  is recovered.
     
Based on the analytic expression for $\cF_{C_0}^{(0)}(\tau,z)$, the maximal charge per excitation level, $\frak{q}_{\rm max}(n)$, can easily be extracted at a given order $n$ in $q$. See Figure~\ref{f:plot_1B_F1_100}.
We confirm the general behaviour as announced already in Section~\ref{sec_chargesummary}, namely that there exist superextremal states whose charges lie on a sublattice of the charge lattice and  satisfy the Sublattice Weak Gravity bound. 

Let us now interpret the expansion coefficients in (\ref{F0-expansion-numbers-1B}) in terms of a nearly tensionless, weakly coupled heterotic string.
The left-moving excitation number $n=0$ is associated with the tachyonic mode of the heterotic string, which is not physical because of lack of level-matching.
The sector at $n=1$ is therefore the first physical sector and corresponds to the $U(1)$-refined index of the massless modes.
This sector must reproduce the massless sector of the heterotic string, which in turn has to agree with the massless sector of the F-theory model itself.
More specifically, at the massless level, the $U(1)$-refined elliptic genus must reproduce the chiral index which counts
 the difference of anti-chiral and chiral fermions,
\be \label{chargedindex}
(n_- - n_+)_{\frak{q}} =  2 (n_V + 3 + n_T - n_H)_{\frak{q}}\,.
\ee
Here $n_V$, $n_T$ and $n_H$ denote the numbers of vector multiplets, tensor multiplets, and hypermultiplets, respectively. The universal contribution of $+3$ comes from the gravitino. 
The subscript indicates that we must distinguish the states according to the $U(1)$ charges $\frak{q}$. Of course,
 the gravitino and the tensor multiplets are always uncharged under the $U(1)$ symmetry we consider.

 The coefficients of the term $q$, $q \xi$ and $q \xi^2$ in (\ref{F0-expansion-numbers-1B}) therefore represent, up to the overall sign flip and the shift of $q$-power in (\ref{FC0-ZK}), the piece in the heterotic elliptic genus counting such massless states with $U(1)$ charge $0$, $1$ and $2$, respectively. 
Indeed, these numbers agree precisely with our expectations based on F-theory/heterotic duality.
Consider first the uncharged sector.
In the dual F-theory, the relevant combination of uncharged massless fields adds up to
\be
2 \, (n_V + 3 + n_T - n_H)_{0} =  2 \,  (h^{1,1}(Y_3) - h^{2,1}(Y_3)) =  \chi(Y_3) \,,
\ee
where $\chi(Y_3)$ is the topological Euler characteristic of $Y_3$.
This follows from the general relations between the Hodge numbers of the elliptic fibration and the massless uncharged spectrum of F-theory on $Y_3$~\cite{Wazir:2001},
\be
h^{1,1}(Y_3) = 1 + {\rm rk}(G) + h^{1,1}(B_2)  = 1 + {n_{V_0}} + n_T + 1 \,, \qquad  h^{2,1}(Y_3) = n_{H_0} - 1 \,.
\ee
The first equation uses that the uncharged vectors correspond to the Cartan generators of the gauge group $G$, whose number agrees with the rank of the gauge group, and the second equation follows from the fact that the uncharged moduli are given by the complex structure moduli plus one universal modulus.
 Explicit computation via PALP using the toric data listed in Appendix \ref{app_44F1} gives the Hodge numbers
 \be
 h^{1,1}(Y_3) = 4 \, \qquad h^{2,1}(Y_3) =112\,,\qquad \chi(Y_3) = -216 \,.
 \ee
In particular, the Euler number agrees with (\ref{F0-expansion-numbers-1B}) if we take into account the overall sign. 
Furthermore, the expansion coefficients in front $q \, \xi$ and $q \, \xi^2$ equal the number of points in the set $C_{II}$ and $C_{I}$ introduced in (\ref{CICIIDef}), which count the number of half-hypermultiplets of charge $1$ and $2$, respectively.
Taking into account also the half-hypermultiplets of charge $-1$ and $-2$, encoded in the coefficients of $q \, \xi^{-1}$ and $q \, \xi^{-2}$, gives the number of charged hypermultiplets.
Since the gauge group is abelian, these are the only charged states.
This fact has the following  geometrical meaning:
By mirror symmetry, the expansion coefficients
represent the BPS numbers for M2-branes wrapping the curves
\bea
q \,  \xi:  & \Gamma_{C_0}(1,1) := C_0 +  C_\mathcal{E} + C^{\rm f}   \\
q \,  \xi^2:   &\Gamma_{C_0}(1,2) := C_0 +  C_\mathcal{E} + 2 C^{\rm f} \,
\eea  
in the dual 5d M-theory compactification.
As mentioned around (\ref{CICIIDef}), the classes $C^{\rm f}$ and $2 C^{\rm f}$ are associated with isolated rational curves localised in the fiber over the base loci $C_{II}$ and $C_{I}$.
The number of  isolated curves in these two classes hence equals the number of points in the sets $C_{II}$ and $C_{I}$. These numbers coincide exactly with the BPS numbers for the curves
$ C_\mathcal{E} + 2 C^{\rm f}$ and $ C_\mathcal{E} + C^{\rm f}$ combined with the base curve $C_0$, i.e. the Gromov-Witten invariants for the two classes $\Gamma_{C_0}(1,2)$ and $\Gamma_{C_0}(1,1)$, respectively. 

In summary, we confirm that the $U(1)$-refined elliptic genus at level $n=1$ precisely counts the index (\ref{chargedindex}) of massless multiplets in F-theory, which agrees, by duality, with the corresponding index in the massless sector of the perturbative heterotic string on the $K3$ surface $\otherK3$. Extrapolating this to higher excitation levels we obtain information about the  tower of charged excitations, at least in the tensionless, perturbative limit.

%----------------------------------------------------------------------------------------------------------------
\subsubsection*{\bf{Model 2: Hirzebruch base $\mathbb F_1$  with extra section, for  $({\rm x}, {\rm y})=(4,6)$}}

This choice of parameters is special in that the class $[b_2]$ in the defining equation (\ref{MP-hypers}) is now trivial.
From the discussion after (\ref{CICIIDef}) this means that the 6d F-theory model has only one type of massless charged hypermultiplets, of charge $\frak{q}_{II}=1$, localised over the point set $C_{II}$. 
%The so-defined $U(1)$ model is hence non-generic. 
We will see that this has interesting consequences also for the excitation numbers of the tensionless heterotic string. 
%Since the general procedure to obtain the BPS numbers is identical to the previous case, we are very brief.
Using the toric data given in in Appendix~\ref{app_46F1},
the generating function~\eqref{F0-expansion} for  low-lying Gromov-Witten invariants can be explicit computed  as follows:
\bea\label{F0-expansion-numbers}
% \mathcal F^{(0)}_{C_0}(\tau,z) &=&-2 + \left(\frac{96}{\xi} + 288 + 96 \xi\right) q \\ \nn
% &&+ \left(- \frac{2}{\xi^4} + \frac{96}{\xi^3} + \frac{10192}{\xi^2} + \frac{69280}{\xi} + 123756  + {69280} \xi + 10192 \xi^2 + 96 \xi^3 - 2 \xi^4\right) q^2 + \cO(q^3) \,,
\mathcal F^{(0)}_{C_0}(\tau,z) &\equiv&   \sum N^{(0)}_{C_0}(n,r)\,q^n\xi^r \\  
&&\!\!\!\!\!\!\!\!\!\!\!\!\!\!\!\!\!\!\!\!\!\!\!\!= -2 + \left(288 + 96 \xi^{\pm 1}\right) q  \nn
+ \left(123756  + {69280} \xi^{\pm 1} + 10192 \xi^{\pm 2} + 96 \xi^{\pm 3} - 2 \xi^{\pm 4}\right) q^2 + \cO(q^3) \,,
\eea
where we recall the notation (\ref{zetanotation}).
As expected,  the expansion coefficient for  $q \, \xi^{\pm2}$ vanishes, because for  $({\rm x}, {\rm y})=(4,6)$
there are no rational curves in the class $2 C^{\rm f}$.
The absence of these excitations hints at a non-trivial cancellation, which
can indeed occur only for special linear combinations of the Jacobi forms in the ansatz which we will now determine.

Since the $U(1)$ fugacity index  for the model with $({\rm x}, {\rm y})=(4,6)$ continues to be given by $m=6-{\rm x}=2$, we make the same ansatz for
the elliptic genus in terms of weak Jacobi forms as before.
Comparison with (\ref{F0-expansion-numbers}) at levels $n=0, 1$ in $q$ fixes the ansatz uniquely as follows:
\bea\label{F0C0-M2}
\mathcal F^{(0)}_{C_0}(\tau,z) &=& -q \,Z_\otherK3(\tau,z) \\
\! &=&\! \frac{q}{\eta(\tau)^{24}}  \left( - \frac{1}{72} E_4(\tau)^2 E_6(\tau) \varphi_{-2,1}(\tau,z)^2 + \frac{7}{432} E_4(\tau)^3 \varphi_{-2,1}(\tau,z) \varphi_{0,1}(\tau,z)  \right. \nn\\
&&\ \ \ \ +  \left. \frac{5}{432} E_6^2(\tau) \varphi_{-2,1}(\tau,z) \varphi_{0,1}(\tau,z) - \frac{1}{72} E_4(\tau) E_6(\tau) \varphi_{0,1}^2(\tau,z) \right)  \,.\nn
\eea
The Gromov-Witten invariants $N^{(0)}_{C_0}(2,r)$ at level $n=2$ in $q$ can be confirmed to be correctly reproduced. 
As before, the elliptic genus   $Z_{K3}(\tau)$ in (\ref{oldellgen}) is recovered in the limit $z=0$.

\begin{figure}[t!]
\centering
\includegraphics[width=12cm] {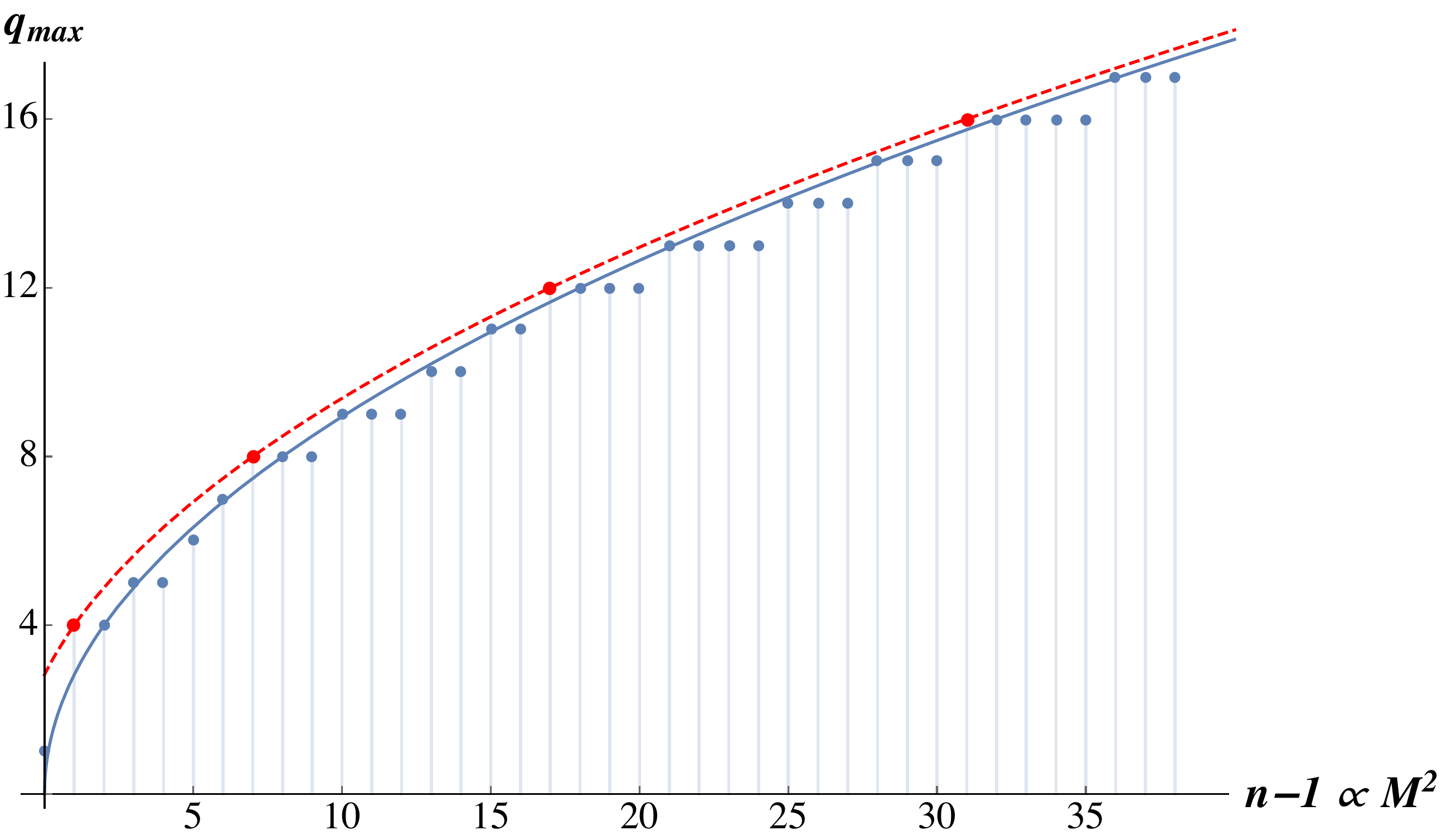}%{newF146.jpg}
\caption{
Maximal charge $\frak{q}_{\rm max}(n)$   for  a ${\rm Bl}_1\mathbb P^2_{112}[4]$ fibration over $\mathbb F_1$ with $(\rm x, \rm y)=(4,6)$. 
Again we oberve that maximally superextremal excitations exist whose charges form a sublattice of index $2m=4$.  However, in contrast to the previous example,
the set of all superextremal states does not populate the full charge lattice.
}
\label{f:plot_1A}
\end{figure}

The maximal charge $\frak{q}_{\rm max}(n)$ can easily be determined at a given level $n$ in $q$. See Figure~\ref{f:plot_1A} 
for a plot..  Evidently it displays the same rough features as for the first example, in particular
a sub-lattice of charges formed by the isolated points peeking above the solid line associated with extremal black holes.

 Furthermore, the elliptic genus as encoded in the Gromov-Witten invariants at order $q$ correctly reproduces the massless spectrum of 
the nearly tensionless heterotic string as defined via the geometry in F-theory.
For the uncharged states this follows from
\be
 h^{1,1}(Y_3) = 4 \, \qquad h^{2,1}(Y_3) =148\,,\qquad \chi(Y_3) = -288 \,,
 \ee
while the expansion coefficients of $q \xi$ and  $q \xi^{-1}$ correctly count the number of half-hypermultiplets of charges $\frak{q} =+1$ and $\frak{q}=-1$, respectively. Hence the number of charged massless hypermultiplets with $|\frak{q}|=1$ is $96$. 

%--------------------------------------------------------------------------------------------------------------------------

%--------------------------------------------------------------------------------------------------------------------------
\subsection{Del Pezzo surface base $B_2=dP_2$ with extra section} \label{sec_dP2example}

We will now discuss an example of a $U(1)$ fibration over the base $dP_2$.
Since this base is not rationally fibered, the structure of the nearly tensionless heterotic string is more subtle, as explained in detail in Section~\ref{sec_tensionlessstrings}. In particular the string won't be perturbative due to the appearance of small instantons which lead
to an extra massless tensor field.

Let us first recall some topological properties of the del Pezzo surfaces $dP_r$. They are constructed by blowing up a $\IP^2$ at $r$ generic points ($r=0, \cdots, 8$). The space $H^2(dP_r, \IZ)$ of its divisor classes is spanned by the hyperplane class $l$ and the blow-up classes $e_i$ ($i=1,\dots,r$). Their non-zero intersection numbers are given as
\beq
l\cdot l = 1 \,,\quad e_i \cdot e_i = -1~(i=1,\cdots,r)\,,
\eeq
and the anti-canonical divisor is
\beq
\bar K= 3l -\sum_{i=1}^r e_i \,.
\eeq
The generators for the Mori cone $\Mcone$ and (the closure of) the K\"ahler cone $\Kcone$ look rather different for different values of $r$. Suffice it here to state them for the case of $r=2$, %(see e.g.~\cite{} for the other cases): 
\bea\label{mori_dP2}
\Mcone(dP_2) &=& {\rm Span}\left<l-e_1-e_2, e_1, e_2\right> \,, \\
\overline{\Kcone(dP_2)} &=& {\rm Span}\left<l, l-e_1, l-e_2 \right> \,.
\eea

Let us now move on to the toric data for ${\rm Bl}_1 \mathbb P^2_{112}[4]$ fibrations over $dP_2$ in terms of the abelian GLSM charges. The class $\beta$ appearing in (\ref{cibetadetf}) is parameterized as
\be
\beta({\rm x}, {\rm y}_1, {\rm y}_2) = {\rm x}l - {\rm y}_1 e_1 -{\rm y}_2 e_2
\ee
and enters the divisor classes of the toric coordinates as displayed in Table~\ref{tb:dP_2}. Both $\beta$ and $2\bar K - \beta$ have to be effective, which implies that  $\beta({\rm x}, {\rm y}_1, {\rm y_2})$ must take values in the following finite range,
\beq
0 \leq {\rm x} \leq 6\,,\quad\quad {\rm x}-4 \leq {\rm y}_{i} \leq {\rm x} \,,~~\text{for}~i=1,2 \,.
\eeq

\begin{table}
\begin{center}
\begin{tabular}{c |ccccc|cccc}
& $\nu_{z_1}$ & $\nu_{z_2}$ & $\nu_{z_3}$ & $\nu_{z_4}$ & $\nu_{z_5}$ & $\nu_u$ & $\nu_v$ & $\nu_w$ & $\nu_s$ \\ \hline
$U(1)_l$ &1&1&1&0&0& 0& ${\rm x}-3$& x & 0 \\ \hline
$U(1)_{l+e_1}$ &0&1& 0 &1&0&0&${\rm x}-{\rm y_1}-2$&${\rm x}-{\rm y_1}$ &0\\ \hline
$U(1)_{l+e_2}$ &1&0&0&0&1&0&${\rm x}-{\rm y_2}-2$&${\rm x}-{\rm y_2}$ &0\\ \hline
$U(1)_U$ &0&0&0&0&0&1&1&2&0\\ \hline
$U(1)_S$ &0&0&0&0&0&0&1&1&1\\ \hline
\end{tabular}\end{center}
\caption{GLSM data for ${\rm Bl}_1 \mathbb P^2_{112}$ fibration over $dP_2$.} 
\label{tb:dP_2}
\end{table}

We are interested in taking the limit in K\"ahler cone where the volume of the height pairing 
\beq
C = b= 6\bar K -2\beta = (18-2 {\rm x}) l - (6 - 2 {\rm y}_1)e_1 - (6  - 2{\rm y}_2 )e_2 
\eeq
associated with the $U(1)$ gauge theory becomes infinite.
To realize the limiting base K\"ahler form~\eqref{baseK}, we need to take a boundary ray $J_0$ of $\Kcone(dP_2)$ such that $\int_{dP_2} J_0 \wedge J_0 = 0$ and $J_0 \cdot C \neq 0$. Given the most general ansatz for $J_0$ in terms of the K\"ahler cone generators,
\beq \label{J0dp2}
J_0 = a l + \sum_{i=1,2} a_i (l-e_i)\,,\quad a, a_i \geq 0 \,,
\eeq
we have
\bea
\int_{dP_2} J_0 \wedge J_0 &=& (a+a_1+a_2)^2 -a_1^2-a_2^2 \\
&=& a^2 +2 a (a_1+a_2) + 2 a_1 a_2\,,
\eea
which only vanishes if $a=0$ and one of the two $a_i$ vanishes. Without loss of generality we may thus take 
\be\label{J0dp2-fixed}
J_0 = l-e_1 \,.
\ee 
Since $(l-e_1) \cdot b = 24 -2{\rm x} - 2 {\rm y}_1$, this ray has non-zero intersection with the height-pairing unless ${\rm x}={\rm y}_1=6$, in which case it is not possible to take a limit with ${\rm vol}(C) \rightarrow \infty$ while keeping ${\rm vol}(B_2) = 1$.

With the choice (\ref{J0dp2}) fixed as~\eqref{J0dp2-fixed}, let us now find the curves of volume zero. 
These correspond to those curves that do not intersect $J_0$.
Given an effective curve of the general form
\beq
C_0=c (l-e_1-e_2) + \sum_{i=1}^2 c_i e_i\,,\quad c, c_i \geq 0\,,
\eeq
its volume, with respect to $J_0$, follows as
\beq
C_0 \cdot (l-e_1) = c_1 \,.
\eeq
The vanishing volume curves in the limit in K\"ahler cone are therefore in the class $C_0 = c (l-e_1-e_2) + c_2 e_2$, with self-intersection number  $C_0 \cdot C_0  =-(c-c_2)^2$. 
In agreement with the general theory of Section \ref{sec_Globalasalimit}, there exists precisely one self-intersection zero curve class $C_0$ (up to multiple wrapping), corresponding to $c = c_2$, and the lowest wrapping number is associated with the curve 
\beq
C_0  \equiv J_0= l-e_1 \,.
\eeq
It is, however, interesting to keep in mind that there may be additional massless charged states from D3 branes wrapping the other possible curves of negative self-intersection obtained by choosing $c \neq c_2$ in the ansatz for $C_0$. This realizes the third point in Section \ref{sec_Globalasalimit}.
The $U(1)$ fugacity index $m$ is parameterised as
\beq
m= \frac12 C_0 \cdot b = 6- {\rm x} + {\rm y}_1 \,.
\eeq
In the sequel we focus on the particular example of a fibration corresponding to the parameters $({\rm x}, {\rm y}_1, {\rm y}_2)=(6,2,2)$, such that the $U(1)$ fugacity index is $m=2$.

By a mirror symmetry computation similar to the one of the previous section, the expansion of the generating function for the 
$U(1)$-refined genus-zero Gromov-Witten invariants up to order $n=3$ in $q$ reads
\bea\label{F0-expansion-numbers-3A}
\mathcal F^{(0)}_{C_0} &=&-2 + \left( 252 + 84 \xi^{\pm 1}\right) q  \\ \nn 
&&\!\!\!\!\!\!\!\!\!\!\!+\left( 116580  + {65164} \xi^{\pm 1} + 9448 \xi^{ \pm 2} + 84 \xi^{\pm 3} -2 \xi^{\pm 4}\right) q^2 \\ \nn
&&\!\!\!\!\!\!\!\!\!\!\!+ \left( 6238536  + {3986964} \xi^{\pm 1} + 965232 \xi^{\pm 2} + 65164 \xi^{\pm 3} +252 \xi^{\pm 4}\right) q^3 \\ \nn
&&\!\!\!\!\!\!\!\!\!\!\!+ \cO(q^4) \,.
\eea
Unlike for the models on the Hirzebruch surfaces, the genus-zero prepotential is expected to be
not a  modular  function by itself, but only a quasi-modular function. 
This is in line with the general discussion of the modular properties of the topological string prepotential in Section~\ref{sec_Ellgenus}. 
The failure of ${\cal F}^{(0)}_{C_0} $  to be modular is measured by the modular anomaly equation~\cite{Klemm:2012sx,Alim:2012ss,Huang:2015sta},
\beq\label{modular-anomaly}
\frac{\partial \mathcal F^{(0)}_{C_0}}{\partial E_2} = \frac{1}{24} \sum_{C_1+C_2= C_0} (C_1 \cdot C_2) \,\mathcal F^{(0)}_{C_1}\mathcal F^{(0)}_{C_2} \,,
\eeq
where the non-trivial $E_2$-dependence appears due to possible splits of $C_0$ into two effective pieces. Given the Mori cone~\eqref{mori_dP2}, it is straightforward to see that $C_0 =l-e_1$ can only split into the sum of $C_1=l-e_1-e_2$ and $C_2=e_2$. Interestingly, both $C_1$ and $C_2$ have self-intersection number $-1$ and hence correspond to an $E$-string, of which $\mathcal F^{(0)}$ is known not to depend on $E_2$. Alternatively, since both $C_1$ and $C_2$ are Mori cone generators, they cannot split into two non-trivial effective pieces. Therefore, $\mathcal F^{(0)}_{C_{i=1,2}}$ cannot have a non-trivial modular anomaly. This means that the RHS of~\eqref{modular-anomaly} does not involve $E_2$ and hence that $\mathcal F^{(0)}_{C_0}$ can depend on $E_2$ at worst linearly. This considerably simplifies the ansatz  (\ref{E2E4E6varphiansatz})   for $\mathcal F^{(0)}_{C_0}$.

Upon requiring that the Gromov-Witten invariants in~\eqref{F0-expansion-numbers-3A} match the ansatz (\ref{E2E4E6varphiansatz}), the coefficients are uniquely fixed and the following generating function is found:
\bea\label{dP2ellgen}
\mathcal F^{(0)}_{C_0}(\tau,z) &=&  -q\,Z_\otherK3(\tau,z)\\
&=&\ \frac{q}{\eta^{24}} \Big( - \frac{23}{1728} E_4^2 E_6 \varphi_{-2,1}^2 + \frac{1}{64} E_4^3 \varphi_{-2,1} \varphi_{0,1} + \frac{19}{1728} E_6^2 \varphi_{-2,1} \varphi_{0,1} - \frac{23}{1728} E_4 E_6 \varphi_{0,1}^2 \nn  \\ 
&& +E_2 (- \frac{1}{1728} E_6^2 \varphi_{-2,1}^2 + \frac{1}{864} E_4 E_6 \varphi_{-2,1} \varphi_{0,1} - \frac{1}{1728} E_4^2 \varphi_{0,1}^2    ) \Big)  \,.\nn
\eea
As expected, when switching off the $U(1)$ background field, $z=0$, the familiar, conformal field theoretic elliptic genus  $Z_{K3}(\tau)$  in  (\ref{oldellgen}) is 
not reproduced, but rather we arrive at\footnote{Note that in this case, the full spin-refined elliptic genus $Z_{C_0}(\tau,\lambda_s,z)$ is not completely fixed by the genus-zero prepotential, but this is not of interest in the  present context. }
\be\label{ZdP2}
Z_\otherK3(\tau,0)\ =\ Z_{K3}(\tau)+\frac{1}{\eta^{24}} \Big(\frac{1}{12}E_2 {E_4}^2-  \frac{1}{12}E_4 E_6\Big)\ =\ \frac2q-420-265968q+\dots
\ee

%--------------------------------------------------------------------------------------------------------------------------
\begin{figure}[t!]
\centering
\includegraphics[width=12cm] {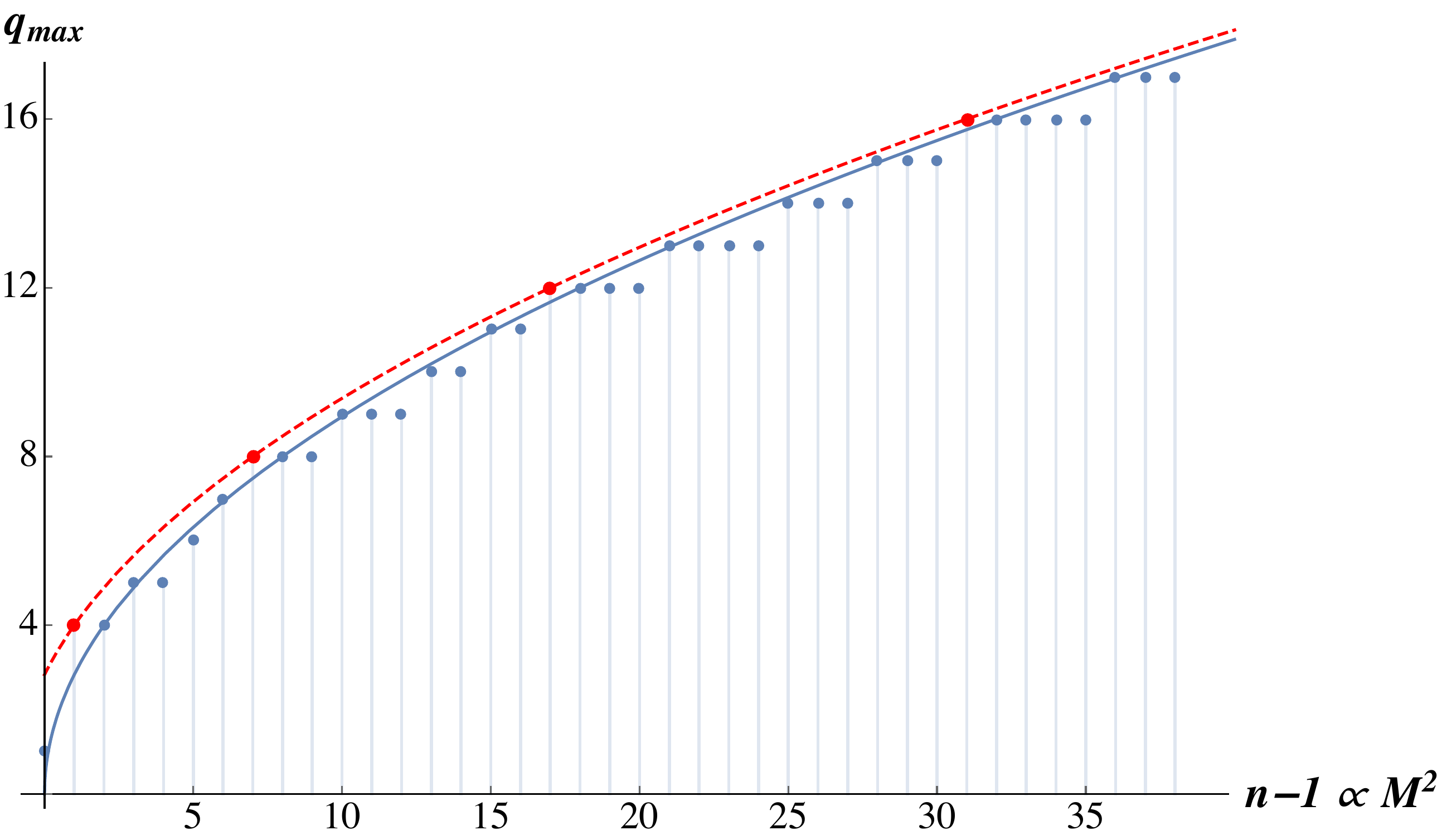}%{dPplot.jpg}
\caption{Maximal charge $\frak{q}_{\rm max}(n)$   for  a ${\rm Bl}_1 \mathbb P^2_{112}[4]$  fibration over $dP_2$, with
$(\rm x, \rm y_1, \rm y_2)=(6,2,2)$.
The general features of this strongly coupled theory are completely in line with the findings of the previous examples, which correspond to weakly coupled strings.
}
\label{f:plot_3A_dP2_100}
\end{figure}
%--------------------------------------------------------------------------------------------------------------------------

This makes it evident that, unlike for models on a Hirzebruch base, we do not obtain a perturbative heterotic dual. 
However, due to existence of a $K3$ fibration of $Y_3$,
 as argued in point 4 of Section~\ref{sec_Globalasalimit}, we can still invoke Type IIA-heterotic duality, albeit with a more complicated heterotic side.
In particular, the heterotic dual  involves one more tensor multiplet, which is associated with a heterotic 5-brane defect which defies a fully
perturbative description. As we see, 
this leaves its imprint in the spectrum of the  heterotic theory. In fact, we observe that the genus-zero contribution 
to the elliptic genus correctly reproduces the chiral index of the F-theory model at the massless level:
The expansion coefficient in eq.~(\ref{F0-expansion-numbers-3A}) at level $q \xi^0$ again agrees with the Euler characteristic of the F-theory 3-fold $Y_3$ because
\be
h^{1,1}(Y_3) = 5 \, \qquad h^{2,1}(Y_3) =131\,,\qquad \chi(Y_3) = -252 \,.
\ee
Furthermore, the expansion coefficient of $q \, \xi$ and $q \, \xi^{-1}$ agrees with the number of $\frak{q}=1$ and $\frak{q}=-1$ half-hypermultiplets, respectively. These are the only massless charged states of the F-theory model. This is because
the fibration we are considering does not contain any charge-two loci $C_I$, like for the Hirzebruch model with $({\rm x}, {\rm y})=(4,6)$ discussed before.
Summing over all massless states gives the chiral index at the massless level,
\be
\sum_\frak{q} (n_- - n_+)_\frak{q} = 2(n_V + n_T + 3 - n_H) =  - 420 \,,
\ee
which reproduces the constant term in eq.~(\ref{ZdP2}), so all is consistent.
As stressed, this result differs from the value $-480$ which pertains to the perturbative heterotic string on K3 in eq.~(\ref{oldellgen}).
Indeed, the model on $dP_2$ under consideration is related by a non-perturbative
tensor transition \cite{Ganor:1996mu,Seiberg:1996vs,Morrison:1996pp} with the model with $({\rm x}, {\rm y})=(4,6)$ on $\mathbb F_1 = dP_1$. 
Starting from $dP_1$ we obtain $dP_2$ by a single blowup, which increases the number of tensors by one. At the same time, the 
 number of vector multiplets is $n_V=1$ in both cases. The 6d gravitational anomaly relation
 \be
 n_H - n_V + 29 n_T = 273
 \ee
requires that the total number of hypermultiplets in going from the $dP_1$ to the $dP_2$ model decreases by $29$. 
In total this indeed leads to the correct shift of the index by
\be
2 (\Delta n_T - \Delta n_H) = 2 (1 - (- 29))  = 60\,.
\ee
We view the fact that the elliptic genus (\ref{dP2ellgen}) correctly reproduces the index of the massless states as evidence that 
it has been computed correctly, and thus that it also encodes the proper charge-to-mass spectrum. 
See Fig.~\ref{f:plot_3A_dP2_100} for a plot, which confirms
again our expectations based on the Weak Gravity Conjecture.
This is reassuring as we lack a completely perturbative description on the heterotic side, due to the tensor transition induced by the putative heterotic 5-brane.

In a similar vein, we have checked whether this non-perturbative heterotic string  forms bound states when wrapped multiple times.  Certainly
the perturbative ones do not ~\cite{Klemm:1996hh,Minahan:1998vr}, 
and we confirm the same behaviour also for the non-perturbative heterotic string discussed in this section.
Concretely, we compute via mirror symmetry the following sequence: 
\bea\label{multipledP2}
\mathcal F^{(0)}_{2C_0} &=&\left( 252 + 84 \xi^{\pm 1}\right) q^2  \\ \nn 
&&\!\!\!\!\!\!+ \left( 6238536  + {3986964} \xi^{\pm 1} + 965232 \xi^{\pm 2} + 65164 \xi^{\pm 3} +252 \xi^{\pm 4}\right) q^3 \ +\  \cO(q^4)\,, \\
\mathcal F^{(0)}_{3C_0} &=&\left( 252 + 84 \xi^{\pm 1}\right) q^3  \\ \nn 
&&\!\!\!\!\!\!+ \left( 161704980  + {110586648} \xi^{\pm 1} + 33904224 \xi^{\pm 2} + 3986964 \xi^{\pm 3} + 116580 \xi^{\pm 4} +84 \xi^{\pm 5}\right) q^4 \  \nn \\ 
&&\!\!\!\!\!\!\!\!+\ \cO(q^5)\,.\nn
\eea
This pattern reproduces the RHS of eq.~(\ref{multiwrap}), which means that all terms above arise from multi-covers of the basic invariants,
$N^{(0)}_{C_0}$.
In other words, the Gopakumar-Vafa BPS invariants associated with the higher wrappings vanish,
 $N^{(0)}_{k\cdot C_0}=0$ for $k>1$, and consequently there are no bound states.

%-------------------------------------------------------------------------------------------

\section{Discussion and Outlook}\label{sec_outlook}

Our analysis has revealed an intricate interplay between various conjectured properties of quantum gravity
 and the geometric realisation of gauge symmetries in string compactifications.
We have demonstrated that whenever a  gauge symmetry  -- abelian or non-abelian -- becomes asymptotically a global symmetry in 
 6d F-theory compactifications on some threefold $Y_3$, while gravity is being kept dynamical, 
there appears a nearly tensionless, critical heterotic string in the theory.
At the level of algebraic geometry,  this is because there must exist a rational curve $C_0$ of self-intersection zero
 in the base of the elliptic 3-fold $Y_3$ whose volume tends to zero in the limit.
Wrapping a D3-brane on this curve gives rise to an effective string which becomes tensionless when $C_0$ shrinks to zero volume.
 This leads to a breakdown of the effective field theory near the point where the global symmetry is reached. Such behaviour is in agreement with by now familiar arguments on the properties of quantum gravity \cite{Banks:1988yz,ArkaniHamed:2006dz,Banks:2010zn}, and similar phenomena have been observed in various other corners of the landscape such as \cite{Heidenreich:2017sim,Grimm:2018ohb,Heidenreich:2018kpg,Andriolo:2018lvp}. 

Our analysis directly addresses the question of having a weakly coupled gauge symmetry in the `open-string' sector.
Somewhat ironically maybe, the crucial ingredient responsible for consistency with quantum gravity is nonetheless a closed string in disguise:
The string which becomes tensionless is dual to the critical heterotic string compactified to six dimensions. We have shown that the tensionless limit can be taken only if the elliptic 3-fold $Y_3$ admits a $K3$ fibration, whose fiber class is the pullback of $C_0$ to $Y_3$. Even though in general this fibration is not compatible with the elliptic fibration of $Y_3$, Type IIA/heterotic duality still allows us to infer the appearance of a nearly tensionless heterotic dual.
 
Our approach to analyze the nearly tensionless string has been to compute its elliptic genus, making use of a number of beautiful results in the literature, 
pioneered in \cite{Klemm:1996hh} and reviewed e.g. in \cite{Heckman:2018jxk};  especially 
concerning the relation between the elliptic genus and the topological string partition function on an elliptically fibered 3-fold $Y_3$~\cite{Haghighat:2013gba, Haghighat:2014pva}. 
At a technical level, we have extended the general techniques to situations in which the 3-fold has general extra rational sections, which translate into global $U(1)$ symmetries of the string. 

While the elliptic genus as such does not directly
 represent a partition function of physical excitations of the string in six dimensions,  it describes a small, moduli-independent sub-sector of left-moving string excitations  that can be paired with appropriate right-moving oscillators such as to provide a subset of the physical spectrum.
The charges of these states are encoded entirely in the left-moving sector and can therefore be read off from the suitably refined elliptic genus as such. 
Our working assumption is that at least in the limit of very weak coupling, we can rely on the perturbative mass relation for the states at given excitation level. Fortunately, the heterotic weak coupling limit coincides with the limit where the gauge symmetry becomes global.
This way we arrive at an intriguing relation between the charges and masses of the states that become light as the gauge group becomes global. We argue  that,
under certain assumptions, the charge-to-mass ratio in particular asymptotically satisfies the (Sub)Lattice Weak Gravity Conjecture bound \cite{ArkaniHamed:2006dz,Heidenreich:2015nta,Heidenreich:2016aqi}.

Further away from the tensionless limit we cannot a priori trust the formula for the masses of the states.
However, it is tempting to speculate that the states remain in the spectrum. This is plausible as they are apparently needed to satisfy the Completeness Hypothesis \cite{Polchinski:2003bq}, according to which every site in the charge lattice should be populated by a physical state. The `perturbative' open string spectrum in (Type IIB language) by itself does not contain arbitrarily high charges, not even at the massive higher excitation levels, because the charges are encoded entirely in the Chan-Paton factors, and this fact remains true taking into account also multi-pronged strings in F-theory.

{
One of our key results is the quantitative confirmation of the Sublattice Weak Gravity Conjecture for six-dimensional heterotic strings, 
at least for the tensionless limit.  The relevant bound with which we need to compare the charge/mass ratio of the excitations 
corresponds to certain extremal Reissner-Nordstr\"om black holes, 
where we additionally take the effect of massless scalar field fluctuations \cite{Heidenreich:2015nta,Palti:2017elp,toappear} into account. Indeed we observe the existence of superextremal string excitations that just minimally lie above this bound and whose charges populate at least a sublattice of the full charge lattice.

As far as the heterotic string is concerned, our findings are completely in line with the original discussion in ref.~\cite{ArkaniHamed:2006dz}. There the heterotic 
string was compactified on $T^6$, while here we consider compactification on $K3$ with a single unbroken $U(1)$ gauge symmetry. As we have argued in section~\ref{sec_genstratmaxcharge}, 
the modular properties of the elliptic genus guarantee that this $U(1)$ gauge symmetry can formally be associated with a single boson compactified on an
$S^1$. It is precisely the charge/mass properties of the winding states in this sector that ultimately forms the common ground between our work and that of  ref.~\cite{ArkaniHamed:2006dz} (as well as \cite{Heidenreich:2015nta,Heidenreich:2016aqi}). 

As far as extremal black holes are concerned, we find it quite remarkable that the balance of gravitational and scalar forces \cite{Palti:2017elp} conspires, in a narrowly consistent manner, with the charge pattern of the heterotic string, which is encoded in the number theoretical properties of weak Jacobi forms.
}

Our focus in this work has been on the role of the particle excitations of the nearly tensionless string in the context of Quantum Gravity conjectures.
At the same time, it is natural to wonder about the implications of the nearly tensionless BPS  {\it string} as such.
First, to argue for a breakdown of the effective field theory in the limit where the gauge symmetry becomes global, it probably suffices 
 to consider the tensionless string.
This string is a priori charged with respect to the 2-form gauge symmetries associated with the tensor multiplets of the 6d $N=(1,0)$ supergravity; however, due to the 6d Green-Schwarz mechanism the shift symmetry of the tensors is gauged with respect to the 1-form $U(1)$ symmetry which becomes global in the limit we are considering. This implies an effective charge also of the string itself. Microscopically this charge can be attributed to the appearance of charged fermionic zero-modes propagating on its world-sheet.

Concerning the Weak Gravity Conjecture, one might alternatively wish to consider not the non-BPS black holes charged under the abelian gauge symmetry, but rather black string solutions of the 6d $N=(1,0)$ supergravity.
 In this case one would have to check for the appearance of super-extremal strings with respect to the charge-to-tension ratio of the black strings. Given that the BPS strings which become tensionless in our limit carry net charge also under the 1-form $U(1)$ symmetry via the Green-Schwarz mechanism, as noted above, it would be interesting to quantify this relation further.

As far as the Completeness Conjecture is concerned, it is rather obvious that the lattice of {\it string} charges is indeed realized explicitly by physical states due to the appearance of BPS saturated strings from wrapped D3-branes. In this sense the Completeness Conjecture with respect to the 1-form $U(1)$ gauge symmetry for the particle-type excitations is far more non-trivial to check, along the lines spelled out in this paper.

On the formal side, one of the outcomes of this article is the determination of the quasi-modular elliptic genus of certain non-perturbative $K3$ compactifications of the heterotic string, which involve extra massless tensor multiplets. This is  the situation in which the $K3$ fibration of the F-theory 3-fold is not compatible with the elliptic fibration. Such configurations are related by a tensor transition to F-theory on Hirzebruch surfaces \cite{Morrison:1996pp}, which has a more conventional perturbative heterotic dual.  We have verified that the proposed elliptic genus of the dual non-perturbative heterotic string correctly captures the chiral index of F-theory massless modes,
for the example of F-theory on a $dP_2$ base. It would  be interesting to analyze this relation further.

Another direction to take would be to relate the 6d F-theory compactifications studied in this work to various other dual setups.
Compactification on an extra torus takes us to a 4d ${\cal N}=2$
compactification of the Type IIA string on the same Calabi-Yau $Y_3$. The `open string $U(1)$' in the F-theory frame becomes a $U(1)$ gauge symmetry in the Ramond-Ramond sector of the 4d theory. The wrapped string states become 4d BPS particles associated with wrapped D2 (and D4) branes in Type IIA. Mirror symmetry then relates this to Type IIB string theory with a $U(1)$ gauge symmetry due to the Ramond-Ramond 4-form $C_4$ with charged BPS states from the sector of wrapped D3-branes. This is exactly the duality frame analyzed in \cite{Grimm:2018ohb}. It might  be illuminating to work out the mirror map in detail and compare our results to the behaviour of this Type IIB theory in the vicinity of 
 infinite distance points in complex structure moduli space. 

Let us end with a few more speculations:
The nearly tensionless critical 6d heterotic string that has been the star of the present analysis, appears to behave very differently from the non-critical 6d strings which are associated with 6d superconformal field theories without gravity. These non-critical strings are due to D3-branes wrapping contractible curves of strictly negative self-intersection on the base. Shrinking such curves to zero volume, while keeping the volume of the base fixed, is possible at finite distance in moduli space.
This is to be contrasted with the critical heterotic string, which is associated with a  non-contractible curve of self-intersection zero: As we have seen, shrinking its volume to zero while keeping the base volume fixed is possible only if we  take a drastic  limit in which another curve approaches infinite volume,  and this limit lies at infinite distance in moduli space.

From the perspective of Swampland Distance Conjecture \cite{Ooguri:2006in}, the fact that this limit is at infinite distance seems to be successfully explained by the appearance of a tower of infinitely many critical string excitations coupling to gravity which become light. This raises the obvious question why the tensionless point of  the non-critical string lies at finite distance in moduli space. 
A priori, of course, the Swampland Distance Conjecture does not need to hold in the reverse, i.e. it does not postulate that {\it every} tower of infinitely many states which becomes massless at a point in moduli space requires this point to be at infinite distance.

 Nonetheless, it is not immediately clear how to reconcile the existence of such a tower of charged states e.g. with the following fact:
If the curve associated with a critical or non-critical string transversally intersects a 7-brane curve, the 7-brane gauge group becomes a flavour symmetry of the string, and its excitations, if present, would be charged with respect to this symmetry.
The limit of shrinking the non-critical string curve is compatible with keeping the transverse 7-brane curve at finite volume.
If the string gives rise to a tower of infinitely many massless charged states in the limit, the running of these states should affect the gauge coupling of the 7-brane theory \cite{Heidenreich:2017sim,Heidenreich:2018kpg,Grimm:2018ohb}. As argued at the end of Section~\ref{sec_ReltoQG} this is exactly the situation for the critical heterotic string; by contrast this is apparently not the case for the non-critical strings as the gauge coupling associated with the transverse 7-brane  stays finite if we contract the curve associated with the string.

 {A bold resolution would be that the non-critical string does not give rise to infinitely many massless 
 physical particle states in six dimensions as we approach the tensionless limit. 
To the best of our understanding this seems to be consistent with all we know about these mysterious strings:
As the curve contracts, the 6d {\it BPS string} becomes tensionless, which is a BPS protected statement. While the wrapped string does give rise to 5d BPS particles, the 6d particle excitations are non-BPS, as stressed several times, and hence are a priori not protected against large corrections to their mass.
In the case of the critical heterotic string we have invoked an explicit duality with a weakly coupled heterotic string to argue for the appearance of massless particle excitations in the tensionless limit, but for non-critical strings no such arguments are immediately available at first sight. If correct, this argument would suggest that (all but possibly a finite number of) the particle excitations of the non-critical string sit at non-zero mass even in the tensionless limit due to non-trivial interaction effects, while the BPS string as such becomes tensionless. In the SCFT limit this mass must of course be infinite as otherwise there would be a scale in the theory. A less radical resolution would be that the asymptotically massless excitations couple both to gravity and to the gauge theories of the other branes in a fundamentally different manner even infinitesimally away from the tensionless point. It would be very desirable to understand this point better.
}

%-------------------------------------------------------------------------------------
%%%%%
\section*{Acknowledgments}
We are indebted to Diego Regalado for collaboration at early stages  of this project, for his collaboration in \cite{Lee:2018ihr}, which partly motivated this work,  and for many important discussions.
 We furthermore thank Francesco Benini, Per Berglund, Cyril Closset, Mirjam Cveti{\v c}, Thomas Grimm, Babak Haghighat, Arthur Hebecker, Chris Hull, Finn Larsen, Sungjay Lee, Dieter~L\"ust, Fernando Marchesano, Miguel Montero, Eran Palti, Gary Shiu, Wati Taylor, Stefan Theisen, Irene Valenzuela, Nick Warner and Fengjun Xu for useful discussions.

\newpage

\begin{appendix}
%%%%%

%-------------------------------------------------------------------------------------
\section{Conventions for Dimensional Reduction} \label{App_DimRed}

Here we collect our conventions for the effective action of 7-branes and D3-branes.
Our starting point is the 10d string frame Einstein-Hilbert action together with the  Dirac-Born-Infeld (DBI) action for the  7-branes and 3-branes, 
whose relevant terms are

\bea
S &=& \frac{2\pi}{\ell_s^8}  \int_{\mathbb R^{1,9}} d^{10}{ x}  \,  e^{-2 \phi}  \sqrt{-{\rm det} g} \,   R   \,       + \frac{2\pi}{\ell_s^8} \int_{\rm D7} d^8{ x}   \, e^{- \phi}  \sqrt{-{\rm det} (g + 2 \pi \alpha' F)}  +  \\
   &&         \frac{2\pi}{\ell_s^4} \int_{\rm D3} d^4{ x}  \, e^{- \phi}   \sqrt{-{\rm det} (g + \ldots)} + \ldots \,.
\eea
In the sequel we will set the string length to unity,
\bea
\ell_s  =2 \pi \sqrt{\alpha'} \equiv 1 \,.
\eea
We dimensionally reduce the Einstein-Hilbert action on the 2-complex dimensional space $B_2$, which serves as base of the elliptic threefold, $Y_3$.
Moreover, we reduce the DBI actions on the curves $C$ and $C_0$ that are wrapped by the 7-brane and the 3-brane, respectively. This produces an 
effective action whose relevant terms read
\bea
S &=&  {2 \pi} \, {\rm vol}(B_2)        \int_{\mathbb R^{1,5}} \, d^{6}x \,  \sqrt{-{\rm det} \, g}     \,      e^{-2\phi}  \,   R   \,    +    \frac{ {\rm vol}(C)}{2\pi}    \int_{\mathbb R^{1,5}} \, d^{6}x \,  \sqrt{-{\rm det} \, g}    \,       e^{- \phi}   \frac{1}{4}  F_{\mu \nu} F^{\mu \nu}      \\
 && + 2 \pi  \, {\rm vol}(C_0)    \int_{\mathbb R^{1,1}}  d^2x \,  e^{- \phi}   \sqrt{-{\rm det} \,  g}  + \ldots \,.
\eea
After rescaling the metric $g \to e^{\frac{\phi}{2}} g$ we obtain the Einstein-Hilbert action in the 
6d Einstein frame. The dilaton dependence both of the 7-brane gauge kinetic term and of the D3-brane tension term is removed by this rescaling.
We can therefore rewrite the relevant part of the effective action as
\be \label{Einsteinframeaction}
S = \int d^{6}x \sqrt{-{\rm det} \, g}   \left( \frac{M_{\rm Pl}^4}{2}    R   +   \frac{1}{4 g^2_{\rm YM}}   F_{\mu \nu}F^{\mu \nu}        \right)   + T \int_{\mathbb R^{1,1}}   \sqrt{-{\rm det} \, g}  + \ldots
\ee
and identify 
\bea
M_{\rm Pl}^4 &=& 4 \pi \, {\rm vol}(B_2) \\
\frac{1}{g^2_{\rm YM}} &=& \frac{1}{2\pi} {\rm vol}(C) \\
T &=& 2 \pi \, {\rm vol}(C_0)  \,.
\eea
Here $T$ denotes the tension of the string obtained by wrapping the D3-brane along $C_0$.

\vfil\eject

%-------------------------------------------------------------------------------------

\section{The Geometric Limit with $g_{\rm YM} \ll M_{\rm Pl}^{-1}$}   \label{app_Jproof}

In this appendix we derive the four key facts characterizing the limit  (\ref{limit-form1}) of the K\"ahler form $J$ of a compact K\"ahler surface presented in Section \ref{sec_Globalasalimit}.

\subsection{The degenerate limit}

As stated already in the main text, the general limiting behaviour of the K\"ahler form $J$ can be summarized in
\begin{claim}
For \eqref{norm} and~\eqref{lim} to be satisfied simultaneously, the K\"ahler form $J$ on $B_2$ must be chosen to lie infinitesimally close to a boundary ray of the K\"ahler cone. 
To be more specific,  the K\"ahler form must asymptote to
\bea \label{baseKB}
J = t J_0 +  \sum_{\nu} s_\nu I_\nu  \,, \ \ \   {\rm as}\ \   t \to \infty\,.
\eea
Here $J_0 \in H^{1,1}(B_2, \IZ)$ is a K\"ahler cone generator satisfying
\be \label{J0J0zeroB}
J_0 \cdot J_0 = 0 \,, \qquad \int_C J_0 \geq 1 \,.
\ee
The remaining K\"ahler cone generators  $I_\nu \in H^{1,1}(B_2, \IZ)$ have  the property that
\be \label{nnusboundB}
\sum_{\nu} n_{ \nu} s_\nu   \leq \frac{1}{\sl{t}} \,,
\ee
where 
\be
n_{ \nu} = J_0 \cdot I_\nu 
\ee
and at least one $n_{ \nu} \neq 0$. In fact 
\be
 \sum_\nu n_{ \nu} s_\nu   \to \frac{1}{\sl{t}}  \qquad  \text{as} \quad  t\to \infty \,.
 \ee
The non-negative expansion parameters $s_\nu$ stay finite as $t \to \infty$ and are chosen such that ${\rm vol}_J(B_2) =1$.
\end{claim}
The first relation in (\ref{J0J0zeroB}) implies  in particular that $J_0$ must lie on the boundary $\partial \Kcone(B_2)$ of the K\"ahler cone of $B_2$. 
The volume of $C$, which governs the inverse gauge coupling, obeys
\bea\label{JConCzeroB}
{\rm vol}_J(C)&=&  t \int_{C} J_0 + \sum_\nu s_\nu \int_{C} I_\nu  \\ \nn
&= &2 m t + s \,,
\eea
where 
\be \label{hatmdefintionforCB}
2m := \int_{C} J_{0} %= \frac{1}{2} C \cdot C_0 
\ee
is a positive integer fixed by the topology and 
\beq
s:=\sum_\nu s_\nu \int_{C} I_\nu 
\eeq 
is a non-negative real number that remains finite in our geometric limit.

To derive this, note first that the K\"ahler form $J$ of any compact K\"ahler surface $B_2$ can be expanded as
\beq \label{Jansatza}
J=\sum_{a}t_a J_a + \sum_{\nu} s_\nu I_\nu \,,\quad\quad t_a, s_\nu \geq 0\,,
\eeq
where $J_a$ and $I_\nu$ are the K\"ahler cone generators in $H^{1,1}(B_2, \IZ)$. We have split the set of K\"ahler parameters into  $t_a$, which go to infinity in the limit (\ref{limit-form1}) we are considering, and $s_\nu$, which remain finite in the limit.
Being  K\"ahler cone generators, the $J_a$ and $I_\nu$ have the important property that
\be \label{Kahlerconeintersection}
J_a \cdot J_b \geq 0 \,, \qquad J_a \cdot I_\nu \geq 0 \,, \qquad I_\nu \cdot I_\mu \geq 0 \,.
\ee
Here, for notational simplicity, the integration of forms over the base has been expressed in terms of intersection products. 
By assumption, at least one of the $J_a$ has a non-zero integral over $C$ so that $\int_C J \to \infty$ as $t_a \to \infty$, as intended.
% being large and finite, respectively, in the limit. 

Let us now start imposing various constraints on $J$.
First in order to have 
\beq
{\rm vol}_J (B_2) = \sl{\frac12}J \cdot J = 1 
\eeq
in the limit $t_a \to \infty$, the $J_a$ must obey a trivial intersection form,
\beq\label{ortho}
J_a \cdot J_b = 0\,.
\eeq
This follows from the fact that all $t_a$ and $s_\nu$ are non-negative, together with (\ref{Kahlerconeintersection}), which makes cancellations between infinite terms impossible.

Now, (\ref{ortho}) implies that there can be only one K\"ahler cone generator $J_a$ which appears with a coefficient $t_a \to \infty$ in the ansatz (\ref{Jansatza}), i.e. 
we can write 
\bea \label{Jansatz2}
J = t J_0 +  \sum_{\nu} s_\nu I_\nu 
\eea
with $J_0 \in H^{1,1}(B_2, \IZ)$ a K\"ahler cone generator satisfying
\be
J_0 \cdot J_0 = 0 \,, \qquad \int_C J_0 \geq 1 \,.
\ee
In (\ref{Jansatz2}) the index $\nu$ runs over the remaining generators of the K\"ahler cone different from $J_0$.

The reason why only one K\"ahler cone generator can appear with a large K\"ahler parameter is that in order for any two elements $J_a$ and $J_b$ to satisfy 
\be \label{J1J2constraint}
J_a \cdot J_a = 0 \,, \qquad J_b \cdot J_b = 0 \,, \qquad J_a \cdot J_b = 0
\ee 
on a compact K\"ahler surface, we must have $J_b = \lambda  \, J_a$ for some $\lambda \in \mathbb R$ as will be shown below. Hence there is only one choice of K\"ahler cone generators which can appear with a coefficient $t_a$ which goes to infinity.
To see this, recall that on any compact K\"ahler surface we can consider a basis $\{\omega_0, \omega_i\}$ of $H^{1,1}(B_2)$ with intersection form
\be
\omega_0 \cdot \omega_0 = +1 \,, \qquad \omega_0 \cdot \omega_i =0 \,, \qquad \omega_i  \cdot \omega_j = - \delta_{ij} \,
\ee
because the intersection form is known to have signature $(+1,-1, \ldots,-1)$.
Let us expand
\be
J_a = a \,   \omega_0 + a_i \,  \omega_i \,, \quad   J_b = b \,  \omega_0 + b_i  \, \omega_i
\ee
and introduce a vector notation $(\vec{a})_i = a_i$ and $(\vec{b})_i = b_i$. 
Then (\ref{J1J2constraint}) is equivalent to requiring that
\be \label{a2b2constraint}
a^2 = \vec{a} \cdot \vec{a} \,, \qquad b^2 = \vec{b} \cdot \vec{b} \,, \qquad a b = \vec{a} \cdot \vec{b} \,.
\ee
Hence
\be
|a b|^2 = |\vec{a} \cdot \vec{b}|^2  \leq |\vec{a}^2| |\vec{b}^2| = a^2 b^2 
\ee
with equality only for $\vec{b} = \lambda  \, \vec{a}$. The only possibility to satisfy (\ref{a2b2constraint}) is then to take also $b = \lambda \, a$ and hence $J_a = \lambda J_b$, which proves the claim. 

Having established the parametrization (\ref{Jansatz2}) of the K\"ahler form, 
we note that the argument after Assertion 2 below establishes that there exists a holomorphic curve $C_0$ with class 
$\left[C_0\right]=J_{0}$. 
Being a holomorphic curve, its volume must be strictly positive inside the K\"ahler cone of $B_2$, i.e. away from the asymptotic limit $t \to \infty$ for the K\"ahler form (\ref{Jansatz2}). 
Since  $\int_{C_0}J_0 = J_0 \cdot J_0 =0$, this requires that there must exist at least one $I_\nu$ with a non-zero intersection $n_{ \nu}:=J_{0} \cdot I_\nu$. 
%In particular since the  $I_\nu$ are among the generators the K\"ahler cone, this non-vanshing intersection must be bigger than or equal to one. 
Then, the volume of $B_2$ is given as
\beq\label{volB}
{\rm vol}_J(B_2) = \sl{t} \sum_{\nu} s_\nu (J_{0} \cdot I_\nu) + \sl{\frac12} \sum_{\nu,\mu} s_\nu s_\mu (I_\nu \cdot I_\mu) \geq \sl{t} \sum_{\nu} n_{ \nu} s_\nu  \,,
\eeq
where in the last step the non-negative terms that do not involve $t$ have been dropped. 
Requiring that ${\rm vol}_J(B_2) =1$ implies the bound
\be \label{nnusbound-a}
\sum_{\nu} n_{ \nu} s_\nu   \leq \frac{1}{\sl{t}} 
\ee
on the parameters $s_\nu$ in the ansatz (\ref{Jansatz2}), and in fact $ \sum_\nu n_{ \nu} s_\nu   \to \frac{1}{\sl{t}}$ as $ t\to \infty$.

It is important to note that $\sum_{\nu} n_{ \nu} s_\nu$ has to asymptote precisely to $\frac{1}{\sl{t}}$. In particular it cannot become {\it parametrically} smaller than $1/t$. For this to happen, we would need $s_{\nu_0} s_{\mu_0} (I_{\nu_0} \cdot I_{\mu_0})$ to be of order one (so that ${\rm vol}_J(B_2)$ can stay finite) for at least one pair of generators $I_{\nu_0}$ and $I_{\mu_0}$ with $\nu_0 \neq \mu_0$, or that at least one  $s_{\nu_0}^2 (I_{\nu_0} \cdot I_{\nu_0})$ is of order one.
Since all $s_\nu$ remain finite as $t \to \infty$, as proved above, this requires both $s_{\nu_0}$ and $s_{\mu_0}$ to be of order one, which in turn implies that 
$J_0 \cdot I_{\nu_0}$ and  $J_0 \cdot I_{\mu_0}$ must both vanish in order for ${\rm vol}_J(B_2)$ to stay finite. The last requirement, however, is not possible for the same reason as given in the proof of Assertion 3 in section \ref{sec_uniqueness}.

\subsection{Existence of the curve $C_0$}

The second observation we make is summarized in 

\begin{claim}
{Modulo the technical assumption (\ref{nbarK-equ}),} as the K\"ahler form asymptotes to (\ref{baseKB}),  the base $B_2$ necessarily contains a rational curve $C_0$ with 
\be \label{Claim2equ}
C_0 \cdot C_0 = 0 \,, \qquad \quad C_0 \cdot \bar K = 2 \,, \qquad \quad C_0 \cdot C \neq 0 
\ee
whose volume vanishes in the limit,
\be
{\rm lim}_{t \to \infty}{\rm vol}_J(C_0)  = 0  \quad \quad {\rm as} \quad \quad {\rm lim}_{t \to \infty}{\rm vol}_J(C)  = \infty \,.
\ee
The curve class $C_0$ coincides with the class $J_0$ in (\ref{baseKB}),%\footnote{This is true possibly up to a factor of normalization which ensures that $C_0$ is an integer class.},\
\be \label{C0J0}
C_0 = J_0 \,.
\ee
\end{claim}
Indeed, if there exists an effective curve $C_0$ in this class, then by (\ref{J0J0zeroB}) 
\be \label{C0C00C0Cneq0}
C_0 \cdot C_0 = 0 \,,\qquad C_0 \cdot C \neq 0 \,,
\ee
and it is clear that its volume vanishes for $t \to \infty$ as
\beq\label{volJC0A}
{\rm vol}_J (C_0) =  C_0 \cdot J = J_0 \cdot J = \sum_\nu n_{ \nu} s_\nu \to \frac{1}{\sl{t}} \,.
\eeq
For later purposes note that this implies the following behaviour for the product of the two volumes: 
\beq\label{geom-boundB}
{\rm vol}_J(C) \,{\rm vol}_J(C_0) \to { \sl{2}m} + \frac{s}{\sl{t}} \,.
\eeq

The question is then if an effective curve  exists in the class $J_0$.
By construction, the asymptotic K\"ahler form is in the closure of the K\"ahler cone, $J_0 \in \overline{\Kcone(B_2)}$. For a surface $B_2$, the latter is contained in $\overline{NE(B_2)}$, the closure of the Mori cone $NE(B_2)$ of effective curves on $B_2$.
According to 
Mori's cone theorem, when applied to $B_2$, there exist countably many rational curves $C_i$ satisfying $0<\bar K \cdot C_i \leq 3$ such that 
\beq \label{Morisconetheorem}
\overline{NE(B_2)} = \overline{NE(B_2)}_{\bar K \leq 0} + \sum_i \mathbb R_{\geq 0} \left[C_i\right] \,.
\eeq
The second term on the RHS of (\ref{Morisconetheorem}) is manifestly in the Mori cone, i.e. describes effective curves, while the first term may not be effective (since it is a priori only in the closure of the Mori cone).
It is the subset of $\overline{NE(B_2)}$ with non-positive intersection number with the anti-canonical class $\bar K$ of $B_2$.

Now, recall that the anti-canonical divisor must be effective for there to exist a Calabi-Yau Weierstrass model on $B_2$, $\bar K \in {NE(B_2)}$,  and the asymptotic K\"ahler form is by construction in the closure of the K\"ahler cone, $J_0 \in \overline{\Kcone(B_2)}$. Therefore 
\be
J_0 \cdot \bar K \geq 0  \,.
\ee 
This leaves two possibilities: Either $J_0 \cdot \bar K = 0$ and $J_0  \in \overline{NE(B_2)}_{\bar K \leq 0}$, or $J_0 \cdot \bar K > 0$ and 
$J_0 \in \sum_i \mathbb R_{\geq 0} \left[C_i \right]$.
In the latter case, $J_0$ is an effective curve class itself, and the genus of the associated holomorphic curve $C_0$ can be computed by the Hirzebruch-Riemann-Roch theorem as
\be
2g(C_0) - 2 = C_0 \cdot( C_0 + K) = J_0 \cdot (J_0 + K) = - J_0 \cdot \bar K < 0 \,.
\ee
The only possibility is  
\be \label{gC0}
g(C_0)= 0 \,,\quad  {\rm i.e.} \quad \, \, C_0 \cdot \bar K = 2 \,.
\ee
Together with (\ref{C0C00C0Cneq0}), $C_0$ hence has all properties advertised in (\ref{Claim2equ}).

The question is now whether the other scenario, $J_0 \cdot \bar K = 0$, is possible. To address it,
let us first assume that the curve $C$ whose volume goes to infinity as $t \to \infty$ supports a non-abelian gauge algebra $\mathfrak{g}_I$, corresponding to the first case in (\ref{Cdefcases}).
In view of (\ref{Sigmaclass}) and (\ref{Sigmadiv}), this means that the class of the discriminant of the fibration must satisfy the relation
\bea
12 \bar K = [\Sigma] = C + [\Sigma'] \,, 
\eea
where both $C$ and  $[\Sigma'] $ are effective curve classes. 
If $J_0 \cdot \bar K = 0$, while at the same time $J_0 \cdot C > 0$ (which is required by the assumption that ${\rm vol}_J(C) \rightarrow \infty$ as $t \rightarrow \infty$), then necessarily $J_0 \cdot [\Sigma'] < 0$. But this is not possible because  $\Sigma'$ is an effective curve corresponding to the residual components of the discriminant locus. 

It remains to exclude $J_0 \cdot \bar K = 0$ also for the second case in (\ref{Cdefcases}), where $C$ corresponds to the height pairing $b = - \pi_\ast (\sigma(S) \cdot \sigma(S))$ of a rational section.
Consider an F-theory model with a single abelian gauge group factor, and suppose that there exists $n >0$ such that  
\be \label{nbarK-equ}
n \,  \bar K   =  b  + \delta    \,,\qquad \delta \quad \text{effective}.
\ee
Then the same argument as above shows that $J_0 \cdot \bar K = 0$ requires $J_0 \cdot \delta <0$, in clash with the effectiveness of $\delta$. 
The question is then if for every $Y_3$ with a rational section, such an $n$ can be found.
While we cannot present a general proof for arbitrary $B_2$, this property is realized in all models we know with abelian gauge symmetries in F-theory.
Consider first the so-called 
Morrison-Park model \cite{Morrison:2012ei}, which is a rather general (though not the most general \cite{Klevers:2016jsz,Morrison:2016xkb,Raghuram:2017qut}) type of an elliptic fibration with a Mordell-Weil group of rank one.
In this case, $b = 6 \bar K - 2 \beta$ with $\beta$ an effective class such that $2\bar K - \beta$ is also effective. Hence (\ref{nbarK-equ}) is true with $n=6$ and $\delta =  2 \beta$.
In the presence of additional non-abelian gauge algebras, $b$ changes by the subtraction of multiples of the divisor classes  $\Sigma_I$ associated with $\mathfrak{g}_I$ from $b$, so that $\delta$ becomes even more effective. 

On the other hand, more general $U(1)$ models are known to exist in which the height pairing does exceed the specific bound $b \leq 6 \bar K$. This comes with the inclusion of higher charges beyond $\frak{q}=1$ and $\frak{q}=2$. As shown in \cite{Morrison:2016xkb}, all of these models are related to a non-generic Weierstrass model of the form \cite{Morrison:2012ei}, which is however not Calabi-Yau because the class of  $f$ and $g$ is $4 \bar K+4 {D}$ and  $6 \bar K + 6 {D}$, respectively, with ${D}$ a priori an effective class. The elliptic Calabi-Yau 3-fold describing the $U(1)$ is related to this model by a non-minimal transformation which changes the class of $f$ and $g$ to $4 \bar K$ and $6 \bar K$, as required for an elliptic Calabi-Yau.
The height pairing of the section is in this case 
\be
b = 6 \bar K + 4  {D} \,.
\ee
However, extra constraints on ${D}$ may arise by requiring that the Weierstrass model has a smooth birational resolution.
Explicit examples of this type were constructed in \cite{Klevers:2016jsz,Raghuram:2017qut} with charged matter with $\frak{q}=1,2,3$, and in all these types of fibrations  (\ref{nbarK-equ}) is satisfied for some higher value of $n$. 

For favorable examples of base spaces, on the other hand, (\ref{nbarK-equ}) is readily proved to hold for any conceivable rational section and associated height pairing:
Indeed, for any base $B_2$ where the effective divisor $\bar K$ lies strictly inside the Mori cone (as opposed to merely on its boundary), given any effective divisor $b$ we can find an integer $ n >0$ such that $n \bar K - b$ is effective. This is simply because by assumption $n \, \bar K$ has a nontrivial positive component along each generator of the Mori cone, which can be made arbitrarily large by scaling up $n$. 
This is  in particular the case for  all Hirzebruch surfaces $\mathbb F_a$ and del Pezzo surfaces $dP_{r}$ ($r\leq 8$). 

In the sequel we are making the technical assumption that (\ref{nbarK-equ}) indeed holds. 
In this case, we have proved rigorously that in the limit where an abelian gauge symmetry becomes global, a rational curve $C_0$ with $C_0^2 = 0$ attains zero volume. Note that for non-abelian gauge symmetries, we have proved the analogous statement even without any further assumptions. 
Importantly, the curve $C_0$ has non-trivial  intersection with the curve $C$ supporting the non-abelian or abelian gauge symmetry.

\subsection{Uniqueness of $C_0$ and K3-fibration} \label{sec_uniqueness}

{If $h^{1,1}(B_2) \geq 3$ one might wonder if there are more curves on $B_2$ whose volume tend to zero in the limit $t \to \infty$. The answer is that $C_0$ is the only curve class of self-intersection zero with this property, and
every additional curve class with asymptotically vanishing volume must have negative self-intersection. This is summarized in
\begin{claim}
Every  curve class $\tilde C$ with asymptotically vanishing volume in the limit~\eqref{baseK} must satisfy $\tilde C \cdot \tilde C \leq 0$, and such curve satisfies $\tilde C \cdot \tilde C = 0$ if and only if $\tilde C  = C_0$ up to a muliplicative factor.
\end{claim}
This is a simple consequence of the well-known lightcone-like structure of the intersection form on the compact K\"ahler surface $B_2$. On $B_2$ we can always find a basis 
\beq
\{\omega_0, \omega_i\} \,,\quad i=1, \ldots n_T 
\eeq
of $H^{1,1}(B_2)$ with diagonal intersection form 
\be
\omega_{0} \cdot \omega_{0} =1 \, \qquad \omega_{i} \cdot \omega_{j} = - \delta_{ij} \, \qquad \omega_{0} \cdot \omega_{i} =0 \,.
\ee
We can use the freedom to define the basis in order to set w.l.o.g. $J_0 = j (\omega_0 + \omega_1)$.  The condition that ${\rm vol}_J(\tilde C) \to 0$ in the limit $t \to \infty$ for (\ref{baseK}) requires $J_0 \cdot \tilde C=0$. Expanding $\tilde C = a_0 \omega_0 + \sum_{i=1}^{n_T} a_i \omega_i$, 
this implies $a_0=a_1$ so that $\tilde C = a_0 (\omega_0 + \omega_1) + \sum_{i=2}^{n_T} a_i \omega_i$. We therefore have 
\beq
\tilde C \cdot \tilde C = -\sum_{i=2}^{n_T} a_i^2 \leq 0\,,
\eeq 
where the equality, $\tilde C \cdot \tilde C = 0$, holds if and only if $\tilde C = a_0 (\omega_0 + \omega_1) = \frac{a_0}{j} J_0$. Hence the class of $\tilde C$ is proportional to that of $C_0$, as claimed.

The curve $C_0$ has the following important property: If we restrict the elliptic fibration $\pi: Y_3 \rightarrow B_2$ to $C_0$, we obtain an
elliptically fibered surface
\be
\widehat C_0= \pi^{-1}(C_0) \in H_4(Y_3) \,,
\ee
which is an elliptic $K3$ surface. 
To see this, note that the curve $C_0$ intersects the discriminant divisor $\Sigma$ of the elliptic fibration $Y_3$ in 
isolated points whose numbers is computed, via (\ref{gC0}), to be
\be
C_0 \cdot \Sigma = 12  \, C_0 \cdot \bar K = 24 \,. 
\ee
The elliptically fibered surface $\widehat C_0$ therefore has {\it generically} 24 $I_1$-fibers at the location of the $24$ intersection points of $C_0$ with the discriminant.
This uniquely identifies $\widehat C_0$ as a K3 surface. 
This surface plays an important role because of our final
\begin{claim}
If the base $B_2$ is compatible with a limit of the form (\ref{baseK}), the Calabi-Yau 3-fold $Y_3$ admits not only an elliptic fibration (\ref{ellfibrationpi}) over  $B_2$, but in addition a $K3$ fibration 
\bea \label{K3fibrrho}
\rho :\quad K3 \ \rightarrow & \  \ Y_{3} \cr 
& \ \ \downarrow \cr 
& \ \  C'
\eea
over a curve $C'$, where the class of the K3-fiber is $\widehat C_0 \in H_4(Y_3)$. 
In general, this $K3$ fibration is not compatible with the elliptic fibration (\ref{ellfibrationpi}).
\end{claim}

For the proof we recall the criterion of Ooguiso  \cite{Ooguiso} in order for a  given Calabi-Yau 3-fold to be $K3$ fibered. 
The criterion states that this is the case if $Y_3$ possesses a nef (numerically effective) divisor $D$ with the property that  $D \cdot D = 0$ and $c_2(Y_3) \cdot D > 0$.
The class of the nef divisor $D$  then serves as the class of the generic $K3$ fiber of $Y_3$ (up to a multiplicative factor).
In the present case we would like to identify the class of $D$ with the class of the divisor $\widehat C_0$. 

First, note that manifestly $\widehat C_0 \cdot \widehat C_0=0$:  Since $\widehat C_0$ is the pullback under $\pi$ of $C_0$ to $Y_3$, the only possible non-zero intersection is with a section $S$, but $\widehat C_0 \cdot \widehat C_0 \cdot S = C_0 \cdot_{B_2} C_0=0.$ To check if $\widehat C_0$ is nef, recall that a divisor is nef if and only if it lies in the dual cone to $\overline{NE(Y_3)}$, the closure of the Mori cone of effective curves on $Y_3$. 
This dual cone is precisely $\overline{\Kcone(Y_3)}$, the closure of the K\"ahler cone. In the present case we are interested in the divisor given by the pullback of $C_0 = J_0$, with $J_0$ one of the generators of the K\"ahler cone of the base. Its pullback to $Y_3$ must then lie in the K\"ahler cone of $Y_3$, and hence $\widehat C_0$ is nef. 

It remains to check for strict positivity of the product 
\be
\widehat C_0 \cdot_{Y_3} c_2(Y_3) = C_0 \cdot_{B_2} \pi_\ast(c_2(Y_3))    \stackrel{?} >0 \,.
\ee
Let us first consider a smooth Weierstrass model over a base $B_2$, assuming that it exists. By standard computation, 
\be \label{Weiersmoothc2}
c_2(Y_3)  = 11 c_1^2(B_2) + c_2(B_2) + 12 S_0 \cdot  c_1(B_2) \,,
\ee
where $S_0$ is the zero-section. 
Hence we confirm
\be
\widehat C_0 \cdot_{Y_3} c_2(Y_3) =   12 \widehat C_0 \cdot S_0 \cdot  c_1(B_2)   =12  \, C_0 \cdot_{B_2} \bar K > 0
\ee
and therefore $Y_3$ is $K3$ fibered. This makes use of the fact $\widehat C_0$ is the pullback of a base divisor, whence the first two terms in (\ref{Weiersmoothc2}) do not contribute, and the last step relies on the important property (\ref{Claim2equ}) of $C_0$.
In presence of an extra abelian or non-abelian gauge symmetry, we are not to consider a smooth Weierstrass model $Y_3$, but a singular Weierstrass model and its resolution $\hat Y_3$.
In this case, Theorem  2.1 in \cite{Esole:2018tuz} applies to all crepant smooth resolutions of a singular Weierstrass model, which ensures that 
\be
\widehat C_0 \cdot_{Y_3} c_2(\hat Y_3) =12 \,  C_0 \cdot_{B_2} \bar K > 0 \,
\ee
still holds. 
The only possible complication can occur if no such crepant resolution exists, i.e. if the partially resolved space $\hat Y_3$ is left with $\mathbb Q$-factorial terminal singularities in the elliptic fiber over codimension-two points on $B_2$ \cite{Arras:2016evy,Grassi:2018rva}. 
Since these are localised at isolated points on the base $B_2$, however, they cannot affect the result for the intersection of the divisor $C_0$ with $c_2(\hat Y_3)$.

%-------------------------------------------------------------------------------------

\section{The Ring of Weak Jacobi Forms} \label{app_Jacring}

In this appendix we collect a few well-known  and useful  properties of weak  Jacobi forms. These
have been introduced in \cite{EichlerZagier} and are reviewed e.g. in \cite{Dabholkar:2012nd}.
The relation between Jacobi forms and the topological string prepotential  has been noticed as early as in ref.~\cite{Kawai:1998md}, 
which also includes a review of many of their properties.  

\subsubsection*{Weyl invariant Jacobi forms}

Given a Lie algebra $\mathfrak g$ with Cartan subalgebra $\mathfrak{h}$, a Weyl invariant Jacobi form $\varphi_{w, {\bf  m}}(\tau , {\bf z})$ with weight $w \in \mathbb N$ and index ${\bf m} \in \mathbb N$ is a holomorphic function
\be
\begin{aligned}
\varphi_{w, {\bf  m}}:   \mathbb H \times \mathfrak h_{\mathbb C}    &\rightarrow \mathbb C  \\
(\tau , {\bf {z}}) &\mapsto \varphi_{w, {\bf  m}}(\tau, {\bf {z}}) 
\end{aligned}
\ee
which satisfies the following four properties \cite{Wirth}: 
\begin{itemize}
\item It transforms under an $SL(2,\mathbb Z)$ transformation of the modular parameter $\tau$ as 
\bea \label{modulartrafo1}
\varphi_{w, {\bf m} }\left( \frac{a \tau + b}{ c \tau + d}, \frac{\bf {z}}{ c\tau + d} \right) &=& (c \tau +d)^w e^{2 \pi i  \frac{{\bf m} \,  c}{c \tau + d} \frac{( {\mathbf {z}} , {\mathbf {z}})}{2} } \varphi_{w, {\bf m}} (\tau, {\bf {z}}) \,.
\eea
Here $(\mathbf {z} , \mathbf {z})$ is the invariant bilinear form on ${\mathfrak h}_{\mathbb C}$, which is normalised in such a way that the norm square of the smallest co-root is 2.
\item 
It is quasi-periodic in shifts of the elliptic parameter by elements of the $\lambda$, $\mu$ in the co-weight lattice of $\mathfrak g$,
\bea \label{quasiperiod}
\varphi_{w, {\bf m}}\left( \tau , {\bf {z}} + \lambda \tau + \mu \right) &=& e^{-2 \pi i  {\bf m} (\frac{({\mathbf {z}},{\mathbf {z}})}{2} \tau  + 2 \frac{ (\lambda , \mathbf {z}) }{2} )  }   \varphi_{w, {\bf m}} (\tau, {\bf {z}}) \,.
\eea
\item
It is invariant under the action of the Weyl group $W_{\mathfrak{g}}$ of  $\mathfrak{g}$,
\bea
 \varphi_{w, {\bf  m}}(\tau, s {\bf {z}})  = \varphi_{w, {\bf  m}}(\tau,  {\bf {z}}) \,, \qquad s \in W_{\mathfrak{g}} \,.
\eea
\item The form $\varphi_{w, {\bf  m}}(\tau, {\bf {z}})$ can be expanded as
\be
 \varphi_{w, {\bf  m}}(\tau, {\bf {z}})  = \sum c(n,r) q^n y^r \,, \qquad q = e^{2 \pi i \tau}, \, \,  y = e^{2 \pi i {\bf z}} \,,
\ee 
where the quasi-periodicity (\ref{quasiperiod}) implies that
\be
c(n,r) = C(4 n {\bf m} - r^2, r) 
\ee
for some function $C$.
\end{itemize}
The ring of Weyl invariant Jacobi forms associated with Lie algebra $\mathfrak{g}$ is usually denoted by
\be
J^{}_{\ast,\ast}(\mathfrak{g})  = \oplus_{l,m} J^{}_{l,m}(\mathfrak{g}) \,.
\ee
For even weight it is a freely generated (for $\mathfrak{g} \neq \mathfrak{e}_8$) polynomial ring over the ring of modular forms. A modular form of weight $w$ is a holomorphic function on $\mathbb H$
transforming under an $SL(2,\mathbb Z)$ transformation as 
\be
f\left(\frac{a \tau + b}{ c \tau + d}\right) = (c \tau + d)^w f(\tau) \,.
\ee
An important role for us is played by the Eisenstein series $E_{2k}(\tau)$ for $k \geq 1$,  which is defined as
\be
E_{2k}(\tau)= 1 + \frac{2}{\zeta(1 - 2 k)} \sum_{n=1}^\infty   \frac{n^{2k-1} q^n}{1- q^n} \,.
\ee

For $k \geq 2$, $E_{2k}(\tau)$ is a modular form of weight $w=2k$, and $E_4(\tau)$ and $E_6(\tau)$ generate the space of modular forms. 
On the other hand, $E_2(\tau)$ is only a quasi-modular modular form of weight $2$ because it suffers a modular anomaly 
\be \label{E2quasimodularity}
E_2\left(\frac{a \tau + b}{ c \tau + d}\right) = (c \tau + d)^2 E_2(\tau) + \frac{6 c}{\pi i}   \, (c \tau + d) \,.
\ee

\subsubsection*{Jacobi forms}

For the special case $\mathfrak{g} = \mathfrak{su}(2)$, the elliptic parameter of a Weyl invariant Jacobi form takes values in $\mathfrak{h}_{\mathbb C} \simeq \mathbb C$. Taking into account the normalization of the invariant bilinear form a Weyl invariant Jacobi form for $\mathfrak{g} = \mathfrak{su}(2)$ hence reduces to a weak Jacobi form.
Indeed, by definition a Jacobi form of weight $w$ and index $m$ is a holomorphic function from $\mathbb H \times  \mathbb C $   to $\mathbb C$
with the analogous transformation behaviour  
\bea
\varphi_{w, { m} }\left( \frac{a \tau + b}{ c \tau + d}, \frac{{z}}{ c\tau + d} \right) &=& (c \tau +d)^w e^{2 \pi i  \frac{{ m} \,  c}{c \tau + d} {z}^2  } \varphi_{w, { m}} (\tau, { {z}}) \,,\\
\varphi_{w, {m}}\left( \tau , { {z}} + \lambda \tau + \mu \right) &=& e^{-2 \pi i  { m} ({z}^2 \tau  + 2  \lambda  {z}) )  }   \varphi_{w, {m}} (\tau, { {z}})\, \quad \lambda, \mu \in \mathbb Z \,. \label{quasiperiod-2}
\eea
A Jacobi form  $\varphi_{w, {m}}(\tau, { {z}})$ is called 
\begin{itemize}
\item holomorphic Jacobi form if $c(n,r) = 0$ unless $4 {m} n \geq r^2$,
\item Jacobi cusp form  if $c(n,r) = 0$ unless $4 {m} n > r^2$,
\item weak Jacobi form if $c(n,r) = 0$ unless $ n \geq 0$ \,.
\end{itemize}
One furthermore defines a weakly (or nearly) holormorphic Jacobi form with similar properties except that  $c(n,r) = 0$ unless $ n \geq n_0$ for negative integer $n_0$.
More information can also be found for instance in \cite{Kawai:1998md,Dabholkar:2012nd}. 
 
The ring $J_{2k,\ast}(\mathfrak{su}(2))$ is freely generated by the Weyl invariant Jacobi forms $\varphi_{0,1}(\tau, z)$ and $\varphi_{-2,1}(\tau, z)$  with coefficients being polynomials in $E_4(\tau)$ and $E_6(\tau)$ \cite{EichlerZagier} (see also e.g. the Appendix A.2 of \cite{Gu:2017ccq} or  section 4.3 of \cite{Dabholkar:2012nd}). 
Here $\varphi_{0,1}(\tau, z)$ and $\varphi_{-2,1}(\tau, z)$ can be defined in terms of the Dedekind function $\eta(\tau) = q^{\frac{1}{24}} \prod_{n=1}^\infty (1 - q^n)$ and the Jacobi theta functions as 
\bea 
&& \varphi_{0,1}(\tau, z) = 4 \left(  \frac{\vartheta_2(\tau,z)^2}{\vartheta_2(\tau,0)^2}  + \frac{\vartheta_3(\tau,z)^2}{\vartheta_3(\tau,0)^2}  + \frac{\vartheta_4(\tau,z)^2}{\vartheta_4(\tau,0)^2}  \right) \\
&& \varphi_{-2,1}(\tau, z) = \frac{\vartheta_1(\tau,z)^2}{\eta^6(\tau)} \,. \label{phim21}
\eea
For given weight and elliptic index, only a finite number of coefficients need to be fixed in the expansion in terms of this generating system. 

An alternative presentation is
\bea \label{phi0102Eisen1}
\varphi_{-2,1}(\tau, z) = \frac{\varphi_{10,1}(\tau, z)}{\Delta} \,, \qquad \quad \varphi_{0,1}(\tau, z) = \frac{\varphi_{12,1}(\tau, z)}{\Delta}
\eea
where the discriminant function, given in terms of the the Dedekind eta-function,
\be
\Delta(\tau)  = \eta^{24}(\tau)\,,
\ee
is a modular function of weight $12$.  The specific Jacobi functions $\varphi_{10,1}(\tau, z)$ and $\varphi_{12,1}(\tau, z)$ can be expressed in terms of the Eisenstein-Jacobi series $E_{4,1}(\tau,z)$ and $E_{6,1}(\tau,z)$  as \cite{EichlerZagier}
\bea  \label{phi0102Eisen2}
\varphi_{10,1}(\tau, z) &=& \frac{1}{144} \left( E_6(\tau) E_{4,1}(\tau,z) - E_4(\tau) E_{6,1}(\tau,z) \right) \\
\varphi_{12,1}(\tau, z) &=& \frac{1}{144} \left( E_4^2(\tau) E_{4,1}(\tau,z) - E_6(\tau) E_{6,1}(\tau,z) \right)  \,.
\eea
We will discuss the Eisenstein-Jacobi series  in more detail in Appendix \ref{app_EJmaxcharge}.

Finally, note that $\varphi_{0,1}(\tau, z) $ and $\varphi_{-2,1}(\tau, z) $ have a polynomial expansion in terms of $z$ whose coefficient functions are polynomials in $E_2(\tau)$, $E_4(\tau)$ and $E_6(\tau)$  \cite{Gu:2017ccq}:
\bea
\varphi_{-2,1}(\tau, z)  &=& - z^2 + \frac{E_2(\tau)}{12} z^4 + \frac{E_4(\tau)-5 E_2(\tau)^2 }{1440} z^6 + \ldots    \label{varphim21exp1}\\
\varphi_{0,1}(\tau, z)   &=& 12 - E_2(\tau) z^2 + \frac{E_2(\tau)^2 + E_4(\tau)}{24} z^4 + \ldots \,.
\eea

%-------------------------------------------------------------------------
\subsubsection*{Modular anomalies and quasi-modularity}

A  weak Jacobi form of weight $w$ and index  $m$ can in general be expanded as a power series in the elliptic parameter \cite{EichlerZagier}
\be
\varphi_{w,m}(\tau, {z}) = \sum_{j=0}^\infty  \phi_j(\tau)  {z}^j \,,
\ee
where the coefficient functions $\phi_j(\tau) $ are {\it quasi-modular} functions of weight $j+w$.
As pointed out e.g. in \cite{Gu:2017ccq}, as a consequence of the quasi-modularity (\ref{E2quasimodularity}) of $E_2(\tau)$ the product 
\be
\Phi_{w,0}(\tau, {z} )  = e^{\frac{\pi^2}{3} m {z}^2 E_2(\tau)} \varphi_{w,m}(\tau, {z})
\ee
 behaves like a modular function, i.e. like a Jacobi form of weight $w$ and index $0$. 
Since $\Phi_{w,0}(\tau, {z} )$ is  modular in $\tau$, the deviation from modularity in the quasi-modular expansion coefficients $\phi_j(\tau)$ of $\varphi_{w,m}(\tau, {z})$ is therefore cancelled by the $E_2(\tau)$ dependent prefactor in $\Phi_{w,0}(\tau, {z} )$. It follows that the only source of quasi-modularity in $\phi_j(\tau)$ can be due to monomials in $E_2(\tau)$, which are cancelled term by term in $\Phi_{w,0}(\tau, {z})$. 

Another conclusion is that $\varphi_{w,m}(\tau, {z})$ satisfies the differential equation \cite{Gu:2017ccq}
\be
0 = \frac{\partial}{\partial E_2} \Phi_{w,0}(\tau, {z} )  =  \frac{\partial}{\partial E_2}  e^{\frac{\pi^2}{3} m {z}^2 E_2(\tau)} \varphi_{w,m}(\tau, {z}) = e^{\frac{\pi^2}{3} m {z}^2 E_2(\tau)} \left(  \frac{\partial}{\partial E_2}  + \frac{(2\pi)^{2}}{12} m {z}^2  \right) \varphi_{w,m}(\tau, {z}) \,.
\ee
This differential equation allows us to read off the elliptic index of a weak Jacobi form.
Conversely, a power series in ${z}$ with quasi-modular expansion functions satisfying periodicity under ${z} \rightarrow {z} +1$ and the equation
\be \label{partialE2equ}
 \left(  \frac{\partial}{\partial E_2}  + \frac{(2\pi)^{2}}{12} m {z}^2  \right) \varphi_{w,m}(\tau, {z}) = 0
\ee
is a weak Jacobi form of index $m$. 

%----------------------------------------------------------------------------
\section{Maximal Charges via Eisenstein-Jacobi Forms} \label{app_EJmaxcharge}

In this appendix we prove the key fact that the maximal charge per excitation level $n$ as encoded in the elliptic genus obeys a bound of the form (\ref{chargemass}).
This makes use of various properties of the Eisenstein-Jacobi forms that we will collect below. The relevance  of the Eisenstein-Jacobi forms is that, as described in section \ref{sec_genstratmaxcharge}, the elliptic genus can always be expressed via an ansatz of the form (\ref{E2E4E6varphiansatz}). The charge dependence of the states is encoded in the weak Jacobi forms $\varphi_{-2,1}(\tau, z)$ and $\varphi_{0,1}(\tau, z)$ introduced in Appendix \ref{app_Jacring}, which in turn have an expression in terms of the Eisenstein-Jacobi series as given in
(\ref{phi0102Eisen1}). 

The Eisenstein-Jacobi forms $E_{4,m}(\tau,z)$ and $E_{6,m}(\tau,z)$ are holomorphic Jacobi forms of index $m$. The coefficients $c(n,r)$ in the series 
\be
E_{w,m}(\tau,z) = \sum_{n,r} c(n,r) \, q^n \, \xi^r\,,\qquad  q=e^{2\pi i \tau}\!, \,\xi=e^{2\pi i z},
\ee
have the property that \cite{EichlerZagier}
\bea\label{non-zero-c}
\begin{aligned}
&c(n,r) \neq  0&   \quad {\rm for } \quad     r^2 < 4 m n    \\
&c(n,r) = 
\begin{cases}
    1& \text{if } r \equiv 0~({\rm mod}~2m) \\
    0              & \text{otherwise}
\end{cases} &\quad {\rm for}\quad r^2=4mn \\
%c(n,r)  = 1  \quad    {\rm when}  \quad   &r^2 = 4 m n      \qquad {\rm} \qquad {\rm for}  \quad   r = 2 \, k \, m \qquad k \in \mathbb N \,.
\end{aligned}
\eea
An analytic expression for the $c(n,r)$ can be found in \cite{EichlerZagier} in terms of Cohen's function. Since the Eisenstein-Jacobi forms are invariant under $\xi \leftrightarrow \xi^{-1}$, we will mostly focus on non-negative powers of $\xi$.   

Eq.~\eqref{non-zero-c} implies that the maximal $\xi$-power $r_{\rm max}(n)$ which appears in the expansion multiplying a term $q^n$ is 
\bea\label{rmax}
\begin{aligned}
&r_{\rm max}(n)=  \floor[\big]{\sqrt{4\,m\,n}} &   \quad {\rm if }~     4mn \text{ is {\it not} a perfect square}   \\
&r_{\rm max}(n) = 
\begin{cases}
    \sqrt{4\,m\,n}                 & \text{if } n=m\, l^2 \text{ for}~l\in \mathbb N_0 \\
    \sqrt{4\,m\,n} -1             & \text{otherwise}      
\end{cases} &  \quad {\rm if}~ 4mn \text{ is a perfect square} \\ 
\end{aligned}
\eea
In other words, the maximal $\xi$-power is
\beq\label{rmax2}
r_{\rm max}(n)=\floor[\big]{\sqrt{4\,m\,n}}-a(n)\,,
\eeq
where the shift function $a$ is generically zero. To be precise, $a(n)$ is given as
\beq\label{shiftfn}
a(n)=
\begin{cases}
    1              & \text{if $4mn$ is a perfect square and $\frac{n}{m}$ is not a square of an integer} \\
    0              & \text{otherwise}\, 
\end{cases} 
\eeq
Note first that $r_{\rm max}(n)$ is a non-decreasing function of $n$ even in the presence of a non-generic shift by $-1$. Furthermore, one can immediately observe that $r_{\rm max}(n)$ continues to have a plateau-like structure as $n$ increases. An intuition for the latter property is gained from the fact that $r_{\rm max}(n)$ is generically given as $\floor[\big]{\sqrt{4\,m\,n}}$. Since the values of $\sqrt{4\,m\,n}$ at $n+\Delta n$ and $n$ differ by
\beq
\sqrt{4\,m\,(n+\Delta n)} - \sqrt{4\,m\,n} = \sqrt{4\,m\,n}\Big(\sqrt{1+\frac{\Delta n}{n}} - 1\Big) = \sqrt{m \frac{\Delta n}{n}} \Big(1+ \cO(\frac{\Delta n}{n})\Big) \,,
\eeq
which can be made arbitrarily small for a small $\frac{\Delta n}{n}$, it follows that $r_{\rm max}(n)$ stays constant in the interval $\left[n, n+\Delta n\right]$ if $\frac{\Delta n}{n}$ is small enough. 

Now, given the expressions~\eqref{rmax2} and~\eqref{shiftfn}, we have the following lower bound for $r_{\rm max}(n)$, 
\be\label{conservative}
%r_n = \floor[\big]{ \sqrt{4 \,  m \,  n} } \geq  \sqrt{4 \,  m \,  n} - 1 =  \sqrt{ \,  m \,  n}  + (\sqrt{ \,  m \,  n} -1)  \geq  \sqrt{ \,  m \,  n} \,.
r_{\rm max}(n) = \floor[\big]{ \sqrt{4 \,  m \,  n} } -a(n) \geq  \sqrt{ 4\,  m \,  n} -1 \,. 
\ee
We will then use the symbol ``$\gtrsim$'' to denote an ``approximate'' asymptotic inequality, 
\beq\label{gtrsim1}
r_{\rm max}(n) \gtrsim \sqrt{4 \,m\,n} \,,
\eeq
in the sense that for any small $\hat \epsilon >0$ there exists a finite $n_0=n_0(\hat \epsilon)$ such that\footnote{The small parameter $\hat \epsilon$ is related to the small parameter $\epsilon$ used in (\ref{betansharper}) and the following discussion in the main text via $\hat \epsilon = 4 m \epsilon$.} 
\beq\label{gtrsim2}
r_{\rm max}(n) > \sqrt{(4m-\hat\epsilon)n} \,\quad\quad \text{if}~n> n_0(\hat \epsilon)\,.
\eeq

Let us now consider the product of two Eisenstein-Jacobi series $E_{w_1,m_1} E_{w_2,m_2}$, which is itself a Jacobi form of weight $w_1 + w_2$ and index $m_1 + m_2$ with expansion
\be
E_{w_1,m_1} E_{w_2,m_2} =  \sum_{n_1,r_1} c_1(n_1,r_1) \, q^{n_1} \, \xi^{r_1}    \sum_{n_2,r_2} c_2(n_2,r_2) \, q^{n_2} \, \xi^{r_2} = \sum_{\substack{n, r}} \,  c(n,r) q^n \xi^r      \,.
\ee
For the series expansions of $E_{w_1, m_1}$ and $E_{w_2,m_2}$, we will denote by $r_{\rm max}^{(1)}(n)$ and $r_{\rm max}^{(2)}(n)$, respectively, the maximal $\xi$-powers which appear in the expansions multiplying a term $q^n$. 
Then, for a given $n$, there is a non-trivial contribution to $c(n,r)$ from the product of the following two terms,
\beq
q^n \xi^{r_{\rm max}^{(1)}(n)} \,,\quad\quad q^0 \xi^0 \,,
\eeq
in the two expansions, respectively. This implies that the maximal $\xi$-power for the series expansion of the product obeys
\be
r_{\rm max}(n) \geq r_{\rm max}^{(1)}(n) \geq \sqrt{4\,m_1\,n}-1 \,.  
\ee 
Similarly, we also have
\be
r_{\rm max}(n) \geq r_{\rm max}^{(2)}(n) \geq \sqrt{4\,m_2\,n}-1 \,,  
\ee
and hence, eventually, the following inequality holds,
\be\label{rmax_prod}
r_{\rm max}(n) \geq {\rm max}(r_{\rm max}^{(1)}(n), r_{\rm max}^{(2)}(n)) \geq \sqrt{4\,({\rm max}\{m_1,m_2\})\,n}-1 \,.
\ee
In formulating these bounds, we are assuming that no complete cancellations can occur amongst the maximal $\xi$-power terms at given $n$, arising from multiplying two monomials $q^{n_1}\xi^{r_1}$ and $q^{n_2}\xi^{r_2}$ with $n_1+n_2=n$, for the case where more than one maximal term exists. Such cancellations could in principle occur for specific values of $n$, spoiling the bound~\eqref{rmax_prod} for the $r_{\rm max}(n)$ of the product.
However, explicit analysis of the behaviour for various examples suggests that this coincidence does not happen in generic situations. 
On the other hand, we are aware of at least one instance where such a cancellation does happen for a specific model at a specific value of $n$, namely the model described by $({\rm x}, {\rm y})=(4,6)$ in Section~\ref{subsec_F1base}, at $n=1$, as explained in more detail in the main text. 

In general, given a generic linear combination of various products of Eisenstein-Jacobi and Eisenstein series of the form
\beq\label{gen-lc}
\sum_{I} c_I \prod_{i} E_{w_{I,i}, m_{I,i}} \prod_{i'}E_{w'_{I,i'}}\,,
\eeq
we must have
\bea\label{rmax-cons}
r_{\rm max}(n)   \geq \sqrt{4\,({\rm max}_{I}\{{\rm max}_{i}\{m_{I,i}\}\}) n} -  1\,,
\eea
where $m_{I,i}$ denote the indices of the $E_{w_{I,i},m_{I,i}}$ appearing in each product summand. 
In particular, the following very conservative bound  holds,
\be\label{rmax-verycons}
r_{\rm max}(n) \geq \sqrt{4\,n}-1 \,,
\ee
where we have used the fact that the indices of the Eisenstein-Jacobi series obey $m_{I,i} \geq 1$.
Again, all of this is true as long as there is no complete cancellation amongst individual contributions to the term $q^n \xi^r$. 

Now, it is interesting to observe that the bound~\eqref{rmax-cons} can  be hugely improved in the asymptotic regime as
\beq\label{estim2}
r_{\rm max}(n) \gtrsim \sqrt{4\,m\,n}\,,
\eeq 
where $m=\sum_i m_{I,i}$ is the index of any product of Eisenstein-Jacobi series. 
Once again, the symbol ``$\gtrsim$'' denotes an approximate asymptotic inequality in the sense defined in eqs.~\eqref{gtrsim1} and~\eqref{gtrsim2}, where the asymptotic bound~\eqref{estim2} is now claimed for a generic linear combination~\eqref{gen-lc}. To see how this improved bound arises, let us first note that the shift of $-1$ may be ignored in the asymptotic regime. Then, we have the obvious bound 
\beq\label{estim3}
r_{\max}(n) \gtrsim  {\rm max}\Big\{\sum_i \sqrt{4\,m_i\,n_i}\,\Big|\,\sum_i n_i = n, \,0\leq n_i\leq n, \,n_i \in \mathbb N_0\Big\}\,,
\eeq
provided that there are no cancellations of the aforementioned kind. Note that we are denoting $m_{I,i}$ by $m_i$ for simplicity. 
To estimate the RHS of~\eqref{estim3} we first consider the  function
\beq
f(\{\rho_i\})=\sum_i \sqrt{m_i \rho_i} \,,
\eeq
and find that the maximum of $f$ in the region $0\leq \rho_i \leq 1$ under the constraint $\sum_i \rho_i = 1$ is 
\beq
f_{\rm max}=\sqrt{\sum_im_i}=\sqrt{m}\,.
\eeq
One can see this by demanding that all the partial derivatives of $f$ with respect to $\rho_{i>1}$ vanish upon taking $\rho_1=1-\sum_{i>1}\rho_i$, that is,
\beq
0=\frac{\partial f}{\partial \rho_i} = -\frac{\sqrt{m_1}}{2\sqrt{\rho_1}} + \frac{\sqrt{m_i}}{2\sqrt{\rho_i}}\,,\quad i>1\,.
\eeq
This means that an extremum, which turns out to be the maximum, occurs at $\rho_i=\frac{m_i}{m}$ with the extremal value
\beq
f_{\rm max}=\sum_i \sqrt{\frac{m_i^2}{m}} = \frac{\sum_i m_i}{\sqrt{m}}=\sqrt{m}\,,
\eeq
as claimed. Since $\sum_i\sqrt{4\,m_i\,n_i} = \sqrt{4n}\sum_i \sqrt{m_i\,\frac{n_i}{n}}=\sqrt{4n} f(\{\frac{n_i}{n}\})$, we may in turn express~\eqref{estim3} as 
\beq\label{better-bound}
r_{\rm max}(n) \gtrsim {\rm max} \Big\{\sqrt{4n} f(\{\rho_i\})\,\Big|\, \sum_i \rho_i=1,\,0\leq \rho_i \leq 1,\, \rho_i n \in \mathbb N_0 \Big\} \gtrsim \sqrt{4n} f_{\rm max} = \sqrt{4\,m\,n} \,.
\eeq
In the first step Eq.~\eqref{estim3} has been rewritten and in the second step we have neglected the fact that $\rho_i n \in \mathbb N_0$, thereby obtaining an approximate asymptotic inequality. The latter inequality actually saturates for $n$'s that are an integer multiple of $m$ but in general, it only works as an asymptotic bound.

Note finally that when we multiply a (product or linear combination of products of) Eisenstein-Jacobi series by a modular function $f(\tau) = a_n q^n$, the conservative bounds~\eqref{rmax-cons} and~\eqref{rmax-verycons}, as well as the improved, asymptotic bound~\eqref{better-bound}, continue to hold as long as $a_0 \neq 0$.  If the expansion of the function $f(\tau)$ starts with a value of $ - n_0 < 0$, then the bounds get even more improved. For example, for the conservative bound~\eqref{rmax-cons} we have the following improvement,
\beq
r_{\rm max}(n)   \gtrsim \sqrt{4\,({\rm max}_{I}\{{\rm max}_{i}\{m_{I,i}\}\}) (n+n_0)} -  1 \geq \sqrt{4\,({\rm max}_{I}\{{\rm max}_{i}\{m_{I,i}\}\}) n} -  1\,,
\eeq 
because of a shift of all levels by $n_0>0$, and similarly, the other bounds also get improved.  All of this again assumes that no cancellations happen at a given
 level, $n$.

%-------------------------------------------------------------------------------------------------------------

\section{Analysis of $U(1)$ Models with elliptic bases $\mathbb F_1$ and  $\mathbb F_2$} \label{sec_HirzebruchF2}

In this appendix we collect some mathematical details that are needed for the computation of the Gromow-Witten invariants
via mirror symmetry, which in turn will determine the $U(1)$-refined
 elliptic genera for the bases $B_2 = \mathbb F_a$ with $a=1,2$.
The actual computations proceed along the steps described in Section~\ref{subsec_F1base}.

\subsection{Hirzebruch base $\mathbb F_1$  with extra section, for $({\rm x}, {\rm y})=(4,4)$} \label{app_44F1}

We begin by providing the geometric details underlying the results (\ref{F0-expansion-numbers-1B}) for the computation of the lowest-lying Gromov-Witten invariants of the $U(1)$ fibration with $({\rm x}, {\rm y})=(4,4)$ over the base space $\mathbb F_1$.

To compute the Gromov-Witten invariants $N_{C_0}^{(0)}(n,r)$ for some  low values of~$n$, we employ the machinery of mirror symmetry for Calabi-Yau three-folds in the framework of toric geometry~\cite{Hosono:1993qy,Hosono:1994ax}. The required input are the Mori cone generators and the triple intersection numbers, both of which can easily be obtained via PALP~\cite{Kreuzer:2002uu,Braun:2012vh}. Based on the toric data in Table~\ref{tb:F_a}
one finds three possible triangulations, one of which is compatible with a flat elliptic fibration.\footnote{A necessary condition for this is that there must exist a $2\times4$ block of zeros in the $4\times8$ Mori cone matrix since the two fibral curves should not intersect with the four toric divisors that are pulled back from the four base toric divisors. This criterion singles out the phase presented here.}
The Mori cone and the intersection numbers associated with this phase are as follows:
\beq
\begin{array}{lclrrrrrrrrrr|}
l^{(1)}&=&(&0,& 0,& -1,& 0,& 0,& 0,& 1,& 1) \\
l^{(2)}&=&(&1,& 0,& 0,& 1,& 0,& 0,& 0,& -2) \\
l^{(3)}&=&(&0,& 1,& 0,& -1,& -1,& 1,& 0,& 0) \\
l^{(4)}&=&(&0,& 0,& 2,& 0,& 1,& 0,& 0,& -1)
\end{array}
\eeq
\bea
\cI&=&-3 J_1^2 J_3+3 J_1 J_2 J_3+3 J_1 J_3^2+14 J_2 J_3^2+80 J_3^3-J_1^2 J_4\\ \nn
&&+J_1 J_2 J_4-J_1 J_4^2-2 J_2 J_4^2+8 J_4^3 \,.
\eea
Here, the Mori cone generators $l^{(i)}$ are described in terms of their intersection numbers with the $8$ toric divisors $d_i = \{{\nu_i} = 0\}$, with the classes of $\nu_i$ given in Table ~\ref{tb:F_a}.
The toric divisors are in turn expressible via the $4$ basis elements $J_i$ of $H^{1,1}(X, \IZ)$  as
\beq 
\begin{array}{lllll|}
d_1=J_1+J_2 \,,&
d_2=J_2\,,&  
d_3=-4J_1-4J_2+J_3-2J_4\,,&~ 
d_4=J_1\,,\\
d_5=-2J_1-3J_2+J_3-J_4\,,& d_6=J_2\,,& d_7=J_3\,,& d_8=J_4 \,. 
\end{array}
\eeq

This data is sufficient to obtain \cite{Hosono:1993qy,Hosono:1994ax} the low-degree Gromov-Witten invariants in the basis of curves $l^{(i)}$. In order to extract the relevant invariants $N_{C_0}^{(0)}(n,r)$ from them, we need to express $\Gamma_{C_0}(n,r)=C_0 + n C_\mathcal{E} + r C^{\rm f}$ in terms of $l^{(i)}$. The correct identification is
\beq
C_0 = l^{(2)} \,,\quad
C_\mathcal{E} =3 l^{(1)} + 2 l^{(4)} \,,\quad
C^{\rm f} =l^{(1)}+ l^{(4)} \,,
\eeq
leading to 
\beq
\Gamma_{C_0}(n,r)= (3n+r) l^{(1)} + l^{(2)} + (2n + r)l^{(4)} \,.
\eeq
Using standard methods of mirror symmetry, the lowest expansion terms of the genus-zero prepotential~\eqref{F0-expansion} can then 
be computed with the result as given in (\ref{F0-expansion-numbers-1B}). 

%---------------------------------------------------------------------------------------------------

\subsection{Hirzebruch base $\mathbb F_1$  with extra section, for $({\rm x}, {\rm y})=(4,6)$} \label{app_46F1}

The elliptic fibration with $({\rm x}, {\rm y})=(4,6)$ over $\mathbb F_1$ admits two inequivalent triangulations, one of which is compatible with the existence of a flat fibration.
The Mori cone and the intersection numbers of this topological phase  are 
\beq
\begin{array}{lclrrrrrrrrrr|}
l^{(1)}&=&(&0,& -1,& {\color{white}+}0,& {\color{white}+}0,& {\color{white}+}0,& {\color{white}+}0,& {\color{white}+}1,& {\color{white}+}1) \\
l^{(2)}&=&(&1,& 0,& -1,& 1,& 0,& 0,& 0,& -1) \\
l^{(3)}&=&(&0,& 0,& 1,& 0,& 1,& 0,& 0,& -2) \\
l^{(4)}&=&(&0,& 2,& 0,& 0,& 0,& 1,& 0,& -1)
\end{array}
\eeq
\bea
\cI&=&2 J_1^2 J_2 + 6 J_1 J_2^2 + 16 J_2^3 + 3 J_1^2 J_3 + 12 J_1 J_2 J_3 + 
 32 J_2^2 J_3 \\ \nn 
 &&~+ 21 J_1 J_3^2
 + 64 J_2 J_3^2 + 120 J_3^3 + J_1^2 J_4 - 
 3 J_1 J_4^2 + 8 J_4^3 \,.
\eea
The Mori cone generators $l^{(i)}$ are described in terms of their intersection numbers with the $8$ toric divisors $d_i = \{{\nu_i} = 0\}$
\beq 
\begin{array}{lllll|}
d_1=-2 J_1-J_2+J_3-J_4 \,,&
d_2=2 J_2-J_3\,,&  
d_3=3 J_1+J_2-J_3+J_4\,,& 
d_4=-2 J_1-J_2+J_3-J_4\,,\\
d_5=J_1\,,& d_6=J_2\,,& d_7=J_3\,,& d_8=J_4 \,. 
\end{array}
\eeq
The correct identification with the base curve $C_0$ as well as the fibral curves (\ref{intfibrala1}) and  (\ref{intfibrala2})  is
\beq
C_0 = l^{(3)} \,,\quad
C_\mathcal{E} =3 l^{(1)} + 2 l^{(4)} \,,\quad
C^{\rm f} =l^{(1)}+ l^{(4)} \,,
\eeq
leading to 
\beq
\Gamma_{C_0}(n,r)= (3n+r) l^{(1)} + l^{(3)} + (2n + r)l^{(4)} \,.
\eeq
This data in turn determines the lowest order Gromov-Witten invariants as given in (\ref{F0-expansion-numbers}).

%------------------------------------------------------------------------------------
\subsection{Hirzebruch base $\mathbb F_2$  with extra section, for $({\rm x}, {\rm y})=(2,4)$}   \label{app_24F2}

%-------------------------------------------------------------------------------
\begin{figure}[t!]
\centering
\includegraphics[width=12cm] {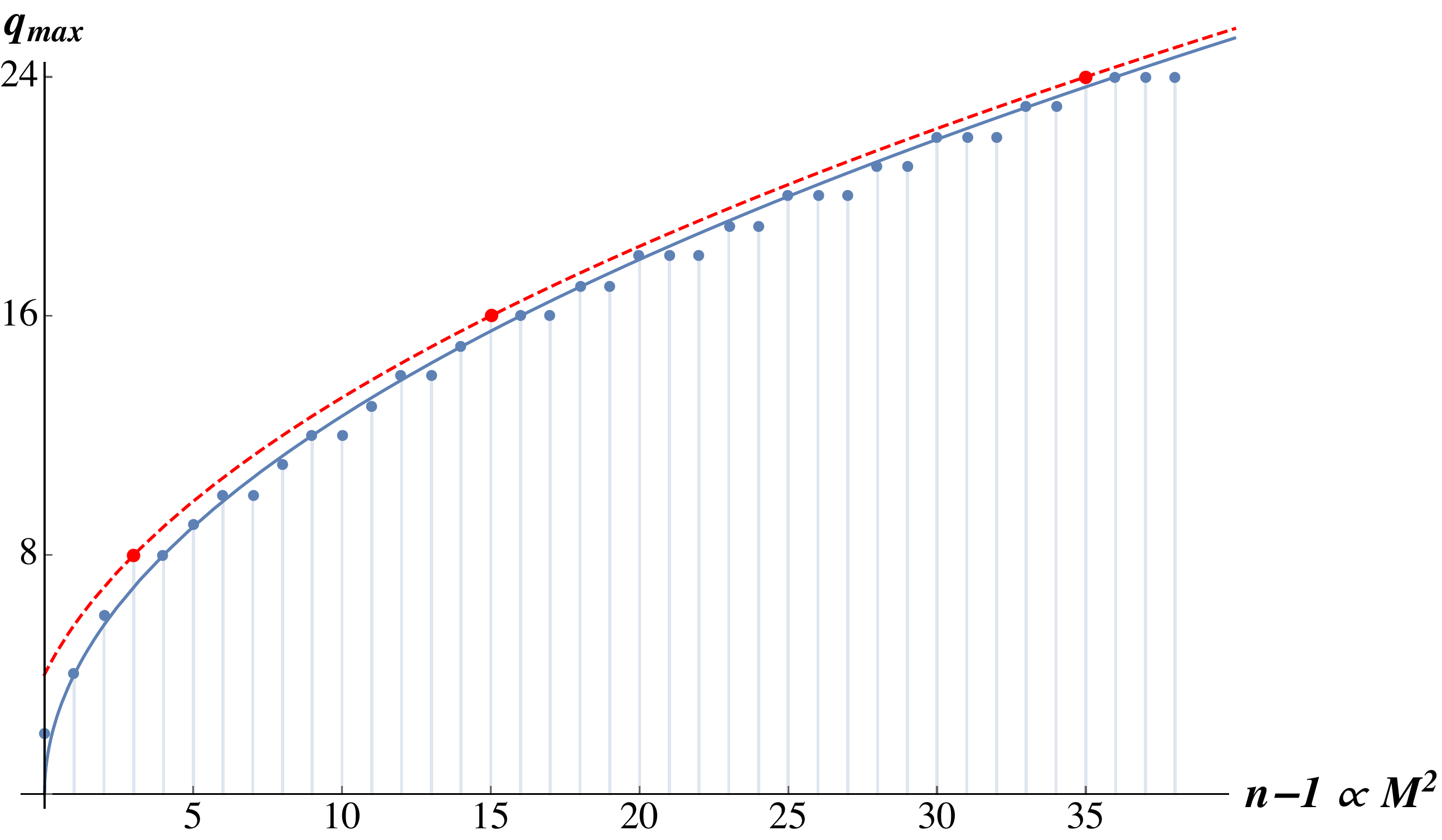}%{newF2.jpg} 
\caption{Maximal charge $\frak{q}_{\rm max}(n)$ per excitation level for ${\rm Bl}_1\mathbb P^2_{112}[4]$ fibration over $\mathbb F_2$ with $(\rm x, \rm y)=(2,4)$. 
The maximally superextremal  states form a sublattice of spacing $2m=8$.  
}\label{f:plot_2B_F2_100}
\end{figure}
%-------------------------------------------------------------------------------

Following the analogous procedure as in Sections~\ref{app_44F1} and~\ref{app_46F1},
we obtain the following expansion of the genus-zero generation function
for an $U(1)$ elliptic fibration with $({\rm x}, {\rm y})=(2,4)$ over $\mathbb F_2$:
 \bea\label{F0-expansion-numbers-2B}
\mathcal F^{(0)}_{C_0}  &=& -2 + 
%\left(\frac{16}{\xi^2}+\frac{128}{\xi} + 192 + 128 \xi + 16 \xi^2 \right) q   
 %\ \   %\\
%&&~+ \left(\frac{228}{\xi^4} + \frac{4992}{\xi^3} + \frac{26880}{\xi^2} + \frac{65664}{\xi} + 87360  + {65664} \xi + 26880 \xi^2 + 4992 \xi^3 +228 \xi^4\right) q^2 + \cO(q^3) \,.\nn
\left( 192 + 128 \xi^{\pm 1} + 16 \xi^{\pm 2} \right) q   
 \\
&&~+ \left(87360  + {65664} \xi^{\pm 1} + 26880 \xi^{\pm 2} + 4992 \xi^{\pm 3} +228 \xi^{\pm 4}\right) q^2 + \cO(q^3) \,.\nn
\eea

The fugacity index of the generating function is computed as
\beq
m=6-{\rm x}=4\,,
\eeq
so that the general ansatz for $\mathcal F^{(0)}_{C_0}$ is
\be
\mathcal F^{(0)}_{C_0} =  - \frac{q}{\eta^{24}}   \Phi_{10,4} \,,
\ee
where $\Phi_{10,4}$ is a weak Jacobi form of weight $10$ and index $4$. 
It is thus expanded as a general linear combination of suitable monomials $E^{a_1}_4 E_6^{a_2} \varphi^{a_3}_{0,1}  \varphi^{a_4}_{-2,1} $ with $4 a_1 + 6 a_2 - 2 a_4 = 10$ and $a_3 + a_4 =4$ for $a_i \geq 0$.
Upon requiring that the GW invariants in~\eqref{F0-expansion-numbers-2B} be obtained for order $n=0,1$ in $q$, the coefficients in the linear combination ansatz are fixed and the full generating function is found as follows:
\bea
\mathcal F^{(0)}_{C_0} &=& -q\,Z_\otherK3(\tau,z) \\
&=&\frac{q}{\eta^{24}} \Big(- \frac{1}{31104} E_4^3 E_6 \varphi_{-2,1}^4 - \frac{1}{15552}E_6^3 \varphi_{-2,1}^4 + \frac{1}{10368} E_4^4 \varphi_{-2,1}^3 \varphi_{0,1} + \frac{1}{3456} E_4 E_6^2 \varphi_{-2,1}^3 \varphi_{0,1} \nn\\ 
&&- \frac{1}{1728} E_4^2 E_6 \varphi_{-2,1}^2 \varphi_{0,1}^2 + \frac{7}{31104} E_4^3 \varphi_{-2,1} \varphi_{0,1}^3 + \frac{5}{31104} E_6^2 \varphi_{-2,1} \varphi_{0,1}^3 -\frac{1}{10368} E_4 E_6 \varphi_{0,1}^4 \Big)  \,.\nn
\eea
The maximal charges per excitation level are plotted in Figure \ref{f:plot_2B_F2_100}, which again confirms  the expectations based on the Sublattice Weak Gravity Conjecture.

%-------------------------------------------------------------------------------
\subsection{Hirzebruch base $\mathbb F_2$  with extra section, for $({\rm x}, {\rm y})=(4,8)$}  \label{app_48F2}

The fibration with $({\rm x}, {\rm y})=(4,8)$ over $\mathbb F_2$ gives rise to a non-generic model in which $[b_2]=0$ and hence cancellations at the level of states with $n=1$, $\frak{q}=2$ are expected, in line with the general discussion of Section~\ref{subsec_F1base}.
Indeed, the expansion of the generation function up to order $n_0=2$ in $q$ is found as follows:
\bea\label{F0-expansion-numbers-2A}
% \mathcal F^{(0)}_{C_0} &=&-2 + \left(\frac{96}{\xi} + 288 + 96 \xi\right) q \\ 
% &&~+ \left(- \frac{2}{\xi^4} + \frac{96}{\xi^3} + \frac{10192}{\xi^2} + \frac{69280}{\xi} + 123756  + {69280} \xi + 10192 \xi^2 + 96 \xi^3 - 2 \xi^4\right) q^2 + \cO(q^3) \,,\nn
% \eea
\mathcal F^{(0)}_{C_0} &=&-2 + \left(288 + 96 \xi^{\pm 1}\right) q \\ 
&&~+ \left(123756  + {69280} \xi^{\pm 1} + 10192 \xi^{\pm 2} + 96 \xi^{\pm 3} - 2 \xi^{\pm 4}\right) q^2 + \cO(q^3) \,,\nn
\eea
exhibiting, as expected, no states at $n=1$, $\frak{q}=2$.  This is exactly the same expression as for the model with $({\rm x}, {\rm y})=(4,6)$  on $\mathbb F_1$ presented in Section~\ref{subsec_F1base}, so the same conclusions apply.

}
\end{appendix}

\newpage
\bibliography{papers}
\bibliographystyle{JHEP}

\end{document}